\documentclass[aps,preprint,prl,groupedaddress,showpacs,amsmath,amssymb]{revtex4-1}
\usepackage[latin9]{inputenc}
\usepackage{times}
\usepackage{comment}
\usepackage{textcomp}
\usepackage{amstext}
\usepackage{graphicx}
\usepackage{hyperref}
\usepackage[dvipsnames]{xcolor}
\usepackage{CJK}
\usepackage{hyperref}
\newcommand{\ket}[1]{\left\vert{#1}\right\rangle}
\newcommand{\bra}[1]{\left\langle{#1}\right\vert}

\newcommand{\ve}{\boldsymbol}
\makeatletter

\newcommand{\lyxmathsym}[1]{\ifmmode\begingroup\def\b@ld{bold}
  \text{\ifx\math@version\b@ld\bfseries\fi#1}\endgroup\else#1\fi}

\usepackage{float}% Include figure files
\usepackage[normalem]{ulem}
\usepackage{color}
\renewcommand{\ve}{\boldsymbol}
\makeatother

\begin{document}
\begin{CJK*}{UTF8}{fs} 
\begin{abstract}
{\bf  Magnetic monopoles play a central role in various areas of fundamental physics, ranging from electromagnetism to topological states of matter. 
  While their observation is elusive in high-energy physics, monopole sources of artificial gauge fields have been recently identified in synthetic matter.
  String theory, a potentially unifying framework that encompasses quantum mechanics, promotes the conventional \emph{vector} gauge fields of electrodynamics to \emph{tensor} gauge fields, and predicts the existence of more exotic \emph{tensor monopoles} in 4D space. 
  Here we report on the characterization of a tensor monopole synthesized in a 4D parameter space by the spin degrees of freedom of a single solid-state defect in diamond. 
  Using two complementary methods, we characterize the tensor monopole by measuring its quantized topological charge and its emanating Kalb-Ramond field.
  By introducing a fictitious external field that breaks chiral symmetry, we further observe an intriguing transition in the spectrum, characterized by spectral rings protected by mirror symmetries. Our work represents the first detection of tensor monopoles in a solid-state system and opens up the possibility of emulating exotic topological structures inspired by string theory.}
\end{abstract}

\title{A synthetic monopole source of Kalb-Ramond field in diamond}

\author
{Mo Chen \CJKfamily{gbsn}(陈墨),$^{1,2,}$\footnote{Present address: Institute for Quantum Information and Matter and Thomas J. Watson, Sr., Laboratory of Applied Physics, California Institute of Technology, Pasadena, CA 91125, USA} Changhao Li,$^{1,3,}$\footnote{MC and CL contributed equally to this work.} Giandomenico Palumbo,$^{4,6}$ Yan-Qing Zhu,$^4$ Nathan Goldman,$^4$ and Paola Cappellaro$^{1,3,5,}$\footnote{pcappell@mit.edu}}
\affiliation{$^{1}$Research Laboratory of Electronics, Massachusetts Institute of Technology, Cambridge, MA 02139, USA}
\affiliation{$^{2}$Department of Mechanical Engineering, Massachusetts Institute of Technology, Cambridge, MA 02139, USA}
\affiliation{$^{3}$Department of Nuclear Science and Engineering, Massachusetts Institute of Technology, Cambridge, MA 02139, USA}
\affiliation{$^{4}$Center for Nonlinear Phenomena and Complex Systems, Universit\'e Libre de Bruxelles, CP 231, Campus Plaine, B-1050 Brussels, Belgium}
\affiliation{$^{5}$Department of Physics, Massachusetts Institute of Technology, Cambridge, MA 02139, USA}
\affiliation{$^6$School of Theoretical Physics, Dublin Institute for Advanced Studies, 10 Burlington Road, Dublin 4, Ireland}

\maketitle

\end{CJK*}
%----------------ABSTRACT----------------------------------
%\pacs{}

\section*{Introduction}
Our current understanding of fundamental physical phenomena relies on two main pillars, general relativity and quantum field theory. Their mutual incompatibility, however, poses critical limitations to the formulation of a unifying theory of all fundamental interactions. String theory proposes a powerful and elegant formalism to unify gravitational and quantum phenomena, providing a concrete route to quantum gravity~\cite{Zwiebach2004}. Within this scenario, conventional point-like particles are replaced by extended objects, such as closed and open strings, and conventional vector gauge fields are promoted to tensor (Kalb-Ramond) gauge fields~\cite{Kalb1974,Henneaux1986}. In direct analogy with the Dirac monopole~\cite{Dirac1931}, tensor gauge fields can emanate from point-like defects called tensor monopoles. In four spatial dimensions, the tensor monopole charge is quantized according to the topological Dixmier-Douady ($\mathcal{DD}$) invariant~\cite{Bouwknegt2000,Nepomechie85,Orland1982}, which generalizes the Chern number associated with the Dirac monopole.

Experimental evidence of magnetic monopoles is still lacking in high-energy physics experiments. However, synthetic monopoles associated with effective gauge fields have recently been detected in ultracold matter~\cite{Schroer2014,Rey2014,Roushan2014,Sugawa2018,Yu2019,Tan19}. Besides, momentum-space monopoles play a central role in topological matter, in particular, to characterize 3D Weyl semimetals. Recently, the notions of tensor monopoles and $\mathcal{DD}$ invariants were shown to arise in 3D chiral topological insulators~\cite{Palumbo2018,Palumbo2019} and in higher-order topological insulators~\cite{Dubinkin2020}.

Here, we exploit the high controllability of engineered qutrit systems in view of revealing exotic gauge structures, originally introduced in the context of string theory~\cite{Kalb1974,Henneaux1986}. We consider the spin triplet ground state of a single NV center, which can be mapped into a 3-level Weyl-type Hamiltonian $\hat H_{4D}(q_x,q_y,q_z,q_w)$ defined over a 4D parameter space
\begin{equation}\label{eq:Ham_parametrization}
\hat{H}_{4D}=
\begin{pmatrix}
0 &q_x-i q_y&0\\
q_x+i q_y&0&q_z+i q_w\\
0&q_z-i q_w&0
\end{pmatrix}.
\end{equation}
Here, the parameters $\ve{q}\!=\!(q_x,q_y,q_z,q_w)$ can be expressed in terms of the experimentally controllable parameters $(H_0, \alpha, \beta, \phi)$ explicated below, through $q_x\!+\!i q_y\!=\!H_0 \cos(\alpha)e^{i \beta}$, $q_z\!+\!i q_w\!=\!H_0 \sin(\alpha)e^{i\phi}$, where $\alpha \in [0,\pi/2]$ and $\beta, \phi \in [0,2\pi)$. This Hamiltonian hosts a three-fold degenerate point in the spectrum, located at the origin $\ve{q}\!=\!0$. This singularity is topologically protected by chiral symmetry $\{\hat{H}_{4D}, U\}=0$, where $U=\textrm{diag}(1,-1,1)$, and is a good candidate for a synthetic monopole source of tensor gauge field, as we now explain.

First recall that a nodal point in a 3D parameter space is associated with an effective Dirac monopole~\cite{Palumbo2018,Yu2019,Tan19}. In this scenario, the Berry-curvature field emanates radially from the node, and its flux through a 2-sphere enclosing it is quantized, characterized by the Chern number~\cite{WuPRD1975}. 
In 4D space, the topological charge associated with a nodal point is provided by a similar invariant, which now involves the flux of a radial 3-form curvature over a 3-sphere that surrounds the node~\cite{Palumbo2018}.
The 3-form curvature $\mathcal{H}_{\mu\nu \lambda}$ is well known in the context of p-form electromagnetism~\cite{Henneaux1986}, where it derives from a 2-form gauge field:~the Abelian and antisymmetric Kalb-Ramond (KR) field $B_{\mu \nu}$~\cite{Kalb1974},
\begin{equation}\label{eq:relation_H_B}
    \mathcal{H}_{\mu\nu \lambda} = \partial_\mu B_{\nu \lambda} + \partial_\nu B_{\lambda \mu}+\partial_\lambda B_{\mu \nu}.
    \end{equation}
This KR field plays an important role in string theory, as it naturally couples to extended objects~\cite{Kalb1974,Henneaux1986}.

Similarly to monopoles associated with \textit{vector gauge fields} in 3D space, the tensor KR field $B_{\mu\nu}$ gives rise to \textit{tensor} monopoles with distinct topological properties~\cite{Orland1982,Nepomechie85,Bouwknegt2000,Palumbo2018,Palumbo2019,Dubinkin2020}.  
These exotic monopoles are point-like sources of the generalized ``magnetic'' field $\mathcal{H}_{\mu\nu \lambda}$, and their topological charge is obtained by measuring the corresponding flux over a 3-sphere surrounding them,
\begin{equation}\label{eq:Tensor_Monopole}
    \mathcal{DD}=\frac{1}{2\pi^2}\int_{S^3} \mathcal{H}_{\mu\nu\lambda} \, dq^\mu \wedge dq^\nu\wedge dq^\lambda .
    \end{equation}
This topological invariant is known as the Dixmier-Douady ($\mathcal{DD}$) invariant~\cite{Orland1982,Nepomechie85,Bouwknegt2000,Palumbo2018,Palumbo2019,Dubinkin2020}, and generalizes the well-known Chern number. We point out that the field $\mathcal{H}_{\mu\nu\lambda}$ radially emanates from the singularity in 4D space, hence providing an observable and unambiguous signature of tensor monopoles.

We now explain how these exotic gauge structures can be measured in engineered systems. First, we note that 
the KR field $B_{\mu\nu}$ in Eq.~\eqref{eq:relation_H_B} can be reconstructed from the eigenstates of the Hamiltonian in Eq.~\eqref{eq:Ham_parametrization}~\cite{Palumbo2018,Palumbo2019}. While state-tomography could be performed to reconstruct these states and the related tensor fields, this approach is resource-intensive. Here we provide two alternative methods to experimentally measure the curvature $\mathcal H_{\mu\nu\lambda}$. The first approach builds on a relation between the 3-form curvature and the Fubini-Study quantum metric $g_{\mu\nu}$~\cite{Palumbo2018,Palumbo2019}
\begin{equation}\label{eq:relation_g_H}
    \mathcal{H}_{\mu\nu \lambda} = \epsilon_{\mu \nu \lambda}(4\sqrt{\det(g_{\bar{\mu} \bar{\nu}}})),
    \end{equation}
where $\epsilon_{\mu\nu \lambda}$ is the Levi-Civita symbol, and $\bar{\mu}, \bar{\nu} = \{q_x,q_y,q_z \}$ for $\mathcal{H}_{xyz}$ (and similarly for the other components). The metric tensor $g_{\mu\nu}$, which defines the distance between nearby states $\ket{u(\ve{q})}, \ket{u(\ve{q}+d\ve{q})}$~\cite{Ozawa2018,Yu2019,Tan19,Kolodrubetz2017,SOM}, thus allows for a measurement of the tensor monopole field.
The second approach builds on our experimental parametrization $(H_0, \alpha, \beta, \phi)$, which  expresses the 3-form curvature in Eq.~\eqref{eq:relation_H_B} as~\cite{SOM}
    \begin{equation}\label{eq: H_fromConnection}
    \mathcal{H}_{\alpha\beta\phi}=-\frac{1}{2}(\mathcal{F}_{\alpha\beta}+\mathcal{F}_{\phi\alpha}),
    \end{equation}
where $\mathcal{F}_{\mu\nu}$ is the standard (2-form) Berry curvature. We will refer to the latter as the Berry curvature, not to be confused with the 3-form curvature $\mathcal{H}_{\mu\nu\lambda}$.
Importantly, both the metric tensor $g_{\mu\nu}$ in Eq.~\eqref{eq:relation_g_H} and the Berry curvature  $\mathcal{F}_{\mu\nu}$ in Eq.~\eqref{eq: H_fromConnection} can be experimentally extracted from spectroscopic responses upon modulating the parameters $\mu,\nu$~\cite{Ozawa2018,Yu2019,Tan19,SOM}.

In our experiment, we  exploit these two different probes of the 3-form curvature to demonstrate two unique signatures of the tensor monopole field:~its quantized topological charge and its characteristic radial behavior in 4D parameter space.

\section*{Results}
\subsection*{Engineering the tensor monopole}

To synthesize the 4D Hamiltonian in Eq.~\eqref{eq:Ham_parametrization}, we use the ground triplet states of a single NV center in diamond
at room temperature. An external magnetic field, $B=490$G, is applied along the N-V axis to lift the degeneracy between $\ket{m_s=\pm 1}$. At this magnetic field, optical illumination of a 532nm laser polarizes both the NV electronic spin and the native $^{14}$N nuclear spin via polarization transfer in the excited-state~\cite{Jacques09}. Hence, we can restrict ourselves to the electronic part of the Hamiltonian in the following.
We apply a dual-frequency microwave pulse~\cite{Mamin2015}, on-resonance with the $\ket{m_s=0}\leftrightarrow\ket{m_s=\pm 1}$ transitions.
In the doubly rotating frame, and upon the rotating wave approximation, we reproduce the minimal tensor monopole model in Eq.~\eqref{eq:Ham_parametrization}, where $\beta, \phi$ are the phases of the two microwave tones, and $H_0\cos\alpha, H_0\sin\alpha$ their corresponding microwave amplitudes~\cite{SOM}.

This Hamiltonian has three eigenstates $\ket{u_{\{-,0,+\}}}$, 
with eigenvalues $\{\epsilon_-,\epsilon_0,\epsilon_+\}$ in ascending order. 
Precise modulations of the microwave frequencies, amplitudes and phases grant us full access to the 4D parameter space spanned by $(H_0, \alpha, \beta, \phi)$. 
Using this parametrization, we note that the system is rotationally symmetric about $\beta, \phi$. Therefore, the measurable geometric quantities (the metric tensor and Berry curvature) are independent of $\beta, \phi$,
\begin{equation}
g+i\frac{\mathcal{F}}{2}\!=\!\begin{pmatrix}\renewcommand{\arraystretch}{1.6}
 \frac{1}{2} & \frac{i\sin(2\alpha )}{4} & -\frac{i\sin(2\alpha )}{4} \\
 -\frac{i\sin(2\alpha )}{4} & \frac{\cos^2(\alpha)[2-\cos^2(\alpha)]}{4}   & -\frac{\sin ^2(2\alpha )}{16} \\
 \frac{i\sin(2\alpha )}{4} & -\frac{\sin ^2(2\alpha )}{16} & \frac{\sin^2(\alpha)[2-\sin^2(\alpha)]}{4} 
\end{pmatrix}.
\end{equation}

As a demonstration of our engineered system, we initialize the NV in the $\ket{m_s=0}$ state and let it evolve under the target Hamiltonian (with  $\beta\!=\!\phi\!=\!0$). We further choose the microwave amplitudes such that
 the parameters span a hypersphere with fixed radius $H_0\!=\!2$~MHz,  which encloses the tensor monopole at the origin.
For various values of $\alpha$, the resulting dynamics of all three states show excellent agreement with theory, as shown in 
Fig.~S6 of the Supplementary Material~\cite{SOM}.

\subsection*{Measurements through parametric modulations}

We now show how to measure the quantum metric tensor and Berry curvature using weak modulations of the parameters $\mu, \nu\in\{\alpha,\beta,\phi\}$~\cite{Ozawa2018,Yu2019,Tan19}. Considering the modulations $\mu (t)=\mu_0+m_\mu \sin(\omega t + \gamma)$, $\nu (t)=\nu_0+m_\nu \sin(\omega t)$, with $m_\mu, m_\nu\ll 1$,  the Hamiltonian takes the form
\begin{equation}\label{eq:parametric_modulation}
\begin{split}
    \hat{H}\approx & \hat{H}(\alpha_0,\beta_0,\phi_0) + m_\mu \partial_\mu\hat{H}\sin(\omega t + \gamma) + m_\nu \partial_\nu\hat{H}\sin(\omega t).
\end{split}
\end{equation}
When $\gamma\!=\!0$ (resp.~$\pi/2$), we linearly (resp.~elliptically) modulate $\mu, \nu$ and extract the metric tensor (resp.~the Berry curvature), as we now explain.

The parametric modulations coherently drive Rabi oscillations between $\ket{u_-}\leftrightarrow\ket{u_0}$ ($\ket{u_-}\leftrightarrow\ket{u_+}$), when the modulation frequency is tuned on resonance with the energy gap between  ground and excited state, $\omega=\epsilon_0-\epsilon_{-}$ ($\omega=\epsilon_+-\epsilon_{-}$). 
We call the transitions $\ket{u_\pm}\leftrightarrow\ket{u_0}$ ``single quantum (SQ) transitions'', and  the transition $\ket{u_-}\leftrightarrow\ket{u_+}$ ``double quantum (DQ) transition'', following the change in quantum number.
Their Rabi frequencies are directly related to the transition matrix elements when varying one Hamiltonian parameter, $\Gamma_{-, n}^\mu=\vert\bra{u_-}\partial_\mu\hat{H}\ket{n}\vert$, or when  linearly (elliptically) modulating two parameters, $\Gamma_{-, n}^{\mu\nu (\mu\bar{\nu})}=\vert \bra{u_-}\partial_\mu\hat{H}\pm (i) \partial_\nu\hat{H}\ket{n}\vert$, $m_\mu=\pm m_\nu$ ~\cite{SOM}. Here the subscript for the matrix element, $\{-, n\}$, stands for the transition between eigenstates $\ket{u_-}\leftrightarrow \ket{n}$. Finally, we reconstruct the quantum metric tensor and Berry curvature from the relations~\cite{Yu2019,SOM}:
\begin{equation}\label{eq:g_from_Gamma}
\begin{split}
g_{\mu\mu}&=\sum_{n\neq -1}\frac{(\Gamma_{-, n}^\mu)^2}{(\epsilon_{-}-\epsilon_n)^2},\\
g_{\mu\nu}&=\sum_{n\neq -1}\frac{[(\Gamma_{-, n}^{\mu\nu})^2-(\Gamma_{-, n}^{\mu\bar{\nu}})^2]}{4(\epsilon_{-}-\epsilon_n)^2}\quad{}\textrm{(linear)},\\
\mathcal{F}_{\mu\nu}&=\sum_{n\neq -1}\frac{[(\Gamma_{-, n}^{\mu\nu})^2-(\Gamma_{-, n}^{\mu\bar{\nu}})^2]}{2(\epsilon_{-}-\epsilon_n)^2}\quad{}\textrm{(elliptical)}.
\end{split}
\end{equation}

In order to measure the quantum metric tensor and Berry curvature in the experiment, we first initialize the NV in the $\ket{m_s=0}$ state and coherently drive it to the ground eigenstate $\ket{u_-}$ of the Weyl-type Hamiltonian by two microwave pulses. 
The system is then subjected to the linear (elliptical) parametric modulations in Eq.~\eqref{eq:parametric_modulation}, which resonantly drives Rabi oscillations between eigenstates. Finally, either the $\ket{u_-}$ or $\ket{u_0}$ state is mapped back to $\ket{m_s=0}$ by microwave pulses, and optically read out~\cite{SOM}. 

\begin{figure*}[ht] 
    \centering
    \includegraphics[width=1.0\textwidth]{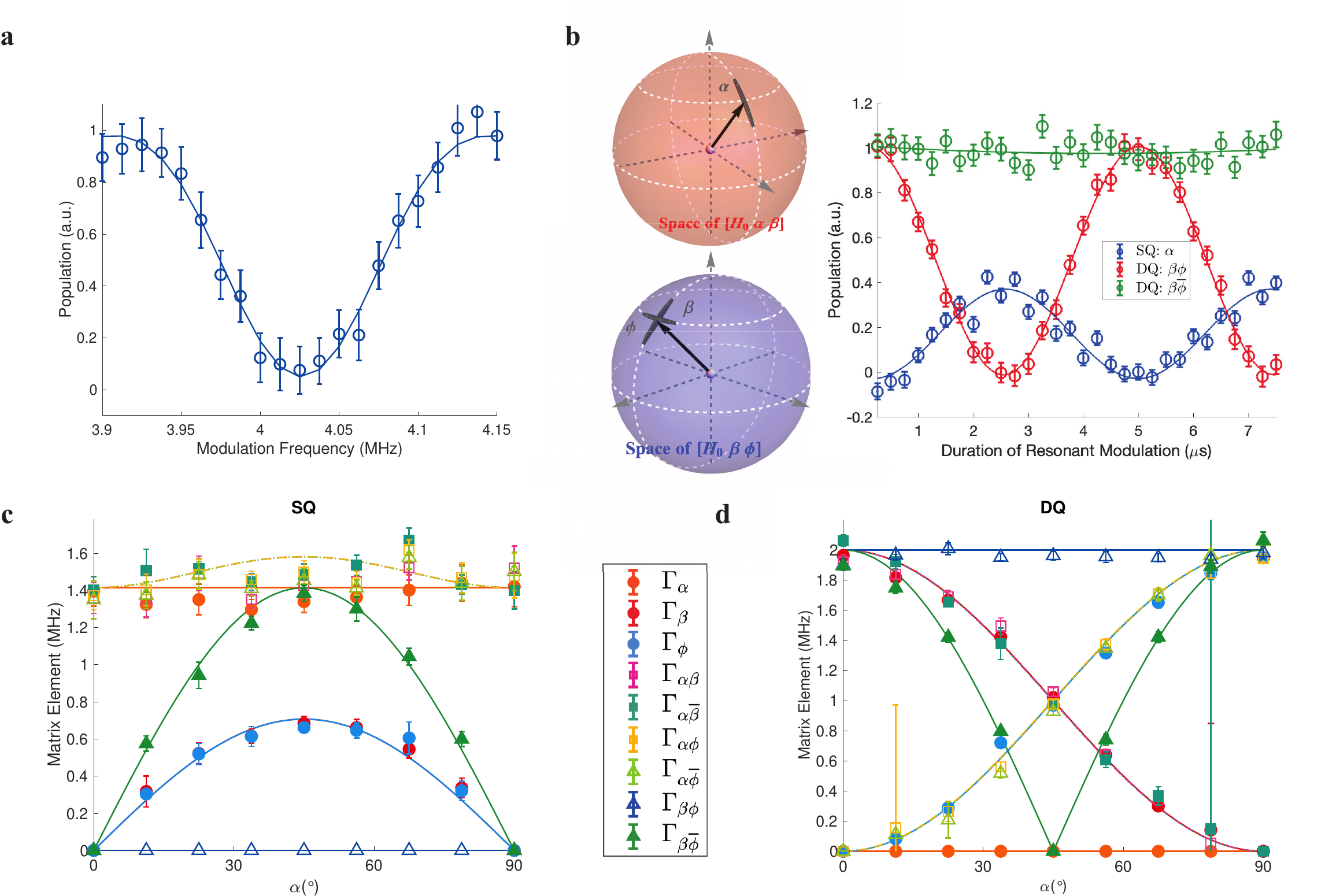}
    \caption{\textbf{Parametric modulations.} \textbf{(a)} Determining the resonance condition for parametric modulation. We fix $\tau=7.5\mu$s, $(m_\alpha, m_\beta, m_\phi)=(0, 1/30, 1/30)$ and sweep the modulation frequency around $4$MHz to find $\omega_r=2H_0$. \textbf{(b)} Examples of coherent Rabi oscillations observed under parametric modulations, for the engineered Hamiltonian at $(\alpha_0=\pi/4, \beta_0=\phi_0=0)$. The Rabi frequencies are used to calculate the matrix elements $\Gamma_{-, m}^{\mu(\nu)}$ (shown in \textbf{(c,d)}). To extract the diagonal components of the metric tensor, we use a single-parameter modulation, as shown, e.g. by the blue curve, representing the SQ transition ($\omega=\omega_r/2$) for $\alpha$ modulation. Due to chiral symmetry, $\vert \Gamma_{-, 0}^{\mu(\nu)}\vert = \vert \Gamma_{+, 0}^{\mu(\nu)}\vert$. We therefore measure the population in the first excited state $\ket{u_0}$, which gives half contrast~\cite{SOM}. The other two curves represent two-parameter modulations resonant with the DQ transition ($\omega=\omega_r$), and possess full contrast. Illustrations of the relevant single- and two-parameter modulations in the Bloch sphere representation are provided on the left. 
    \textbf{(c)} Matrix elements $\vert \Gamma_{-, 0}^{\mu(\nu)}\vert$ measured for SQ transitions at $\omega=\omega_r/2$. Note that many matrix elements are expected from theory to coincide and thus their measured values are superimposed at $\sqrt2$~MHz.   \textbf{(d)} Matrix elements $\vert \Gamma_{-, +}^{\mu(\nu)}\vert$ measured for DQ transitions at $\omega=\omega_r$. Markers are experimental data and solid lines are \textbf{(a,b)} fits and \textbf{(c,d)} theory.}
    \label{fig:fig2_temp}
\end{figure*}

We start our measurements by precisely determining the resonant frequency $\omega_r=\epsilon_{+}-\epsilon_{-}=2 (\epsilon_0-\epsilon_{-})$. As shown in Fig.~\ref{fig:fig2_temp}~(a), we fix the time and sweep the modulation frequency $\omega$ to find the resonance condition. A very weak modulation amplitude reduces power broadening and improves the precision in estimating $\omega_r$. 

We then measure the coherent Rabi oscillations under linear (elliptical) parametric modulations at the calibrated $\omega=\frac{\omega_r}{2}$ ($\omega_r)$ for SQ (DQ) transitions. Examples of SQ and DQ Rabi curves for the quantum-metric measurements are shown 
in Fig.~\ref{fig:fig2_temp}~(b) and Fig.~S8-S12, including both single- and two-parameter modulations for extracting the diagonal and off-diagonal components. We note that no decoherence effect was observed in these parametric modulations thanks to the long coherence time of the NV center. For every combination of modulations $\mu (\mu\nu)$, we measure both the SQ and DQ Rabi frequencies and recover the matrix element $\Gamma_{-, n}^\mu$ ($\Gamma_{-, n}^{\mu\nu}$). All measured matrix elements $\Gamma$ are plotted in Fig.~\ref{fig:fig2_temp}~(c,d) for the quantum-metric tensor and in Fig.~S14 for the Berry curvature, showing good agreement with theoretical predictions.

\subsection*{Revealing the tensor monopole}
As the main results of this work, we reconstruct both the quantum metric and the Berry curvature of our 4D setting, and use them as two complementary approaches to determine the 3-form curvature $\mathcal{H}_{\mu\nu\lambda}$ and its related monopole charge ($\mathcal{DD}$ invariant).

\begin{figure*}[ht] 
    \centering
     \includegraphics[width=0.9\textwidth]{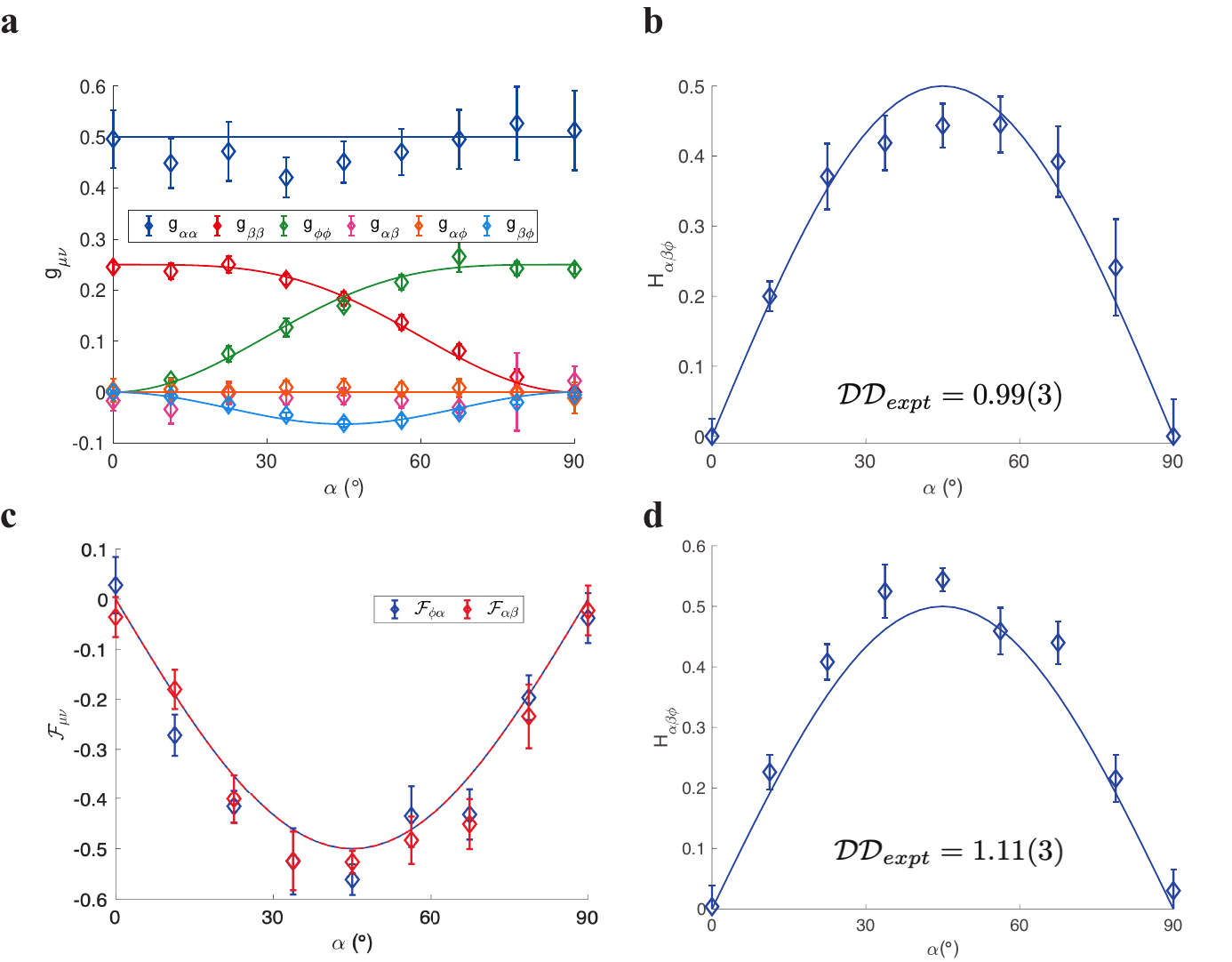}
    \caption{\textbf{Revealing the tensor monopole.} \textbf{(a)} shows all 6 independent components of the metric tensor as a function of $\alpha$. \textbf{(b)} Generalized 3-form curvature $\mathcal{H}_{\alpha\beta\phi}$ with respect to $\alpha$, calculated from the metric tensor in \textbf{(a)} using Eq.~\eqref{eq:relation_g_H}. The topological invariant $\mathcal{DD}_{expt}=0.99(3)$ reveals the existence of a tensor monopole within the hypersphere. \textbf{(c)} Measurements of non-zero Berry curvature as a function of $\alpha$. \textbf{(d)} 3-form curvature $\mathcal{H}_{\alpha\beta\phi}$, calculated from the gauge potential using Eq.~\eqref{eq: H_fromConnection}. The result $\mathcal{DD}_{expt}=1.11(3)$ further confirms the existence of the tensor monopole. Diamonds are experimental data and solid lines are theory. The errorbars are propagated from fitting error of resonant frequencies and Rabi oscillations. }
    \label{fig:g_H}
\end{figure*}

The independent components of the metric tensor, reconstructed using Eq.~\eqref{eq:g_from_Gamma}, are shown in Fig.~\ref{fig:g_H}~(a). The excellent agreement between theory and our experiment demonstrates our exquisite control over the 4D Weyl-type Hamiltonian in Eq.~\eqref{eq:Ham_parametrization}, providing precise information about the quantum geometry of the ground-state manifold.

We then connect the metric tensor to the 3-form curvature using Eq.~\eqref{eq:relation_g_H}. The measured 3-form curvature $\mathcal{H}_{\alpha\beta\phi}$ is shown in Fig.~\ref{fig:g_H}~(b). Using this experimental data, we obtain the quantization of the generalized ``magnetic'' flux over the 3-sphere,
\begin{equation}
\mathcal{DD}_{expt}=\frac{1}{2\pi^2}\int_0^{\frac{\pi}{2}}d\alpha\int_0^{2\pi}d\beta\int_0^{2\pi}d\phi\, \mathcal{H}_{\alpha\beta\phi}=0.99(3),
\end{equation}
which provides an estimation of the $\mathcal{DD}$ invariant in Eq.~\eqref{eq:Tensor_Monopole} and signals the presence of the tensor monopole at the center of our parameter space.

Alternatively, one can identify the tensor monopole via the Berry curvatures, $\mathcal F_{\mu\nu}$, using Eq.~\eqref{eq: H_fromConnection}.
We show the Berry curvature measured through elliptical parametric modulations in Fig.~\ref{fig:g_H}~(c) and the reconstructed 3-form curvature in Fig.~\ref{fig:g_H}~(d). This second approach, which is complementary to the metric-tensor measurement, further confirms the existence of the tensor monopole through the measurement of its quantized charge:
\begin{equation}
\mathcal{DD}_{expt}=\frac{1}{2\pi^2}\int_0^{\frac{\pi}{2}}d\alpha\int_0^{2\pi}d\beta\int_0^{2\pi}d\phi\, \mathcal{H}_{\alpha\beta\phi}=1.11(3).
\end{equation}

Besides its topological charge, the 4D tensor monopole is also fully characterized by its field distribution~\cite{Nepomechie85,Palumbo2018}
\begin{equation}
\mathcal{H}_{\mu \nu \lambda} (\ve{q}) = \epsilon_{\mu \nu \lambda \gamma} q_{\gamma}  \, / (q_x^2+q_y^2+q_z^2+q_w^2)^2,
\end{equation}
which reflects the fact that the curvature field radially emanates from the topological defect in 4D parameter space. As a consequence, the monopole field has a characteristic inverse-cube dependence on the radial coordinate, $\mathcal{H}\sim(1/H_0)^3$. We have verified this additional signature of the tensor monopole, through the experimental determination of the 3-form curvature distribution (Fig.~S15), hence fully confirming the existence of a 4D tensor monopole in our setting.

\subsection*{Transition to topological spectral rings}

We further explore a novel spectral transition, which can be induced by adding a longitudinal field  to the Weyl-type Hamiltonian~\footnote{The more conventional phase transition that is induced by translating the hypersphere is described in~\cite{SOM}}
\begin{equation}
    \hat{H}_{ST}=\hat{H}_{4D} + \textrm{diag}(B_z,0,-B_z)/\sqrt{2}.
\end{equation}
The field is achieved by detuning the dual-frequency microwave driving by equal and opposite amounts.
The field breaks the chiral symmetry but preserves mirror symmetries: $M_1 \hat{H}_{ST}(q_x,q_y,q_z,q_w) M_1^{-1}= \hat{H}_{ST}(-q_x,-q_y,q_z,q_w)$, $M_2 \hat{H}_{ST}(q_x,q_y,q_z,q_w) M_2^{-1}= \hat{H}_{ST}(q_x,q_y,-q_z,-q_w)$, with $M_1=diag(-1,1,1)$, $M_2=diag(1,1,-1)$, keeping the Hamiltonian gapless.

\begin{figure*}[ht] 
    \centering
   \includegraphics[width=1.0\textwidth]{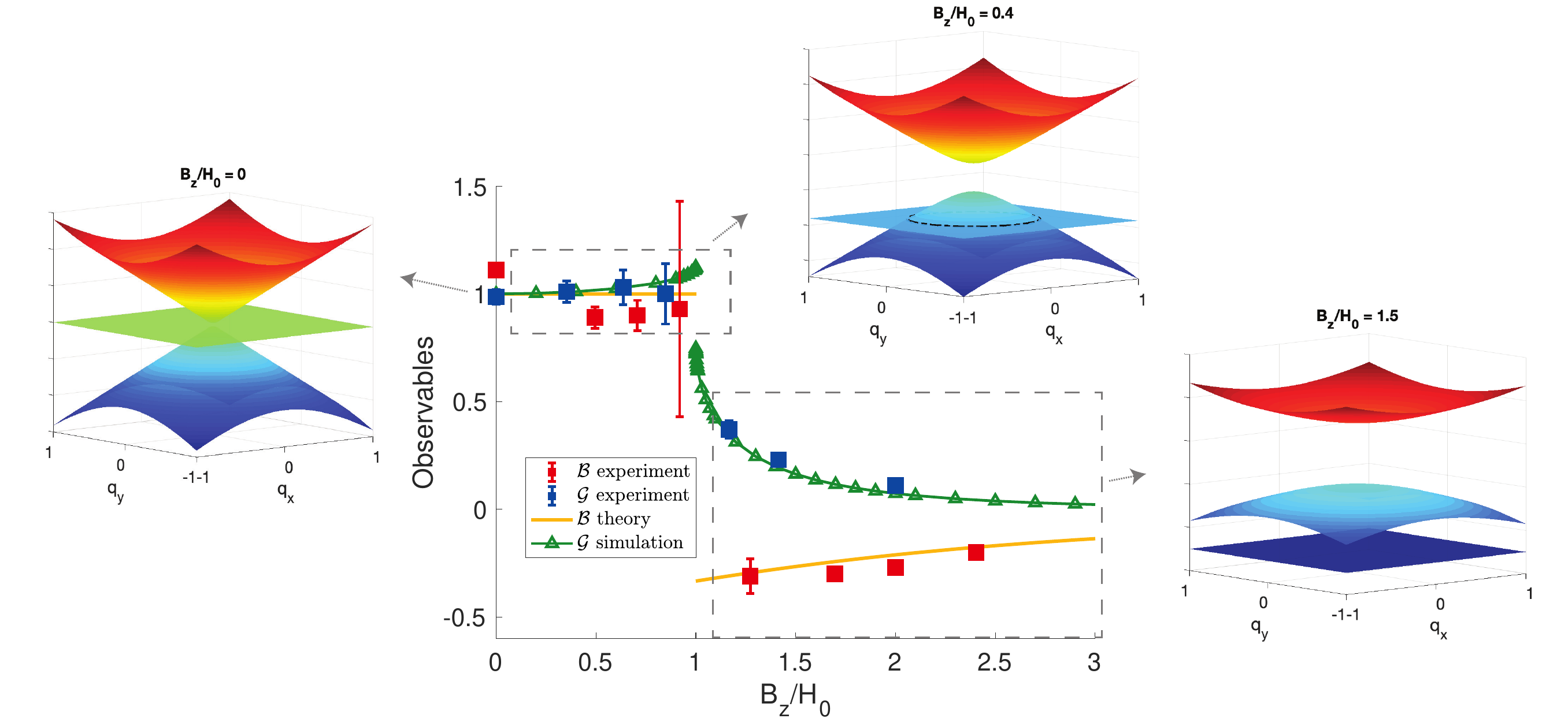}
    \caption{\textbf{Spectral transition triggered by an external field}. The central plot shows experimental data (blue squares) and numerical simulation (green triangles) of the experimental observable $\mathcal{G}$ based on the metric tensor, experimental data (red squares) and analytical result (yellow line) of the observable $\mathcal{B}$ based on the Berry curvature. Both methods shows a sharp response at $B_z=H_0$ when the spectral rings cross the boundary of the integration hypersphere. The experimental observable $\mathcal G$, $\mathcal B$ correspond to the $\mathcal{DD}$ invariant when $B_z=0$ and chiral symmetry is preserved. On the side we show 3 representative energy spectra as the longitudinal field $B_z$ increases ($q_z=q_w=0$). The external field splits the triply degenerate Weyl node (left) into doubly degenerate spectral rings (middle). As the field further increases, the system becomes gapped in the enclosed integration hypersphere (right).
    }
    \label{fig:phase_transition}
\end{figure*}

Upon application of the field, the system undergoes a topological spectral transition from the 4D Weyl-like structure. The new symmetry-protected energy spectrum features a pair of doubly degenerate surfaces in the $\beta-\phi$ space along ($\alpha=0 (\pi/2)$, 
$B_z=H_0$). The spectrum has a more intuitive description in cartesian coordinates,  $(q_x,q_y,q_z,q_w)$, where the field gives rise to two spectral rings in the $q_x-q_y$ ($q_z-q_w$) space along $q_z=q_w=0$ ($q_x=q_y=0$) with radius $B_z$, as shown in Fig.~\ref{fig:phase_transition}. 
We  identify signatures of the spectral rings using two observables inspired by the tensor-monopole measurements, $\mathcal{G}=8\int\epsilon_{\mu\nu\lambda}\sqrt{\det g_{\bar{\mu}\bar{\nu}}}\,d\alpha$ and $\mathcal{B}=-\int(\mathcal{F}_{\alpha\beta}+\mathcal{F}_{\phi\alpha})\,d\alpha$. They represent integration over a hyperspherical surface with radius $H_0$ when viewed in the cartesian coordinate, and converge to the $\mathcal{DD}$ invariant when $B_z=0$. 

As the field strength $B_z$ increases, the two spectral rings expand from the origin and cross the boundary of our integration hypersphere at $B_z=H_0$.  
For various $B_z$, we perform linear and elliptical parametric modulations to reconstruct the metric tensor (Fig.~S16-S21) and the Berry curvature (Fig.~S22-S28), from which we obtain $\mathcal{G}$, $\mathcal{B}$. Remarkably, we observe a sharp response from both experimental observables $\mathcal{G}$, $\mathcal{B}$ at $B_z=H_0$, signaling the topological spectral ring (Fig.~\ref{fig:phase_transition}). The results are in agreement with the simulation for $\mathcal{G}$ and analytical form for $\mathcal{B}$~\cite{SOM},
\begin{equation} \mathcal{B}=\left\{\begin{array}{ll} 
1,&B_z<H_0\\
-\frac12 \left(1-\frac{B_z}{\sqrt{B_z^2+8H_0^2}}\right)&\textrm{otherwise}\end{array}\right..
\end{equation}

These results reveal how exotic spectral transitions can be simulated in our system, upon increasing $B_z$ while keeping $H_0$ fixed (i.e. restricting ourselves to a hypersphere in parameter space):~From the Weyl-type nodes ($B_z=0$) to topological spectral rings  ($B_z<H_0$), characterized by a robust $\mathcal B$ index, and eventually to a gapped spectrum ($B_z>H_0$).

\section*{Outlook}

We demonstrated precise measurements of the quantum metric tensor and Berry curvature over the 4D parameter space provided by a solid-state qutrit. Using these measurements we reconstruct the 3-form curvature, a generalized ``magnetic'' field that was predicted to emanate from nodal points in 4D space. The measured 3-form curvature is shown to exhibit the characteristic radial behavior of a monopole field, which is further confirmed by evaluating its quantized topological charge ($\mathcal{DD}$ invariant). Altogether, our results point towards the first experimental observation of a tensor monopole in a synthetic 4D parameter space.

Our exquisite control over the Weyl-type Hamiltonian illustrates the potential offered by solid-state qudits in the realm of quantum simulation. Interesting perspectives include the fate of tensor monopoles upon coupling the system to other spins or qubits~\cite{Roushan2014}, 
and the study of non-Abelian structures induced by spectral degeneracies and tensor fields~\cite{2102.12471}.

Besides, the Hamiltonian $\hat H_{4D}(\ve{q})$ in Eq.~\eqref{eq:Ham_parametrization} suggests that the physics of tensor monopoles could be investigated in systems of particles moving on a 4D lattice, where $\ve{q}$ would represent the corresponding crystal momenta. Such 4D Weyl lattice systems have been recently proposed~\cite{Palumbo2019,Zhu2020} and could be realized in quantum-engineered systems,  extending the 3D lattice where particles lie with a synthetic dimension~\cite{WangNC2020,OzawaNRP2019}.
 Such 4D Weyl many-body settings are particularly intriguing as they would enable studying the effects of interactions in systems where quasiparticles are effectively coupled to higher-form fields~\cite{ManovitzPRXQuantum2020}.

\subparagraph*{Note added:}
During the preparation of the manuscript, we noticed another experimental work describing the observation of the tensor monopole using superconducting circuits~\cite{Tan2020}.

\bibliography{main} 

\textbf{Acknowledgement}\\
MC, CL and PC thank Mingda Li for helpful discussions. This work is supported in part by NSF Grant PHY1734011. Work in Brussels is supported by the Fonds De La Recherche Scientifique (FRS- FNRS) Belgium and the ERC Starting Grant TopoCold.

\textbf{Author Contributions}\\
MC, CL designed and performed the experiments, analyzed the data with assistance from PC, and input of NG on the quantum-metric measurement. MC, CL, PC discussed and interpreted the results, ran simulations and developed the analytical model describing the spectral transition. MC, CL, GP analyzed the symmetries and spectral structures of the model, with inputs from all authors.  All authors contributed to the writing of the manuscript. PC supervised the overall project.

\textbf{Competing interests}
The authors declare no conflict of interest.

\newpage{}

\end{document}

\begin{CJK*}{UTF8}{} 

\title{A synthetic monopole source of Kalb-Ramond field in diamond:\\ Supplementary Material}% Force line breaks with \\%

\author{Mo Chen \CJKfamily{gbsn}(陈墨)}
\thanks{Present address: Institute for Quantum Information and Matter and Thomas J. Watson, Sr., Laboratory of Applied Physics, California Institute of Technology, Pasadena, CA 91125, USA}
\affiliation{
Research Laboratory of Electronics, Massachusetts Institute of Technology, Cambridge, Massachusetts 02139, USA
}
\affiliation{
Department of Mechanical Engineering, Massachusetts Institute of Technology, Cambridge, Massachusetts 02139, USA
}
\author{Changhao Li}
\thanks{MC and CL contributed equally to this work.}
\affiliation{
Research Laboratory of Electronics, Massachusetts Institute of Technology, Cambridge, Massachusetts 02139, USA
}
\affiliation{
Department of Nuclear Science and Engineering, Massachusetts Institute of Technology, Cambridge, Massachusetts 02139, USA
}

\author{Giandomenico Palumbo}
\affiliation{Center for Nonlinear Phenomena and Complex Systems, Universit\'e Libre de Bruxelles, CP 231, Campus Plaine, B-1050 Brussels, Belgium}
\affiliation{School of Theoretical Physics, Dublin Institute for Advanced Studies, 10 Burlington Road, Dublin 4, Ireland}

\author{Yan-Qing Zhu}
\affiliation{Center for Nonlinear Phenomena and Complex Systems, Universit\'e Libre de Bruxelles, CP 231, Campus Plaine, B-1050 Brussels, Belgium}

\author{Nathan Goldman}
\affiliation{Center for Nonlinear Phenomena and Complex Systems, Universit\'e Libre de Bruxelles, CP 231, Campus Plaine, B-1050 Brussels, Belgium}

\author{Paola Cappellaro}
\thanks{pcappell@mit.edu}
\affiliation{
Research Laboratory of Electronics, Massachusetts Institute of Technology, Cambridge, Massachusetts 02139, USA
}
\affiliation{
Department of Nuclear Science and Engineering, Massachusetts Institute of Technology, Cambridge, Massachusetts 02139, USA
}
\affiliation{
Department of Physics, Massachusetts Institute of Technology, Cambridge, Massachusetts 02139, USA
}
\date{\today}
\begin{abstract}

\end{abstract}

\maketitle
\end{CJK*}

\tableofcontents

\newpage{}

\section{Theoretical description of the model Hamiltonian and  the tensor monopole}
\subsection{Tensor monopole and $\mathcal{DD}$ invariant}

In this section we review the generalization from Chern number to $\mathcal{DD}$ invariant and present the relation between the 3-form curvature and the 2-form connection.

In electromagnetism, the Dirac monopole is closely related to the first Chern number. Similar to Gauss's law for the electric charge, the Dirac monopole is revealed by integral of the well-known Berry curvature over any enclosed manifold containing the monopole:
\begin{equation}\label{eq:Dirac_Monopole}
C_1=\frac{1}{2\pi}\int_{S^2} \mathcal{F}_{\mu\nu} dq_\mu \wedge dq_\nu.
\end{equation}

The Berry curvature $\mathcal{F}_{\mu\nu}$ appears in the quantum geometric tensor (QGT):
\begin{equation}
\chi_{\mu\nu}=g_{\mu\nu}+i\mathcal{F}_{\mu\nu}/2,
\end{equation}
where the real part is the metric tensor. More details on the measurement of the metric tensor is covered in the next section. It has been shown that one can obtain the Berry curvature entirely from the metric tensor $g_{\mu\nu}$~\cite{Palumbo2018}: 
\begin{equation}\label{eq:Berry_connect_to_metric}
\mathcal{F}_{\mu\nu}=2\epsilon_{\mu\nu}\sqrt{\det g_{\bar{\mu}\bar{\nu}}},
\end{equation}
where $\epsilon_{\mu\nu}$ is the Levi-Civita symbol. 

From the viewpoint of metric tensor, it is a natural generalization to have  the generalized 3-form Berry curvature in the 4D parameter space:
\begin{equation}
\mathcal{H}_{\mu\nu \lambda} = \epsilon_{\mu \nu \lambda}(4\sqrt{\det(g_{\bar{\mu} \bar{\nu}}})),
\end{equation}
and the corresponding topological invariant (Dixmier-Douady invariant):
\begin{equation}\label{eq:Tensor_Monopole}
\mathcal{DD}=\frac{1}{2\pi^2}\int_{S^3} \mathcal{H}_{\mu\nu\lambda} dq^\mu \wedge dq^\nu\wedge dq^\lambda.
\end{equation}

Alternatively, the curvature $\mathcal{H}$ can be derived from the generalized 2-form Berry connection $B_{\mu\nu}$ associated with the ground state $\ket{u_-}$:
\begin{equation}\label{eq:relation_H_B}
\mathcal{H}_{\mu\nu \lambda} = \partial_\mu B_{\nu \lambda} + \partial_\nu B_{\lambda \mu}+\partial_\lambda B_{\mu \nu},
\end{equation}
where the 2-form tensor connection can be constructed from the state $\ket{u_-}$~\cite{Palumbo2018,Palumbo2019}:
\begin{equation}\label{eq:connection_state_components}
 B_{\mu \nu} = \Phi \mathcal{F}_{\mu \nu}, \Phi = \frac{-i}{2} \log (u_1 u_2 u_3)
\end{equation}
with $u_{1(2,3)}$ denoting the components of $\ket{u_-}$, $\mathcal{F}_{\mu\nu} = \partial_{\mu}A_{\nu}-\partial_{\nu}A_{\mu}$ being the 2-form Berry curvature. From Eq.~\ref{eq:relation_H_B} and~\ref{eq:connection_state_components}, it follows that in general, the generalized curvature can be obtained by performing state tomography on the eigenstate, upon external perturbations of the parameters. We will provide a simpler form of the 2-form connection and the 3-form curvature for our model, as shown in Sec.~\ref{sec:analytical_DD}.

We note that the 2-form Berry connection $B_{\mu\nu}$ can be more generally constructed from a mixed set of pseudoreal and complex scalar fields $\psi_{1,2,3}$ satisfying the U(1) gauge transformation and linked to the ground state,
\begin{equation}
 B_{\mu \nu} = \frac{i}{3}\sum_{j,k,l=1}^{3} \epsilon^{jkl}\psi_j\, \partial_\mu \psi_k\, \partial_\nu \psi_l = \frac i3\psi_j (\partial_\mu \psi_k\, \partial_\nu \psi_l-\partial_\mu \psi_l\, \partial_\nu \psi_k) 
\label{eq:Bpsi}
\end{equation}
 Choosing for example
\begin{equation}
\psi_1 = -i\log(u_1+u_3),\qquad\psi_2= u_1^*-u_3^*,\qquad\psi_3=u_3-u_1,
\label{eq:psi}
\end{equation}
where $\ket{u_-}=[u_1,u_2,u_3]^T$, yields a gauge-invariant $\mathcal H_{\mu\nu\lambda}$ that gives the same $\mathcal{DD}$ and $\mathcal B$ as  above.

We compare the quantum geometric properties in electromagnetism in 3D and in the tensor gauge field in 4D in Table~\ref{tab:comparison}.

\begin{table*}[]
    \centering
   \begin{tabular}{ c|c|c } 
     \hline
     Quantum geometry & Electromagnetism & Tensor (2-form) gauge field \\ 
     \hline
     Berry connection: $A_{\mu}$,$B_{\mu\nu}$ & vector potential $A_{\mu}$ & tensor potential $B_{\mu\nu}$ \\ 
     \hline
     Berry curvature: $\mathcal{F}_{\mu\nu}$,$\mathcal{H}_{\mu\nu \lambda}$ & magnetic field strength $\mathcal{F}_{\mu\nu}=\partial_{\mu}A_{\nu}-\partial_{\nu}A_{\mu}$ & field strength $\mathcal{H}_{\mu\nu \lambda} =\partial_\mu B_{\nu \lambda} + \partial_\nu B_{\lambda \mu}+\partial_\lambda B_{\mu \nu}$ \\ 
     \hline
      Topological invariant & first Chern number $C_1$ & Dixmier-Douady invariant $\mathcal{DD}$ \\
     \hline
         
    \end{tabular}
    \caption{Ground state quantum geometric properties in electromagnetism in 3D and in the tensor gauge field in 4D.}
    \label{tab:comparison}
\end{table*}

\subsection{Analytical solutions of the metric tensor and generalized Berry curvature ($B_z=0$)}
For convenience, we show the system Hamiltonian here
\begin{equation}\label{eq:H_parameterized}
\hat{H}_{ST}=\hat{H}_{4D} + \textrm{diag}(B_z,0,-B_z)/\sqrt{2},
\end{equation}
\begin{equation}
\hat{H}_{4D}=
\begin{pmatrix}
0 &q_x-i q_y&0\\
q_x+i q_y&0&q_z+i q_w\\
0&q_z-i q_w&0
\end{pmatrix},
\end{equation}
where the parameters $\ve{q}\!=\!(q_x,q_y,q_z,q_w)$ can be expressed in terms of the experimentally controllable parameters $(H_0, \alpha, \beta, \phi)$  through $q_x\!+\!i q_y\!=\!H_0 \cos(\alpha)e^{i \beta}$, $q_z\!+\!i q_w\!=\!H_0 \sin(\alpha)e^{i\phi}$, where $\alpha \in [0,\pi/2]$ and $\beta, \phi \in [0,2\pi)$.
The Hamiltonian in Eq.~\ref{eq:H_parameterized} is parameterized by the three parameters $\alpha,\beta, \phi$ at fixed $H_0$ when $B_z=0$.  The analytical solutions for the quantum metric tensor 
can be easily obtained via exact diagonalization:
\begin{equation}\label{eq:g_analytical}
\begin{split}
    & g= \begin{pmatrix}
    g_{\alpha\alpha} & g_{\alpha\beta} & g_{\alpha\phi}\\
    g_{\beta\alpha} & g_{\beta\beta} & g_{\beta\phi}  \\
    g_{\phi\alpha} & g_{\phi\beta} & g_{\phi\phi}\\
    \end{pmatrix}\\
    & = 
    \begin{pmatrix}
    \frac{1}{2} & 0 & 0\\
    0 & \frac{1}{4}\cos^2 \alpha(2-\cos^2\alpha) & -\frac{1}{16}\sin^2 2\alpha  \\
    0 & -\frac{1}{16}\sin^2 2\alpha & \frac{1}{4}\sin^2\alpha(2-\sin^2 \alpha)\\
    \end{pmatrix}. \\
\end{split}
\end{equation}
The resulting 3-form Berry curvature is:
\begin{equation}\label{eq:H_analytical}
    \mathcal{H}_{\alpha \beta \phi} =\cos \alpha \sin \alpha.
\end{equation}
One can then easily verify that the integral of the above curvature over $\alpha,\beta$ and $\phi$ will yield $\mathcal{DD}=1$.

\subsection{Analytical form of $\mathcal{B}$ calculated from the tensor Berry connection (arbitrary $B_z$)}
\label{sec:analytical_DD}
With $\hat{H}_{ST}$ parametrized by $(H_0, \alpha, \beta, \phi)$, we present the analytical calculations of the experimental observable $\mathcal{B}$ (see main text) in the presence of external field ($B_z\neq 0$). 
We can easily prove that, up to a global phase, the ground state of the Hamiltonian Eq.~\ref{eq:H_parameterized} has the form 
\begin{equation}\label{eq:um_form}
    \ket{u_-}=[e^{-i\beta}v_1,v_2,e^{-i\phi}v_3]^T
\end{equation}
where $v_1,v_2, v_3$ are functions of $\alpha$ only.
Then, we can calculate the vector gauge potential for this special gauge
\[\mathcal A_\alpha=0,\qquad \mathcal A_\beta=v_1^2,\qquad \mathcal A_\phi=v_3^2\]
and the 2-form Berry curvature $\mathcal F_{\mu\nu}=\partial_\mu A_\nu - \partial_\nu A_\mu$
\begin{equation}
\begin{split}
\mathcal F_{\alpha\beta}=\partial_\alpha(v_1^2),\\
\mathcal F_{\phi\alpha}=-\partial_\alpha(v_3^2).
\end{split}
\end{equation}
We repeat Eq.~\ref{eq:relation_H_B} and~\ref{eq:connection_state_components} here
for convenience
\begin{equation}
\mathcal H=\partial_\alpha B_{\beta\phi}+\partial_\phi B_{\alpha\beta}+\partial_\beta B_{\phi\alpha},
\end{equation}
where $B_{\mu\nu}=\mathcal F_{\mu\nu}\Phi$, and
\[\Phi=-\frac i2\log \prod_{i=1}^3 u_i =-\frac i2\log(e^{-i(\phi+\beta)}v_1v_2v_3). \]
Combining all above, we obtain the simplified form for the curvature
\begin{equation}\label{eq:H_in_F}
\begin{split}
\mathcal H_{\alpha\beta\phi}&=\partial_\phi(\Phi \mathcal F_{\alpha\beta})+\partial_\beta (\Phi \mathcal F_{\phi\alpha}) + \partial_\alpha (\Phi \mathcal{F}_{\beta\phi})\\
&=-\frac12 (\mathcal F_{\alpha\beta}+\mathcal F_{\phi\alpha})\\
&=-\frac 12 \frac{d}{d\alpha}(v_1^2-v_3^2),
\end{split}
\end{equation}
where the second line is used in experiments to extract the curvature.

With the alternative definitions in Eq.~\ref{eq:Bpsi} and \ref{eq:psi} we arrive at 
\begin{equation}\label{eq:Halt}
\mathcal H_{\alpha\beta\phi}=
-\frac{2}{e^{-i\beta}v_1+e^{-i\phi}v_3}\left[e^{-i\phi}v_3\frac{d}{d\alpha}(v_1^2) -e^{-i\beta}v_1\frac{d}{d\alpha}(v_3^2)\right],
\end{equation}
which yields the same $\mathcal B$ upon integration. Note that while the choice of the $\psi$'s fields in Eq.~\ref{eq:psi} ensures that $\mathcal H_{\alpha\beta\phi}$ coincides at $B_z=0$ with its value calculated from the QGT, there is much freedom in the choice of the field $\psi$.

As a result, there is no well-defined $\mathcal B$ that can  be considered a good topological number to characterize the emergent topological spectral ring phase protected by mirror symmetries. The classification of topological nodal line semimetals is not yet complete in 3D~\cite{Fang2016}, and still lacking in 4D. Nevertheless, as we have shown, the observable ${\mathcal{B}}$, although not topological, serves as a convenient experimental tool to signal the spectral ring and the associated spectral transition.

We can gain further insight into the observable $\mathcal{B}$ by explicitly evaluating the integral of Eq.~\ref{eq:H_in_F}
\[\mathcal{B}=\frac{1}{2\pi^2}\int_{S_3}\mathcal{H}_{\alpha\beta\phi} d\alpha d\beta d\phi = [v_1^2(0)-v_3^2(0)]-[v_1^2(\pi/2)-v_3^2(\pi/2)]\]
For $\alpha=0,\pi/2$ the Hamiltonian eigenvectors can be calculated easily. We obtain
\[u_1(0)=\left\{\begin{array}{ll} 
-\sqrt{\frac12 \left(1-\frac{B_z}{\sqrt{B_z^2+8H_0^2}}\right)},&B_z<H_0\\
0&\textrm{otherwise}\end{array}\right.\]    
\[u_3(0)=\left\{\begin{array}{ll} 
0,&B_z<H_0\\
1&\textrm{otherwise}\end{array}\right.\]    
\[u_1(\pi/2)=0\qquad u_3(\pi/2)=\sqrt{\frac12 \left(1+\frac{B_z}{\sqrt{B_z^2+8H_0^2}}\right)} \]

Finally we obtain
\begin{equation} \mathcal{B}=\left\{\begin{array}{ll} 
1,&B_z<H_0\\
-\frac12 \left(1-\frac{B_z}{\sqrt{B_z^2+8H_0^2}}\right)&\textrm{otherwise}\end{array}\right.
\end{equation}

We remark that upon breaking chiral symmetry, we choose here a special gauge ($v_2\in \mathbb{R}$) for convenience. Indeed, the 3-form curvature in Eq.~\ref{eq:H_in_F} is not gauge-invariant and the gauge structure we defined is not universal.
Nevertheless, the $\mathcal{B}$ only depends on the eigenvector when $\alpha=0, \pi/2$, where the two-fold degenerate points reside, and as we show analytically, numerically and experimentally, it indeed provides signatures of these non-trivial singularity points.

\subsection{Topological phase transition triggered by manifold displacement}
The most straightforward topological phase transition could be induced by  displacement of the hypersphere along one of the parameter axis , e.g.,  $q_x \rightarrow q_x +\delta_x$. The Hamiltonian can be written as:
\begin{equation}\label{eq:H_displacement}
\hat{H}_{disp}=
\begin{pmatrix}
0 &H_0 \cos\alpha e^{-i\beta}+\delta_x&0\\
H_0\cos\alpha e^{i\beta}+\delta_x&0&H_0\sin\alpha e^{i\phi}\\
0&H_0\sin\alpha e^{-i\phi}&0
\end{pmatrix}.
\end{equation}
We can analytically calculate the 3-form Berry curvature in the presence of displacement:
\begin{equation}
    \mathcal{H}_{\alpha\beta\phi}(H_0, \alpha,\beta,\delta_x)= \frac{H_0^3\cos \alpha \sin\alpha (H_0+\delta_x\cos\alpha \cos \beta)}{(\delta_x^2+H_0^2+2H_0 \delta_x \cos\alpha \cos\beta)^2}
\end{equation}
and evaluate its integral to find the $\mathcal{DD}$ invariant.

\begin{figure*}[ht]
    \centering
    \includegraphics[width=0.6\textwidth]{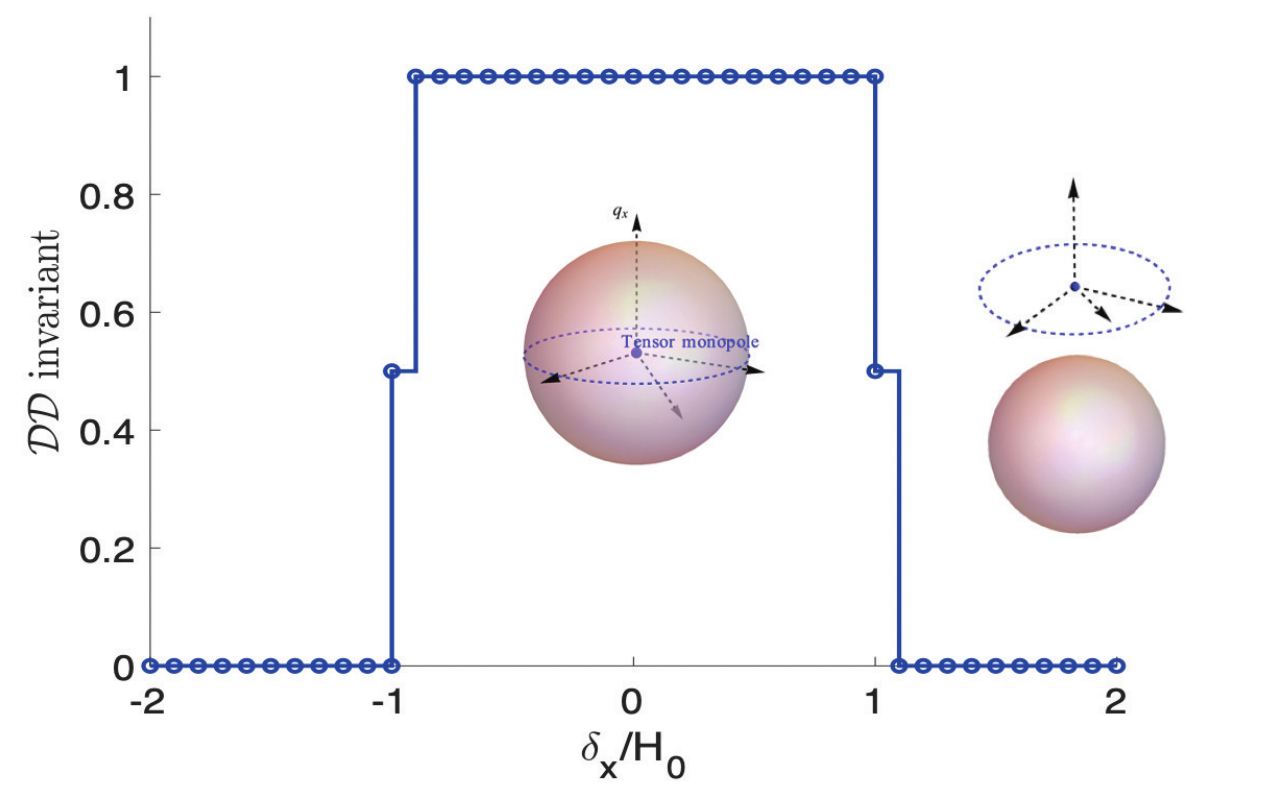}
    \caption{\textbf{Phase transition triggered by a manifold displacement}. When $|\delta_x/H_0| < 1$, the manifold encloses the triply degenerate point and $\mathcal{DD}=1$; when $|\delta_x/H_0| >1$, the monopole is no longer enclosed, leading to a trivial phase $\mathcal{DD}=0$.}
    \label{fig:displacement_transition}
\end{figure*}

The topological phase transition is characterized by the $\mathcal{DD}$ invariant, as shown in Fig.~\ref{fig:displacement_transition}, where $\mathcal{DD}=1\rightarrow 0$ when $|\delta_x/H_0| >1$. 

However, the  $\beta$ dependence of the 3-form curvature introduced by the translation poses a challenge in experiments, as it greatly prolongs the total measurement time. Instead, in our experiment we choose to add a fictitious $z$ field to the system which preserves rotation symmetry about $\beta, \phi$, as discussed in more detail in Sec.~\ref{sec:phase_transition_by_Bz} next.

\subsection{Topological spectral transition triggered by external field}\label{sec:phase_transition_by_Bz}
Having been discussed in the main text, we elaborate in more detail about the topological spectral transitions induced by external field $B_z$. For this purpose, it is useful to rewrite the Hamiltonian (\ref{eq:H_parameterized}) in terms of the Gell-Mann matrices~\cite{Gellmann62} that highlight its symmetries
\begin{equation}
\hat H_{ST}=q_x\lambda_1+q_y\lambda_2+q_z\lambda_6+q_w\lambda_7+\frac{B_z}{2\sqrt{2}}(\lambda_3+\sqrt{3}\lambda_8)
\label{eq:gellmann}	
\end{equation}

Without the external field ($B_z=0$), our Hamiltonian preserves chiral symmetry
\begin{equation}
    \{\hat{H}_{4D}, U\}=0,
\end{equation}
where $U=diag(1,-1,1)$.

\begin{figure*}[h] 
    \centering
    \includegraphics[width=0.42\textwidth]{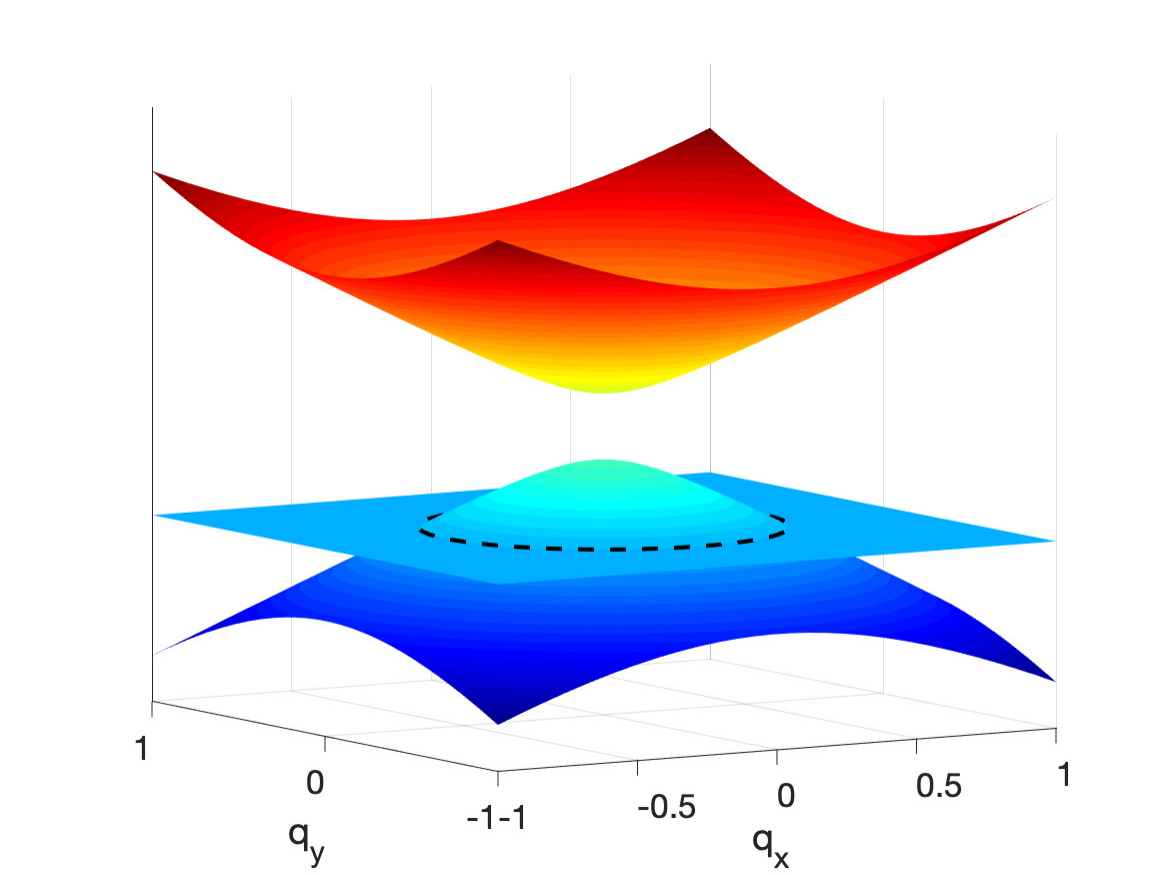}
    \includegraphics[width=0.42\textwidth]{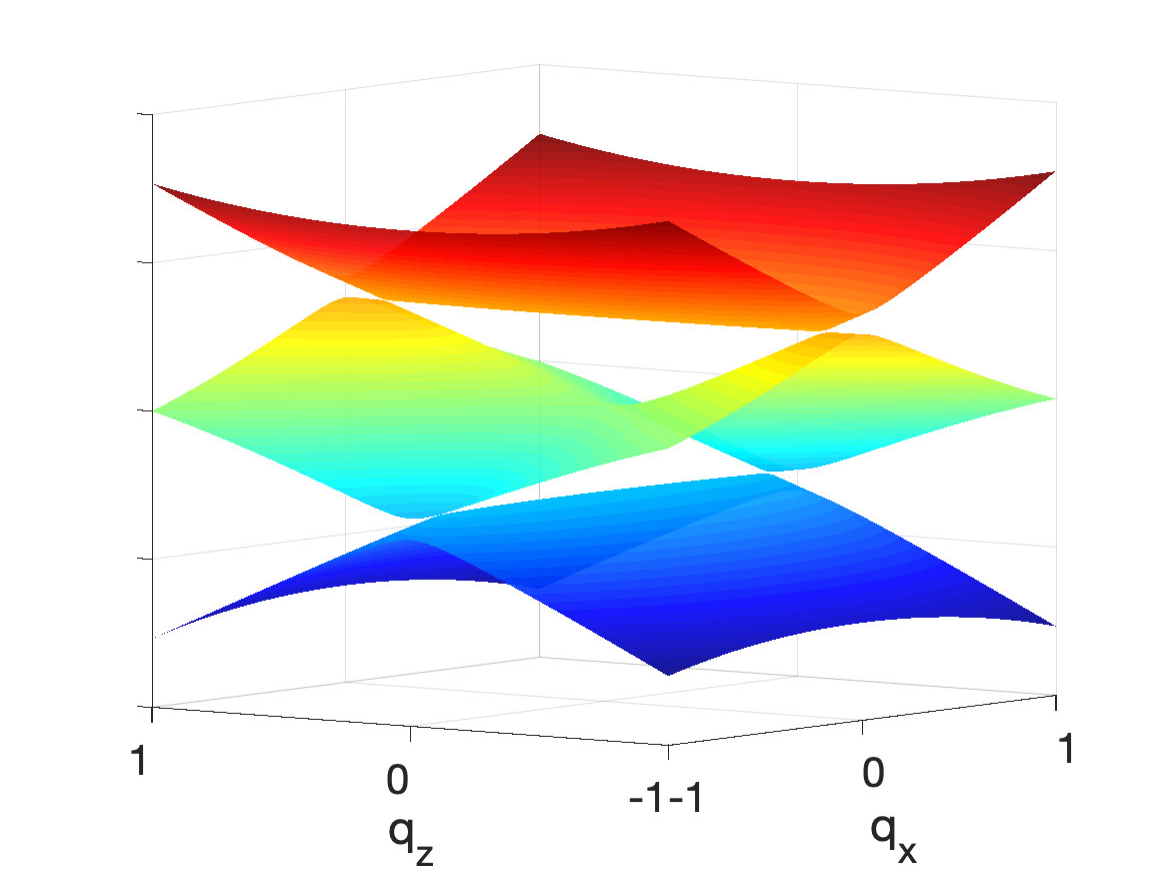}
    \caption{\textbf{Nodal structures} 
    (Left) nodal ring between the lower two bands in the $q_x, q_y$ plane when $q_z=q_w=0$ and $B_z\neq 0$. The nodal rings are protected by the mirror symmetries. (Right) nodal points between the middle band and the lower (upper) band in the $q_x, q_z$ plane when $q_y=q_w=0$ and $B_z\neq 0$. These nodal points are protected by the PT symmetry.}
    \label{fig:nodal_ring}
\end{figure*}

Upon breaking the chiral symmetry ($B_z\neq 0$), the tensor monopole disappears and splits into spectral rings. The nodal structures are two degenerate nodal surfaces spanning the $\beta-\phi$ space along ($\alpha=0, \pi/2, B_z=H_0$). Viewed from the $(q_x,q_y,q_z,q_w)$ parameter space, for example, the eigenvalue analytical forms in the subspace of $q_z=q_w=0$ are
\begin{equation}
   \epsilon_-=\frac{1}{2}\left(\frac{B_z}{\sqrt{2}}-\sqrt{\frac{B_z^2}{2}+4q_x^2+4q_y^2}\right), \quad \epsilon_0=-\frac{B_z}{\sqrt{2}}, \quad 
\epsilon_+=\frac{1}{2}\left(\frac{B_z}{\sqrt{2}}+\sqrt{\frac{B_z^2}{2}+4q_x^2+4q_y^2}\right),
\end{equation}
and the lower two bands become degenerate when $q_x^2+q_y^2=B_z^2$, as shown in Fig.~\ref{fig:nodal_ring}. Indeed, we find that 
there are two nodal rings, one between the middle band and the lower band  (corresponding to $\alpha=0$) and another at $\alpha=\pi/2$, between the middle  and   upper bands:
\begin{equation}
\begin{split}
q_x^2+q_y^2&=B_z^2,\quad{}q_z=q_w=0,\\
\textrm{or}\quad 
q_z^2+q_w^2&=B_z^2,\quad{}q_x=q_y=0.
\end{split}
\end{equation}
Because nodal/spectral rings are more commonly encountered and studied in 3D~\cite{Fang2016} than nodal surfaces, we will refer to the nodal structure in our model as nodal/spectral rings for the convenience of the reader.

As soon as the chiral symmetry is broken by $B_z\neq 0$, our system undergoes a phase transition from the Weyl-type Hamiltonian hosting a tensor monopole to the topological spectral ring phase, protected by two mirror symmetries:
\begin{equation}\label{eq:mirror_sym}
    \begin{split}
        & M_1 \hat H_{ST}(q_x,q_y,q_z,q_w) M_1^{-1}= \hat H_{ST}(-q_x,-q_y,q_z,q_w) \\
        & M_2 \hat H_{ST}(q_x,q_y,q_z,q_w) M_2^{-1}= \hat H_{ST}(q_x,q_y,-q_z,-q_w), \\
    \end{split}
\end{equation}
where $M_1=diag(-1,1,1)$, $M_2=diag(1,1,-1)$.  These mirror symmetries naturally imply inversion symmetry:
\begin{equation}
    U_I \hat H_{ST}(q_x,q_y,q_z,q_w) U_I^{-1}= \hat H_{ST}(-q_x,-q_y,-q_z,-q_w)
\end{equation}
where $U_I=M_1M_2$. The aforementioned nodal rings are protected by these mirror symmetries (see experimental data in Sec.~\ref{sec:exp_phase_transition}).
The symmetries can be further broken by introducing terms that are proportional to $\lambda_{4(5)}$ Gell-Mann matrices,
\begin{equation*}
\lambda_4=
\begin{pmatrix}
0 &0 &1\\
0&0&0\\
1&0&0
\end{pmatrix}, \quad
\lambda_5=
\begin{pmatrix}
0 &0 &-i\\
0&0&0\\
i&0&0
\end{pmatrix},
\end{equation*}
and the degenerate rings will become gapped.

We further remark that there exists PT symmetry for the specific 2D subsystem given by $q_y = q_w = 0$. The nodal points in the $q_x-q_z$ plane between the lower and upper two bands shown in Fig.~\ref{fig:nodal_ring} (right plot) are protected by PT symmetry. 
Introducing the $\lambda_4$ term breaks the mirror symmetries, but preserves the PT symmetry. 
Therefore it gaps the nodal rings into PT-symmetry-protected nodal points, and the system remains gapless due to the PT symmetry. 
An energy gap can fully open only by adding terms, such as the $\lambda_5$ Gell-Mann matrix, that break the PT symmetry. On the same note, we find that for the subsystem given by $q_x = q_z = 0$, the system Hamiltonian satisfies anti-commutation relation with the PT operator and we observe similar nodal points as mentioned above.

\begin{figure*}[h] 
    \centering
    \includegraphics[width=0.42\textwidth]{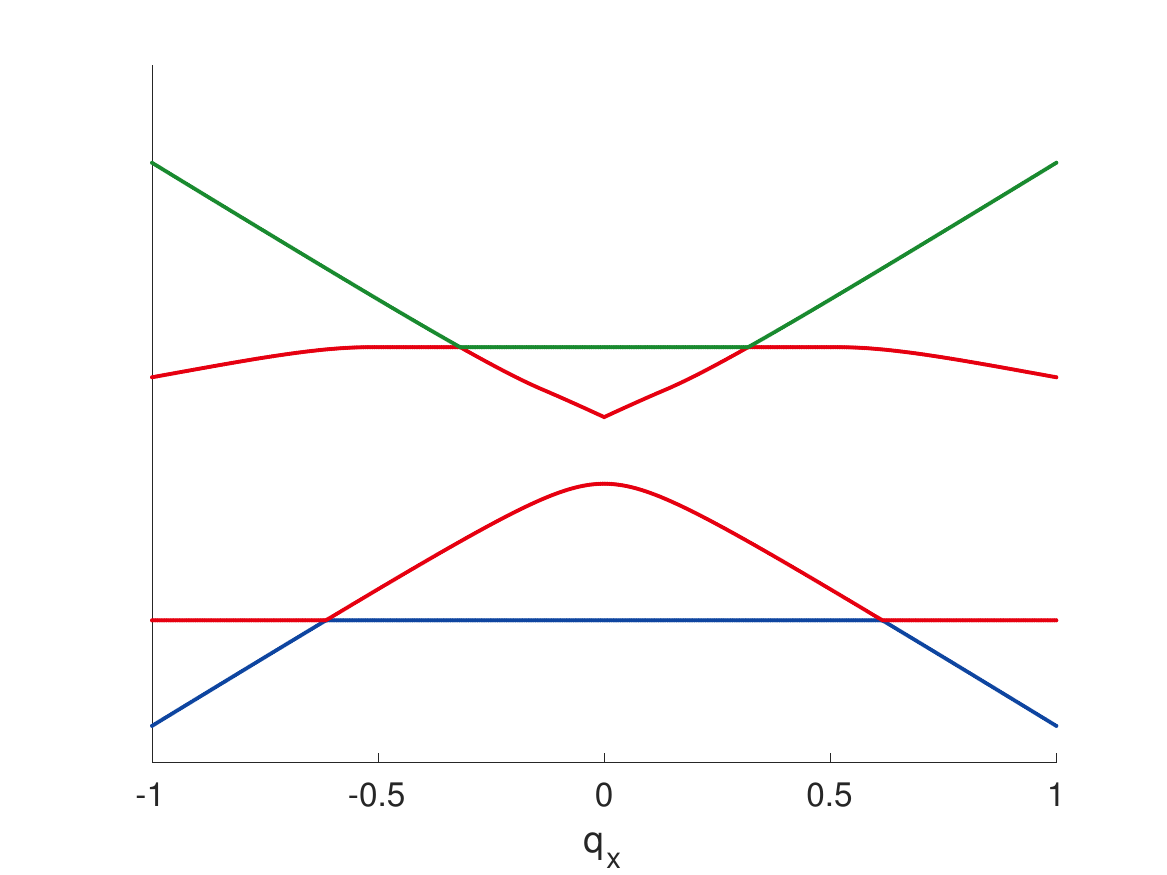}
    \includegraphics[width=0.42\textwidth]{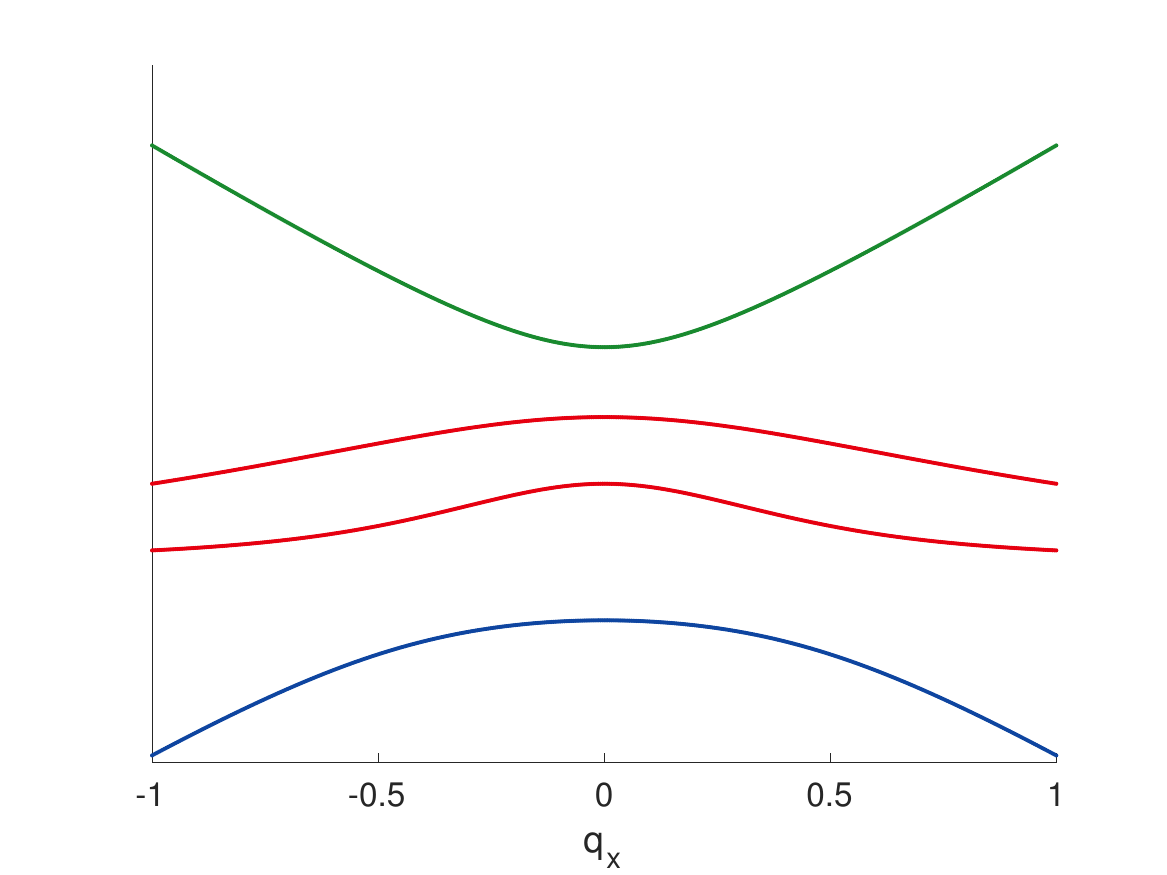}
    \caption{\textbf{Energy spectrum upon broken symmetries} We take the $q_y=q_w=0$ slice, project the energy spectrum to the $q_x$ axis and plot the envelope of the three energy bands. The red lines correspond the envelop of the middle band.  Two examples are (left) broken mirror symmetries, but preserved PT symmetry and (right) broken PT symmetry. The PT symmetry protect the nodal points and keep the system gapless. By introducing terms that break the PT symmetry (such as $\lambda_5$ matrices), we can fully open the gap.}
    \label{fig:broken_symmetry_spectrum}
\end{figure*}

As examples of broken mirror symmetries and broken PT symmetry, we plot in Fig.~\ref{fig:broken_symmetry_spectrum} the energy spectrum projected to the $q_x$ axis, when breaking mirror symmetries by introducing $\lambda_4$ and the PT symmetry by introducing $\lambda_5$. In both cases, we break the chiral symmetry due to the $B_z$ field.

\subsection{System Hamiltonian: equivalence with SG220 linear $\bold{k}\cdot\bold{p}$ model}\label{sec:SG220}

In this section we review the Hamiltonian implemented in our experiments, and point out an interesting relation between this model and a three-band linearized $\bold{k}\cdot\bold{p}$ Hamiltonian~\cite{Bradlyn2016}.

As shown in the main text, our Hamiltonian is:
\begin{equation}\label{eq:H_parameterized}
\hat{H}=H_0
\begin{pmatrix}
0 &\cos\alpha e^{-i\beta}&0\\
\cos\alpha e^{i\beta}&0&\sin\alpha e^{i\phi}\\
0&\sin\alpha e^{-i\phi}&0
\end{pmatrix} + \frac{B_z}{\sqrt{2}} \begin{pmatrix}
1 &0&0\\
0&0&0\\
0&0&-1
\end{pmatrix},
\end{equation}
where $\alpha \in [0,\pi/2]$ and $\beta, \phi \in [0,2\pi)$.
The linearized $\bold{k}\cdot\bold{p}$ Hamiltonian for space group (SG) 220 is given by:
\begin{equation}\label{eq:H_220}
\hat{H}_{220}(\bold{k})=
\begin{pmatrix}
0 &k_y & k_x\\
k_y&0&-k_z\\
k_x&-k_z&0 
\end{pmatrix},
\end{equation}
We note that when we take a slice of our Hamiltonian $\bold{k}=(k_x,k_y,k_z)=(H_0 \sin(\alpha-\pi/4), H_0 \cos(\alpha-\pi/4), B_z/\sqrt{2})$,  Eq.~\ref{eq:H_parameterized} and Eq.~\ref{eq:H_220} have identical eigenvalue spectrums for any $\beta, \phi$.

For the Hamiltonian in Eq.~\ref{eq:H_220}, pairs of two bands are degenerate along $|k_x|=|k_y|=|k_z|$. This corresponds to the condition $B_z=H_0$ and $\alpha=0,\pi/2$ in our model, matching our numerical simulation and experimental results. Due to the rotation symmetry of $\beta$ and $\phi$ in our model, we remark that a pair of doubly degenerate nodal surfaces along $(\alpha=0, \pi/2, B_z=H_0)$ emerge when $B_z\neq 0$ in our 4D parameter space. This corresponds to two nodal rings in the ($q_x, q_y, q_z, q_w$) coordinate. We will discuss this later in Sec.~\ref{sec:phase_transition_by_Bz}.

The equivalence between the SG220 model and a slice of our model presented here implies that the doubly degeneracy induced by the detuning $B_z$ has a correspondence to crystal symmetry-protected fermionic excitations, and the observed phase transition has a non-trivial topological meaning. It would be interesting to use our experimental system to further simulate the topological properties of the SG220 model (and other triple-degenerate point models).

\section{Experimental details}

\subsection{Sample and Experiment Setup}
We used a home-built confocal microscope to initialize and measure a single NV center in an electronic grade diamond sample (Element 6, \Nit concentration $n_N<5$~ppb, natural abundance of \carb). The NV center is chosen to be free from close-by \carb. A magnetic field of $490$~G is applied using a permanent magnet, where excited state level anti-crossing allows polarization transfer from the NV electronic spin to the native \Nit nuclear spin~\cite{Jacques09}. Consequently, a $1\mu s$ green laser excitation polarizes the NV-\Nit system to the $\ket{0,+1}$ state. Throughout the experiment, \Nit remains in $\ket{+1}$. Hence, we can restrict ourselves to the electronic part of the NV Hamiltonian. For the NV of interest, we measure $T_1=3.2$~ms, $T_{2,echo}>700\mu s$.

\subsection{Hamiltonian under dual-frequency microwave driving}\label{sec:DQ_rotating_frame}
We now derive the effective Hamiltonian in the rotating frame of the dual-frequency microwave pulses following Ref.~\cite{Mamin2015}. 

The intrinsic Hamiltonian of NV is $D S_z^2+\gamma_e B S_z$, where $\gamma_e=2.8$ MHz/G is the gyromagnetic ratio, $D=2.87$ GHz is the zero-field energy splitting of the NV ground state, $B =490$ G here is the external field along N-V axis and $S_z$ is the spin-1 $z$ operator. Under the dual-frequency microwave control, the total Hamiltonian is given by:
\begin{equation}
\begin{split}
    \hat H_{NV} =& DS_z^2 + \gamma_{e} B S_z \\
    &+ 2\sqrt{2}[\gamma_{e}B_1 \cos(\omega_1 t + \phi_1)+\gamma_{e}B_2 \cos(\omega_2 t + \phi_2)]S_x
\end{split}
\end{equation}
where $S_x, S_z$ are the spin-1 operators, the first line is the NV spin Hamiltonian and the second line represents the dual-frequency microwave pulse at frequencies $\omega_1, \omega_2$.
In the bare NV frame with basis $\ket{m_s=+1, 0, -1}$, the above Hamiltonian can be written as:
\begin{equation}\label{eq:H_NV_bare}
    \hat H_{NV}=\begin{pmatrix}
    D+ \gamma_{e} B & 2\Big(\begin{aligned} &\gamma_{e}B_1 \cos(\omega_1 t + \phi_1)\\
    &+\gamma_{e}B_2 \cos(\omega_2 t + \phi_2) \end{aligned}\Big) & 0\\
 2\Big(\begin{aligned} &\gamma_{e}B_1 \cos(\omega_1 t + \phi_1)\\
    &+\gamma_{e}B_2 \cos(\omega_2 t + \phi_2) \end{aligned}\Big) &
    0 & 2\Big(\begin{aligned} &\gamma_{e}B_1 \cos(\omega_1 t + \phi_1)\\
    &+\gamma_{e}B_2 \cos(\omega_2 t + \phi_2) \end{aligned}\Big) \\
    0 & 2\Big(\begin{aligned} &\gamma_{e}B_1 \cos(\omega_1 t + \phi_1)\\
    &+\gamma_{e}B_2 \cos(\omega_2 t + \phi_2) \end{aligned}\Big) & D- \gamma_{e} B \\
    \end{pmatrix},
\end{equation}
We now enter the rotating frame defined by the unitary transformation:
\begin{equation*}
    V = \begin{pmatrix}
    e^{-i\omega_1 t} & 0 & 0\\
 0& 1 & 0 \\
    0 &0 & e^{-i\omega_2 t} \\
    \end{pmatrix},
\end{equation*}
the Hamiltonian Eq.~\ref{eq:H_NV_bare} can be rewritten as the same form as $\hat H_{ST}$:
\begin{equation} \label{eq:H_DQ_frame}
    \hat{H}_{NV} = \begin{pmatrix}
    D+ \gamma_{e} B-\omega_1 & B_1 e^{-i\phi_1} & 0\\
 B_1 e^{i\phi_1}& 0 & B_2 e^{i\phi_2} \\
    0 &B_2 e^{-i\phi_2} & D- \gamma_{e} B-\omega_2 \\
    \end{pmatrix}
    = \begin{pmatrix}
   B_z/\sqrt{2} &H_0 \cos\alpha e^{-i\beta}&0\\
H_0\cos\alpha e^{i\beta}&0&H_0\sin\alpha e^{i\phi}\\
0&H_0\sin\alpha e^{-i\phi}&-B_z/\sqrt{2}
    \end{pmatrix}= \hat H_{ST}
\end{equation}
where $B_z = D \pm \gamma_{e} B-\omega_{1(2)}$ corresponds to detunings in microwave frequency, $B_1 = H_0\cos\alpha$, $B_{2}=H_0\sin\alpha$ and $\phi_{1}=\beta$,$\phi_2=\phi$ are the amplitudes and phases of the microwave pulses, respectively. We call this Hamiltonian the double quantum (DQ) Hamiltonian in the following, for its ability to drive the $\ket{m_s=-1}\leftrightarrow\ket{m_s=+1}$ transition, where the quantum number changes by $2$.

When both microwave frequencies are on-resonance $\omega_{1(2)} =  D \pm \gamma_{e} B$, the eigenstates of Eq.\ref{eq:H_DQ_frame} are:
\begin{equation} \label{eq:H_DQ_eigenvectors}
    \ket{u_\pm} = \frac{1}{\sqrt{2}}\begin{pmatrix}
 \frac{ B_1 e^{-i\phi_1}}{\sqrt{B_1^2+B_2^2}}  \\
 \pm1 \\
 \frac{ B_2 e^{-i\phi_2}}{\sqrt{B_1^2+B_2^2}}   \\
    \end{pmatrix},
  \ket{u_0}  = \begin{pmatrix}
 \frac{ B_2 e^{-i\phi_1}}{\sqrt{B_1^2+B_2^2}}  \\
 0\\
 \frac{ -B_1 e^{-i\phi_2}}{\sqrt{B_1^2+B_2^2}}   \\
    \end{pmatrix},
\end{equation}
and the corresponding eigenenergies are $\epsilon_{\pm} = \pm \gamma_{e}\sqrt{B_1^2+B_2^2}$ and $\epsilon_{0}=0$. This Weyl-like Hamiltonian hosts a tensor monopole at the origin.

\subsection{Generation of dual-frequency microwave pulses}
\begin{figure*}[ht] 
    \centering
    \includegraphics[width=0.5\textwidth]{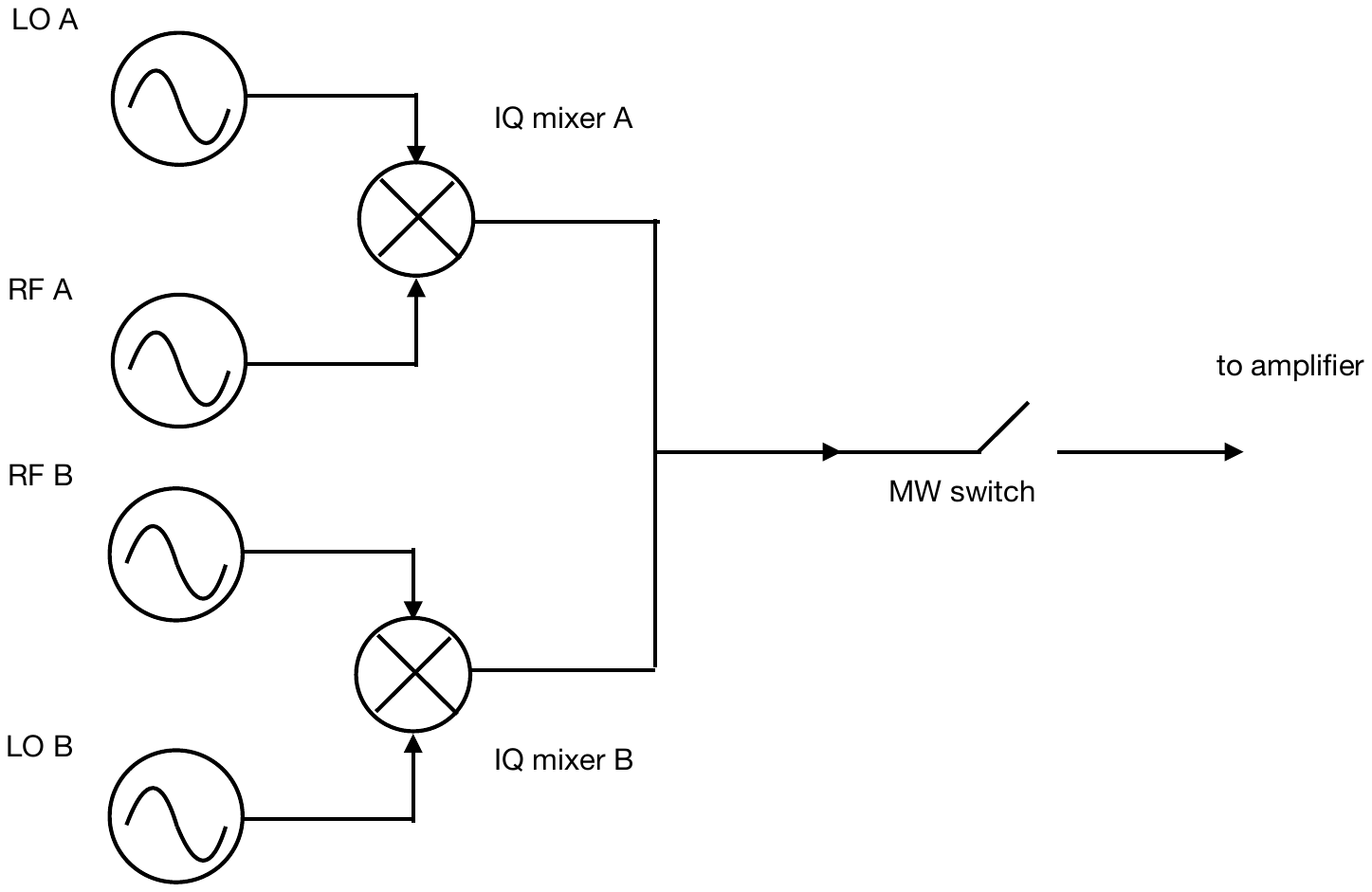}
    \caption{\textbf{Setup schematics}}
    \label{fig:setup_schematics}
\end{figure*}

To achieve precise control over the amplitude and phase of both microwave frequencies, we choose to use frequency modulation with two separate IQ mixers, shown schematically in Fig.~\ref{fig:setup_schematics}. The in-phase (I) and quadrature (Q) RF signals are generated from AWG (Tektronix 5014B) using 3 separate channels. One of the outputs is further split into $0$ and $90^{\circ}$ (Mini-Circuits ZMSCQ-2-90). A two-channel microwave generator (Windfreak SynthHD) generates the Local Oscillator (LO) signals. The IQs and LOs are combined in two IQ mixers (Texas Instrument, TRF370317; Marki Microwave, IQ-0318) that create up-converted single-sideband microwave signals. The output signals are then combined and controlled by a microwave switch (Analog Devices, ADRF5020) before amplified. For brevity, we left out pre-amplifiers in the schematics.

To characterize our engineered Weyl-type Hamiltonian in Eq.~\ref{eq:H_DQ_frame} under dual-frequency microwave driving, we prepare NV in the $\ket{m_s=0}$ state and let it evolve under the DQ Hamiltonian (Eq.~\ref{eq:H_DQ_frame}). When both microwave frequencies are on-resonance, we expect the following time-dependent state evolution
\begin{equation}\label{eq:DQ_state_evolution}
\begin{pmatrix}
c_+(t)\\
c_0(t)\\
c_-(t)
\end{pmatrix}
=\begin{pmatrix}
\frac{-iB_1e^{-i\phi_1}\sin\omega_et}{\sqrt{B_1^2+B_2^2}}\\
\cos\omega_et\\
\frac{-iB_2e^{-i\phi_2}\sin\omega_et}{\sqrt{B_1^2+B_2^2}}
\end{pmatrix},
\end{equation}
with the effective Rabi frequency $\omega_e=\gamma_e\sqrt{B_1^2+B_2^2}$. By measuring the amplitude and frequency of the Rabi oscillation, we can extract both $B_{1(2)}$.

\begin{figure*}[ht] 
    \centering
    \includegraphics[width=0.3\textwidth]{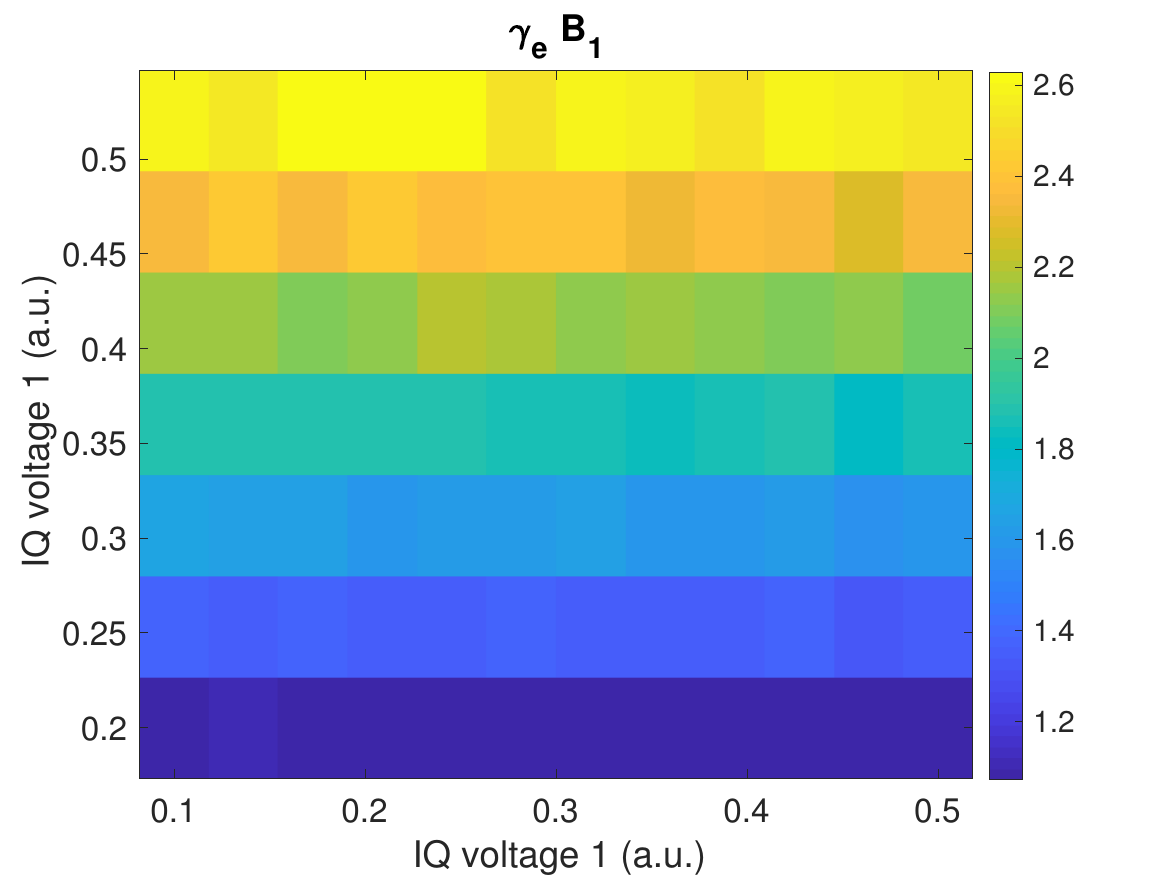}
    \includegraphics[width=0.3\textwidth]{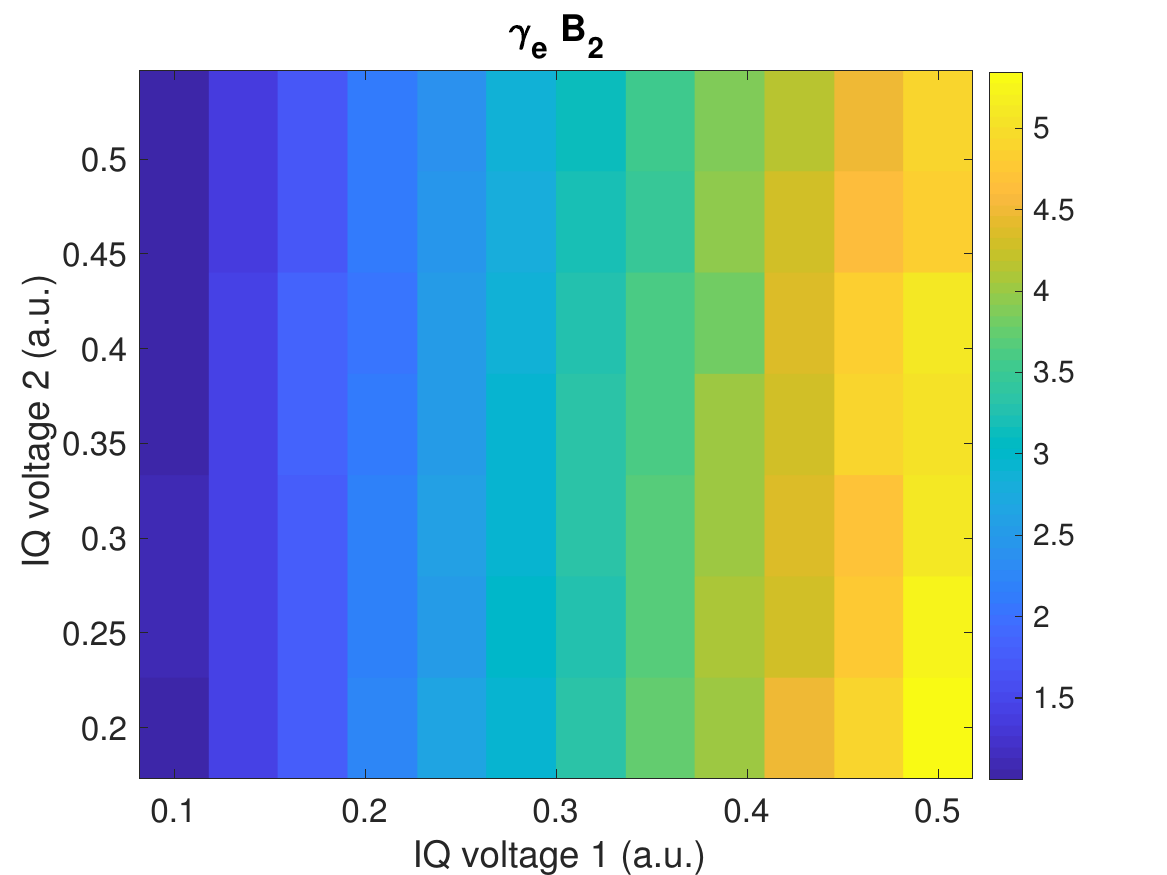}
    \includegraphics[width=0.3\textwidth]{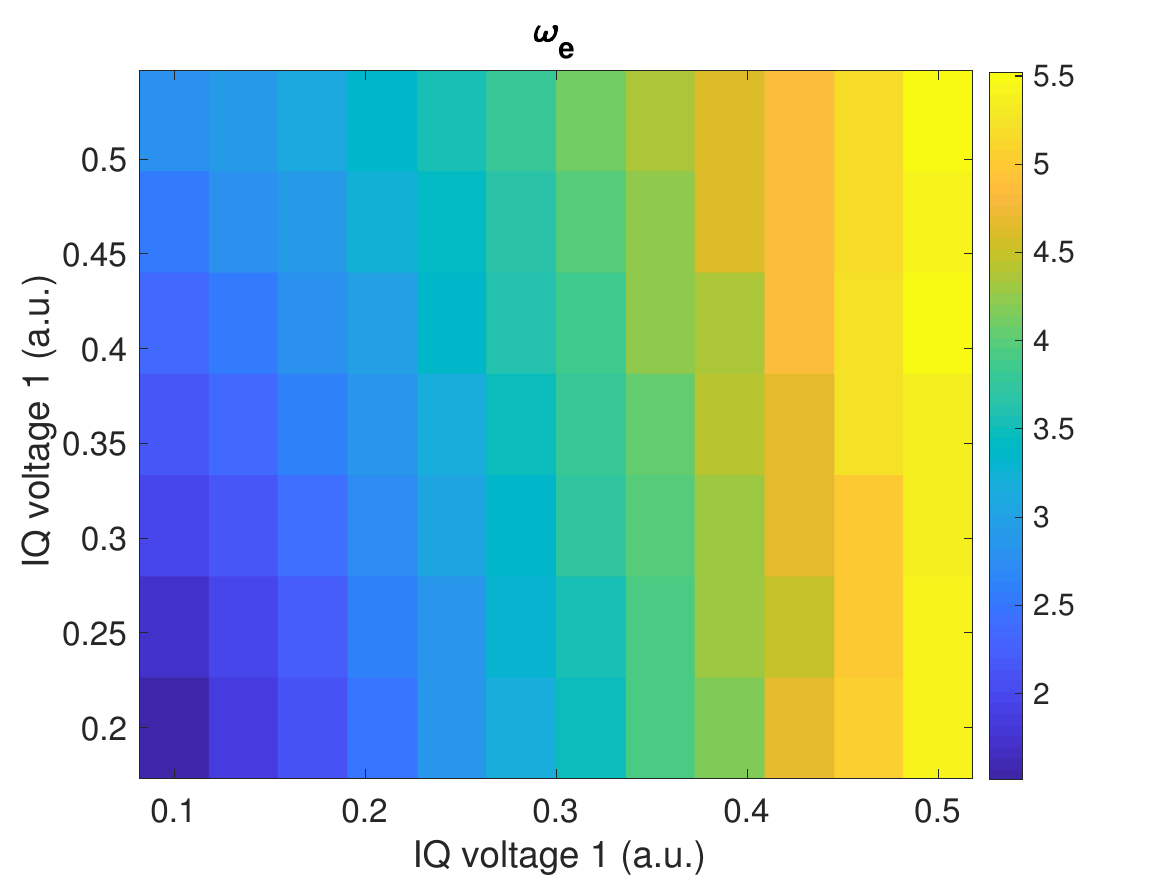}
    \caption{\textbf{IQ voltage calibration} From left to right we show the measured $\gamma_e B_1$ (left), $\gamma_e B_2$ (middle) and $\omega_e$ in units of $(2\pi)$~MHz. $x(y)$-axis represents the IQ voltage for the AWG channels that drive the $\ket{m_s=0}\leftrightarrow \ket{m_s=-1}$ ($\ket{m_s=0}\leftrightarrow \ket{m_s=+1}$) transition.}
    \label{fig:IQ_voltage_calibration}
\end{figure*}

\begin{figure*}[ht] 
    \centering
    \includegraphics[width=0.3\textwidth]{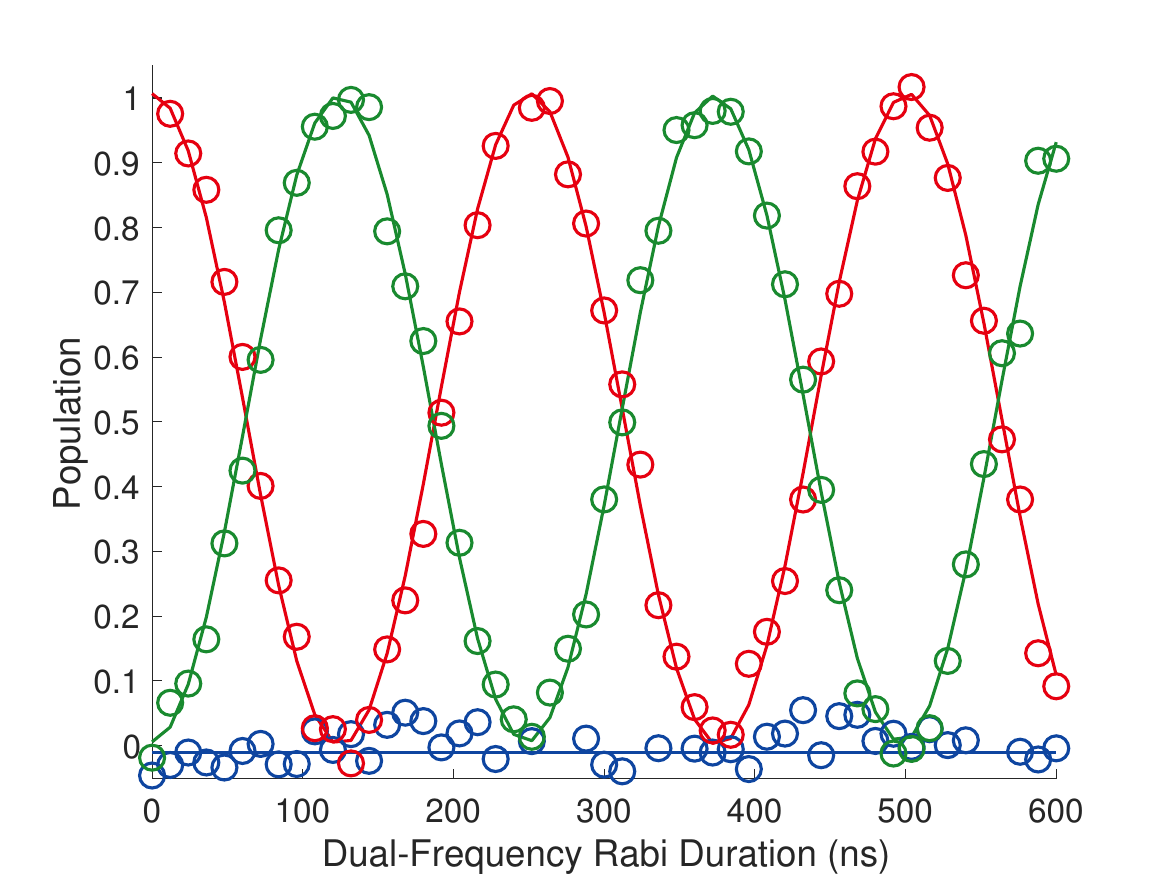}
    \includegraphics[width=0.3\textwidth]{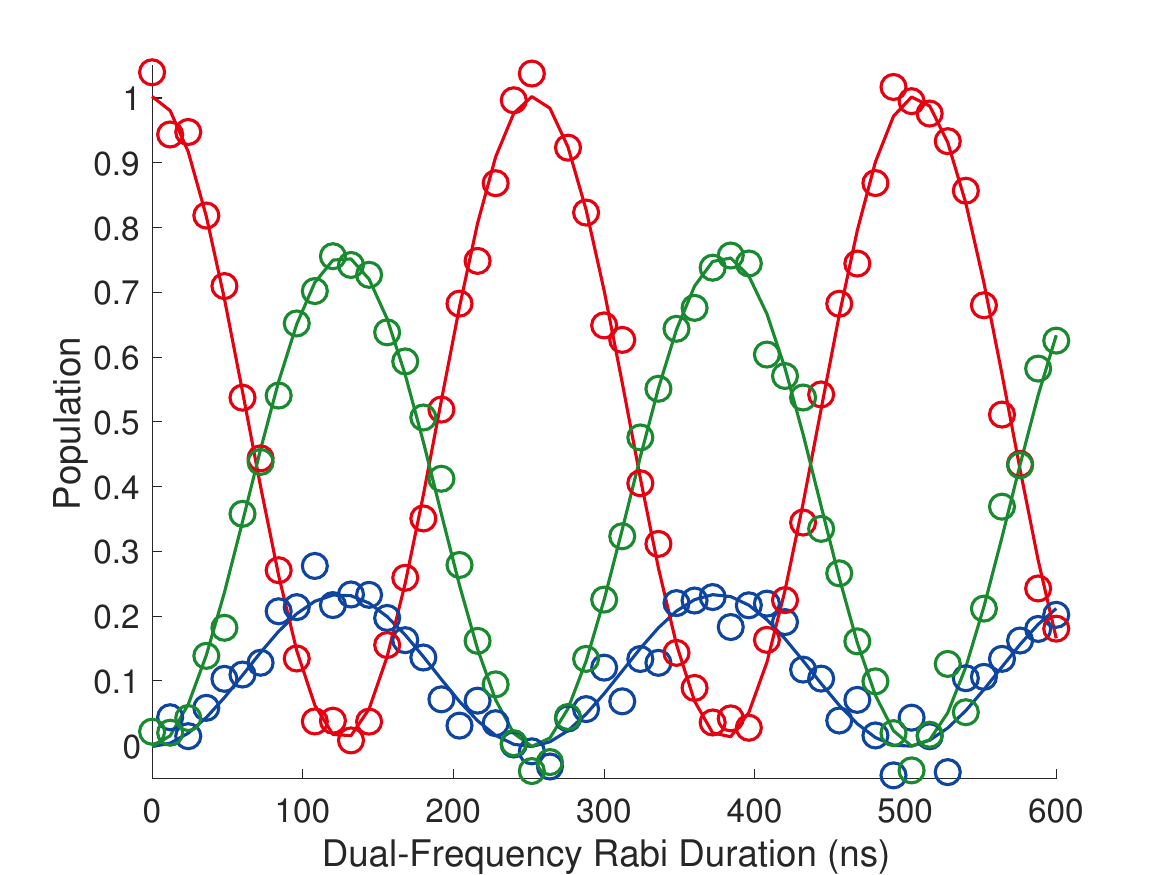}
    \includegraphics[width=0.3\textwidth]{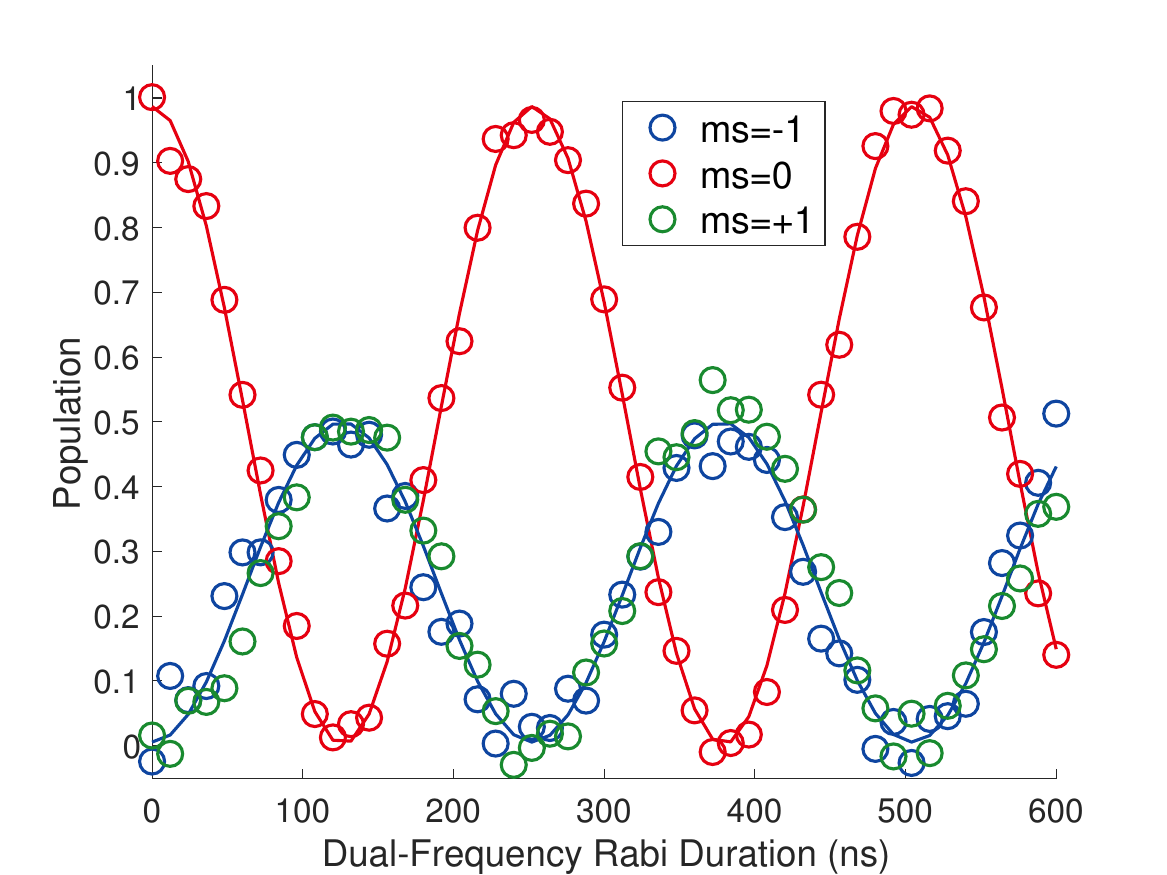}
    \caption{\textbf{Double Quantum Rabi Oscillations.} Evolution of the state $\ket{m_s=0}$  under the DQ Hamiltonian in Eq.~\ref{eq:H_DQ_frame}. The experimental conditions are $\alpha=0$ (left), $\alpha=\pi/6$ (middle), and $\alpha=\pi/4$ (right). The expected oscillation amplitude for the $\ket{m_s=+1}$ ($\ket{m_s=-1}$) state is $\cos^2\alpha$ ($\sin^2\alpha)$), in good agreement with our experiments}
    \label{fig:DQ_Rabi}
\end{figure*}

2D maps of the relationship between IQ voltages and the corresponding Rabi frequencies ($\gamma_e B_1, \gamma_e B_2, \omega_e$) are shown in Fig.~\ref{fig:IQ_voltage_calibration}. We choose to work in the linear regime of the microwave amplifier, where $\omega_e=2$~MHz. A few examples of the state evolution under $\alpha=0, \pi/6, \pi/4$ are shown in Fig.~\ref{fig:DQ_Rabi}. Recalling that we set $B_1=H_0\cos\alpha, B_2=H_0\sin\alpha$, we expect the amplitude of the $\ket{m_s=+1}$ state to be $\cos^2\alpha$, in excellent agreement with the experiments in Fig.~\ref{fig:DQ_Rabi}.

\begin{figure*}[ht] 
    \centering
    \includegraphics[width=0.5\textwidth]{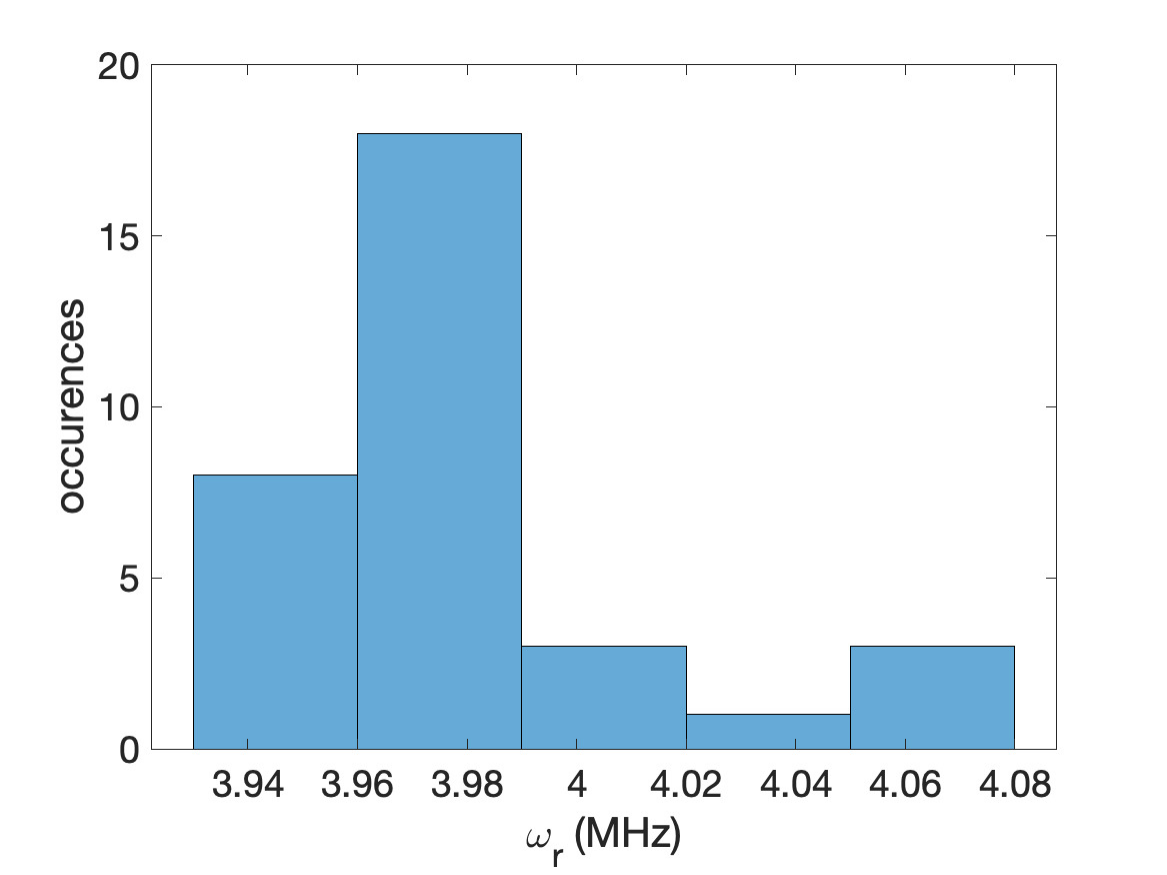}
    \caption{\textbf{Distribution of the resonant parametric modulation frequencies $\omega_r$}}
    \label{fig:omegar_distribution}
\end{figure*}

As a last demonstration, we show in Fig.~\ref{fig:omegar_distribution} a histogram of the resonant parametric modulation frequencies $\omega_r$ measured throughout experiments to measure the tensor monopole. See also Fig.~1(a) in the main text. The fluctuation of $\omega_r$ is well within $2\%$ over the whole $\alpha\in[0,\pi/2]$ range, subject to real experimental conditions including heating due to prolonged microwave driving. This result verifies the (spherical) shape of hypersphere $H_0=2$~MHz we choose in order to reveal the tensor monopole.
 
\subsection{State preparation and readout}\label{sec:state_prep_readout}
At the beginning of every parametric modulation experiment, we polarize the NV into $\ket{m_s=0}$, then apply two microwave pulses to prepare NV into the ground eigenstate $\ket{u_-}$ before subjecting the system to the engineered Weyl-type Hamiltonian. 

In preparing the ground state 
\begin{equation}
\ket{u_-}=\frac{1}{\sqrt{2}}
\begin{pmatrix}
-\cos\alpha e^{-i\beta}\\
1\\
-\sin\alpha e^{-i \phi}
\end{pmatrix}, 
\end{equation}
we first apply a microwave pulse that drives the $-1$ transition $\ket{m_s=0}\leftrightarrow\ket{m_s=-1}$, immediately followed by another pulse for the $+1$ transition $\ket{m_s=0}\leftrightarrow\ket{m_s=+1}$. We set the Rabi frequency of both pulses to be $\omega_{init}$. Then the durations $t_\pm$ and phases $\delta_\pm$ of the two initialization pulses for the $\pm 1$ transitions are
\begin{equation}
\begin{split}
t_-&=\sin^{-1}(\sin\alpha/\sqrt{2})/\omega_{init},\\
\delta_-&=\phi+\pi/2,\\
t_+&=\sin^{-1}(\cos\alpha/\sqrt{2-\sin^2\alpha})/\omega_{init},\\
\delta_+&=\beta+\pi/2.
\end{split}
\end{equation}

Similarly for the first excited state 
\begin{equation}
\ket{u_0}=\begin{pmatrix}
-\sin\alpha e^{-i\beta}\\
0\\
\cos\alpha e^{-i\phi}
\end{pmatrix}
\end{equation}
we have
\begin{equation}
\begin{split}
t_-&=\pi/\omega_{init},\\
\delta_-&=\phi+\pi,\\
t_+&=\pi/2\omega_{init},\\
\delta_+&=\beta.
\end{split}
\end{equation}

When modulating at the DQ frequency $\omega_r=2H_0$, Rabi oscillations occur between these two states, and we apply the inverse mapping to rotate $\ket{u_-}$ back to $\ket{m_s=0}$ for fluorescent readout. When modulating at the SQ frequency $\omega_r=H_0$, the situation is more involved. Due to chiral symmetry of the Weyl-type Hamiltonian, both SQ transitions $\ket{u_0}\leftrightarrow\ket{u_\pm}$ are on-resonance, and they have the same driving strength $\Gamma=\vert\Gamma_{-\lra0}\vert=\vert\Gamma_{+\lra0}\vert$. This leads to an effective DQ Hamiltonian in the eigenbasis of our engineered Weyl-type Hamiltonian. After entering the DQ rotating frame and taking the rotating wave approximation similar to what we did in Eq.~\ref{eq:H_NV_bare}, we obtain:
\begin{equation} \label{eq:SQ_transition_ham}
\hat{H}=\begin{pmatrix}
0&\Gamma e^{-i\phi_1}&0\\
\Gamma e^{i\phi_1}&0&\Gamma e^{i\phi_2}\\
0&\Gamma e^{-i\phi_2}&0
\end{pmatrix},
\end{equation}

where $\phi_{1(2)}$ are the phases associated with $\Gamma_{0\lra\pm}$. We remark that although Eq.~\ref{eq:SQ_transition_ham} is of the same form as Eq.~\ref{eq:H_DQ_frame}, they represent two different Hamiltonians. Eq.~\ref{eq:H_DQ_frame} is the DQ Hamiltonian in the eigenbasis of $\ket{m_s}$, and Eq.~\ref{eq:SQ_transition_ham} is in the eigenbasis of Eq.~\ref{eq:H_DQ_frame}, whose eigenvectors are $\ket{u_{0,\pm}}$. 

Starting from the ground state $\ket{u_-}$, the system under Eq.~\ref{eq:SQ_transition_ham} evolves as:
\begin{equation}
\begin{pmatrix}
c_+(t)\\
c_0(t)\\
c_-(t)
\end{pmatrix}
=\begin{pmatrix}
\frac{(-1+\cos(\sqrt{2}\Gamma t))e^{i(\phi_1-\phi_2)}}{2}\\
\frac{1}{\sqrt{2}}\sin(\sqrt{2}\Gamma t) e^{i\phi_1}\\
\frac{1+\cos(\sqrt{2}\Gamma t)}{2}
\end{pmatrix}.
\end{equation}

For ease of fitting, we map $\ket{u_0}$ back to $\ket{m_s=0}$ and read out the population optically. The matrix element $\Gamma$ of interest is $1/\sqrt{2}m_\mu$ of the fitted Rabi frequency.

We remark that the mapping pulses described in this section do not perform the unitary transformation between the basis $\{\ket{m_s}\}$ and $\{\ket{u}\}$. We emphasize that this unitary transformation could be achieved by three microwave pulses, and is useful in determining the relevant matrix element when the chiral symmetry of the Hamiltonian is broken by the fictitious transverse field, where $\vert\Gamma_{-\lra0}\vert\neq\vert\Gamma_{+\lra0}\vert$, and the modulation frequency is resonant for both SQ transitions. In this case, we prepare the initial state in $\ket{u_0}$, such that it evolves according to Eq.~\ref{eq:DQ_state_evolution}. By measuring both the amplitude and frequency of the Rabi oscillation in $\ket{u_\pm}$ states, we are able to reconstruct the matrix element of interest.

To determine the oscillation amplitudes accurately for all three states, we have to perform three sets of experiments. Each experiment consists of the same state preparation, parametric modulation, and the unitary map back to all three $\ket{m_s}$ states, followed by (i) no operation (ii) $\pi$ pulse between $\ket{m_s=0}\leftrightarrow\ket{m_s=-1}$ and (iii)$\pi$ pulse between $\ket{m_s=0}\leftrightarrow\ket{m_s=+1}$, and then optically read out. The fluorescence signals recorded in each experiments are labelled as $S_i$, the reference fluorescent level for each $\ket{m_s}$ states as measured in separate experiments are $r_{m_s}$, and the final population of each $\ket{m_s}$ states are $n_{m_s}$. From the three sets of experiment, we have
\begin{equation}
\begin{split}
r_+ n_+ + r_0 n_0 + r_- n_- &= S_1,\\
r_+ n_+ + r_- n_0 + r_0 n_- &= S_2,\\
r_0 n_+ + r_+ n_0 + r_- n_- &= S_3.
\end{split}
\end{equation}
It is therefore straightforward to extract the populations accurately
\begin{equation}
\begin{pmatrix}n_+\\n_0\\n_-\end{pmatrix}
=
\begin{pmatrix}
r_+&r_0&r_-\\r_+&r_-&r_0\\r_0&r_+&r_-
\end{pmatrix}^{-1}
\begin{pmatrix}
S_1\\S_2\\S_3
\end{pmatrix}.
\end{equation}

In addition to the parametric modulations, this readout technique is used in e.g. Fig.~\ref{fig:DQ_Rabi} to reveal the accurate populations of all three states.

We remark that this method is only possible when the magnetic field is close to the excited state level anticrossing at $510$~G, where the excited state electron-nuclear spin flip-flops yield distinguishable fluorescent levels for $\ket{m_s=\pm1}$~\cite{Chen15,Hirose16}.

\subsection{Measurement of the Quantum Geometric Tensor}\label{sec:QGT_modulation}

The quantum geometric tensor (QGT) $\chi_{\mu\nu}$ naturally appears when one defines the distance between nearby states $\ket{n(\ve{q})}$ and $\ket{n(\ve{q}+d\ve{q})}$:
\begin{equation}
    ds^{2} \equiv 1-|\bra{n(\ve{q})}{n(\ve{q}+d\ve{q})\rangle}|^2 = dq_{\mu} \chi_{\mu\nu} dq_{\nu} + O(|d\ve{q}|^3),
\end{equation}
where we apply the Taylor expansion about $d\ve{q}=0$ and assume a generic system parametrized by the generalized position $\ve{q}$. Here $\chi_{\mu\nu}$ contains information on the geometry of the manifold for the state $\ket{n}$ and is found to be:
\begin{equation}\label{eq:QGTdef}
\chi_{\mu\nu}^{(n)}=\bra{\partial_\mu n(\ve{q})}(\ve{1}-\ket{n(\ve{q})}\bra{n(\ve{q})})\ket{\partial_\nu n(\ve{q})} = g_{\mu\nu}  + i\mathcal{F}_{\mu\nu}/2,
\end{equation}
where the symmetric part is the metric tensor $g_{\mu\nu}$ and determines the distance between the states, and the anti-symmetric part is the conventional 2-form Berry curvature.

Using time-independent perturbation theory to first order we obtain
\begin{equation}
\begin{split}
&\ket{n^{(1)}}=\ket{\partial_\mu n(\ve{q})}=\sum_{k\neq n} \frac{\bra{k(\ve{q})} \partial_\mu\hat{H} \ket{n(\ve{q})}}{\epsilon_n-\epsilon_k}\ket{k(\ve{q})}
\end{split}
\end{equation}

Then we can plug in Eq.~\ref{eq:QGTdef} and simplify to the following form
\begin{equation}\label{eq:QGTusefulform}
\begin{split}
\chi_{\mu\nu}^{(n)}&=\sum_{k\neq n}\sum_{m\neq n} 
\frac{\bra{k} \partial_\mu\hat{H} \ket{n}^\dagger}{\epsilon_n-\epsilon_k}\bra{k}
(\ve{1}-\ket{n}\bra{n})
\frac{\bra{m} \partial_\nu\hat{H} \ket{n}}{\epsilon_n-\epsilon_m}\ket{m}\\
&=\sum_{m\neq n} 
\frac{\bra{m} \partial_\mu\hat{H} \ket{n}^\dagger}{\epsilon_n-\epsilon_m}
\frac{\bra{m} \partial_\nu\hat{H} \ket{n}}{\epsilon_n-\epsilon_m}\\
&=\sum_{m\neq n} 
\frac{\bra{n} \partial_\mu\hat{H} \ket{m}\bra{m} \partial_\nu\hat{H} \ket{n}}{(\epsilon_n-\epsilon_m)^2}\\
\end{split}
\end{equation}

As discussed earlier, the real part of the QGT, namely the metric tensor, contains all the information about the monopole, and it could be used to reconstruct the 3-form curvature, while the imaginary part, namely the Berry curvature, is connected with the tensor Berry connection. In the following, we show how to experimentally measure the metric tensor and the Berry curvature.

\subsubsection{Metric tensor}
The technique we use is parametric modulation of the system Hamiltonian~\cite{Ozawa2018,Yu2019}. We apply the following linear modulations
\begin{equation}
\begin{split}
\mu_t&=\mu_0+m_\mu \sin\omega t,\\
\nu_t&=\nu_0+m_\nu \sin\omega t.
\end{split}
\end{equation}

If we modulate the parameters weakly, $m_\mu, m_\nu \ll 1$, then
\begin{equation}\label{eq:parametric_modulation}
    \hat{H}\approx \hat{H}(\ve{q}_0) + m_\mu \partial_\mu\hat{H}\sin\omega t  + m_\nu \partial_\nu\hat{H}\sin\omega t.
\end{equation}

When the modulation frequency is resonant with the energy difference between eigenstates, the parametrically modulated Hamiltonian will drive coherent Rabi oscillations between relevant eigenstates. 

Here we consider the case where $\omega=\epsilon_+-\epsilon_-$ is resonant with the double quantum (DQ) transition between $\ket{u_-}\leftrightarrow\ket{u_+}$. Then we have the Rabi frequency 
\begin{equation}
\begin{split}
\Omega_{-\leftrightarrow +}&=\vert\bra{u_-} \hat{H}(\ve{q}_0) + m_\mu \partial_\mu\hat{H}  + m_\nu \partial_\nu\hat{H} \ket{u_+}\vert\\
&=\vert\bra{u_-} m_\mu \partial_\mu\hat{H}  + m_\nu \partial_\nu\hat{H} \ket{u_+}\vert
\end{split}
\end{equation}

To measure the diagonal component $g_{\mu\mu}$ of the metric tensor, we set $m_\nu=0$:
\begin{equation}
\Omega_{-\lra +}^\mu\equiv m_\mu\Gamma_{-\lra +}^\mu=m_\mu\vert\bra{u_-} \partial_\mu\hat{H}\ket{u_+}\vert.
\end{equation}
Similarly, we can obtain the contribution from SQ transition $\Gamma_{-\lra 0}^\mu$ (some complication arises for SQ, which is discussed in Sec.~\ref{sec:state_prep_readout}). Using the alternative form of QGT in Eq.~\ref{eq:QGTusefulform}, we obtain the diagonal components of the metric tensor from experimentally measurable quantities:
\begin{equation}
g_{\mu\mu}=\sum_{m\in{0,+}} \frac{(\Gamma_{-\lra m}^\mu)^2}{(\epsilon_m-\epsilon_-)^2}.
\end{equation}

To measure the off-diagonal components $g_{\mu\nu}$, we modulate both parameters such that $m_\mu=\pm m_\nu$. Then the coherent Rabi oscillation is:
\begin{equation}
\Omega_{-\lra +}^{\mu\pm\nu}=m_\mu\vert\bra{u_-} \partial_\mu\hat{H}\pm \partial_\nu\hat{H}\ket{u_+}\vert.
\end{equation}
Setting $\Gamma_{-\lra +}^{\mu\pm\nu}=\Omega_{-\lra +}^{\mu\pm\nu}/m_\mu$, we have
\begin{equation}
(\Gamma_{-\lra +}^{\mu\nu})^2-(\Gamma_{-\lra +}^{\mu\bar{\nu}})^2=4\vert\bra{u_-}\partial_\mu\hat{H}\ket{u_+} \bra{u_+}\partial_\nu\hat{H}\ket{u_-}\vert
\end{equation}
We thus obtain an expression for the off-diagonal components
\begin{equation}
g_{\mu\nu}=\sum_{m\in{0,+}} \frac{(\Gamma_{-\lra m}^{\mu\nu})^2-(\Gamma_{-\lra m}^{\mu\bar{\nu}})^2}{4(\epsilon_m-\epsilon_-)^2}.
\end{equation}

We remark that $g_{\mu\nu}=g_{\nu\mu}$, and there are in total 6 independent components in a 4D parameter space. The long coherence time ($T_2>1$~ms under multipulse dynamical decoupling) of the NV center allows us to extract the metric tensor from measuring these Rabi oscillations. 

\subsubsection{Berry curvature}
We next summarize the experimental details of measuring the Berry curvature, which is in turn used to get the experimental $\mathcal{DD}$ invariant (Eq.~\ref{eq:H_in_F}). We use the following elliptical modulation of the system Hamiltonian~\cite{Ozawa2018,Yu2019}:
\begin{equation}
\begin{split}
\mu_t&=\mu_0+m_\mu \cos\omega t,\\
\nu_t&=\nu_0+m_\nu \sin\omega t.
\end{split}
\end{equation}
When the modulation amplitude is small, we have
\begin{equation}
       \hat{H} \approx \hat{H}(\bold{q}_0)+m_\mu \partial_\mu \hat{H}\cos\omega t +m_\nu\partial_\nu \hat{H}\sin\omega t,
\end{equation}
and we could then measure $\bra{u_-} \partial_\mu H \pm i \partial_\nu H\ket{m}$, and obtain the imaginary part of the QGT. Next we lay out the exact realization in a 3-level system
For SQ and DQ drive, where we need two different interaction pictures.
\subparagraph{SQ interaction picture}
In general, $\omega_1=\epsilon_+-\epsilon_0\neq\omega_2=\epsilon_0-\epsilon_-$. We define the unitary (more details in Sec.~\ref{sec:DQ_rotating_frame})
\begin{equation}
    V=\begin{pmatrix}e^{-i\omega t}&0&0\\
    0&1&0\\0&0&e^{-i\omega t}.
    \end{pmatrix}
\end{equation}
For the relevant driving term, we have
\begin{equation}
    H_x=\partial_\mu H=\begin{pmatrix} 0&B^*&C^*\\B&0&A^*\\C&A&0\end{pmatrix}.
\end{equation}
After rotating wave approximation, we have
\begin{equation}
\begin{split}
    V^\dagger (H_x\cos(\omega t)) V &=
    \frac{1}{2}\begin{pmatrix}0&B^*&0\\B&0&A^*\\0&A&0\end{pmatrix}\\
     V^\dagger (H_x\sin(\omega t)) V &=
    \frac{i}{2}\begin{pmatrix}0&B^*&0\\-B&0&-A^*\\0&A&0\end{pmatrix}
\end{split}
\end{equation}

To measure the $(3,2)$ matrix element, our choice of the elliptical drive is obvious: 
\begin{equation}
    \partial_\mu H \cos(\omega t) \pm \partial_\nu H \sin(\omega t) \leftrightarrow \vert \bra{u_-} \partial_\mu H \pm i \partial_\nu H \ket{u_0}\vert
\end{equation}

However, note that here $\omega=\epsilon_- -\epsilon_0<0$. In experiment, we always use $\tilde{\omega}=\vert\omega\vert$, therefore here we have
\begin{equation}
    \partial_\mu H \cos(\tilde{\omega} t) \mp \partial_\nu H \sin(\tilde{\omega} t) \leftrightarrow \vert \bra{u_-} \partial_\mu H \pm i \partial_\nu H \ket{u_0}\vert
\end{equation}

\subparagraph{DQ interaction picture}
For the DQ drive, the middle state is not involved. Then it is effectively a two level system
\begin{equation}
    V=\begin{pmatrix}e^{-i\omega t}&0&0\\
    0&1&0\\0&0&e^{i\omega t}.
    \end{pmatrix}
\end{equation}
After rotating wave approximation,
\begin{equation}
\begin{split}
    V^\dagger (H_x\cos(\omega t)) V &=
    \frac{1}{2}\begin{pmatrix}0&0&C^*\\0&0&0\\C&0&0\end{pmatrix}\\
     V^\dagger (H_x\sin(\omega t)) V &=
    \frac{i}{2}\begin{pmatrix}0&0&C^*\\0&0&0\\-C&0&0\end{pmatrix}
\end{split}
\end{equation}

To measure the $(3,1)$ matrix element, our choice of the elliptical drive is : 
\begin{equation}
    \partial_\mu H \cos(\omega t) \mp \partial_\nu H \sin(\omega t) \leftrightarrow \vert \bra{u_-} \partial_\mu H \pm i \partial_\nu H \ket{u_+}\vert
\end{equation}

Combining the SQ and DQ analysis, we see that in both cases, we should have $\cos, -\sin$ elliptical modulations.

\subparagraph{Degenerate SQ transitions}
A special case is when $B_z=0$, then $\omega_1=-\omega_2$. In this case, when we modulate at the SQ transition frequency $\vert\omega_1\vert=\vert\omega_2\vert$, both SQ transitions will turn on and yield an effective DQ transition. In experiment, we start from $\ket{u_0}$. After the modulation, we perform a unitary map between $\ket{u_{-,0,+}}\leftrightarrow\ket{m_s=-1,0,+1}$. The population oscillations should follow
\begin{equation}\label{eq:degenerate_population_oscillation}
    \begin{pmatrix}n_+\\n_0\\n_-\end{pmatrix}=
    \begin{pmatrix}\frac{B_1^2}{B_1^2+B_2^2}\sin^2(\omega_e t)\\\cos^2(\omega_e t)\\\frac{B_2^2}{B_1^2+B_2^2}\sin^2(\omega_e t)\end{pmatrix}.
\end{equation}

In terms of the parametric modulation, we have,
\begin{equation}
    V=\begin{pmatrix}e^{-i\omega t}&0&0\\
    0&1&0\\0&0&e^{i\omega t}.
    \end{pmatrix}
\end{equation}

\begin{equation}
\begin{split}
    V^\dagger (H_x\cos(\omega t)) V &=
    \frac{1}{2}\begin{pmatrix}0&B^*&0\\B&0&A^*\\0&A&0\end{pmatrix}\\
     V^\dagger (H_x\sin(\omega t)) V &=
    \frac{i}{2}\begin{pmatrix}0&B^*&0\\-B&0&A^*\\0&-A&0\end{pmatrix}
\end{split}
\end{equation}

Upon $\cos/\sin$ elliptical modulation, we find the correspondence
\begin{equation}
    \begin{split}
    B_{1}&=\vert \bra{u_+}\partial_\mu H \pm i \partial_\nu H\ket{u_0} \vert = \vert \bra{u_-}\partial_\mu H \pm i \partial_\nu H\ket{u_0} \vert\\
    B_{2}&=\vert \bra{u_-}\partial_\mu H \mp i \partial_\nu H\ket{u_0} \vert
    \end{split}
\end{equation}

\subsection{Coherent Rabi oscillations under parametric modulations}
\begin{figure*}[ht] 
    \centering
    \includegraphics[width=0.4\textwidth]{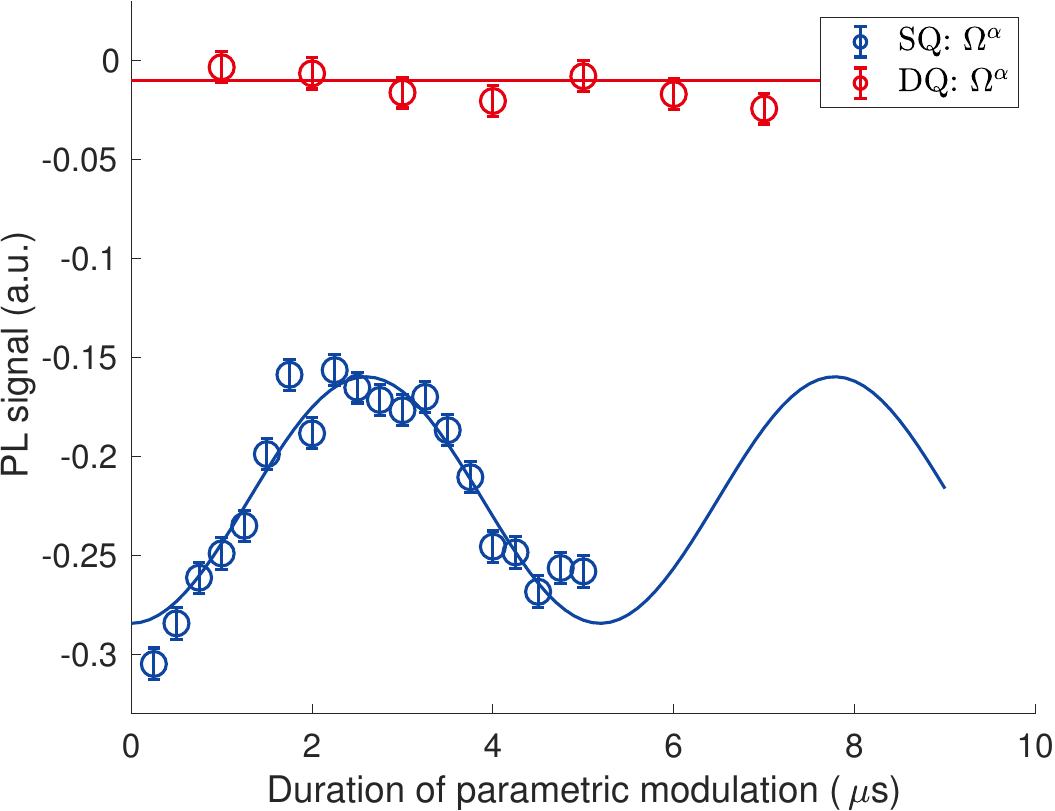}
    \includegraphics[width=0.4\textwidth]{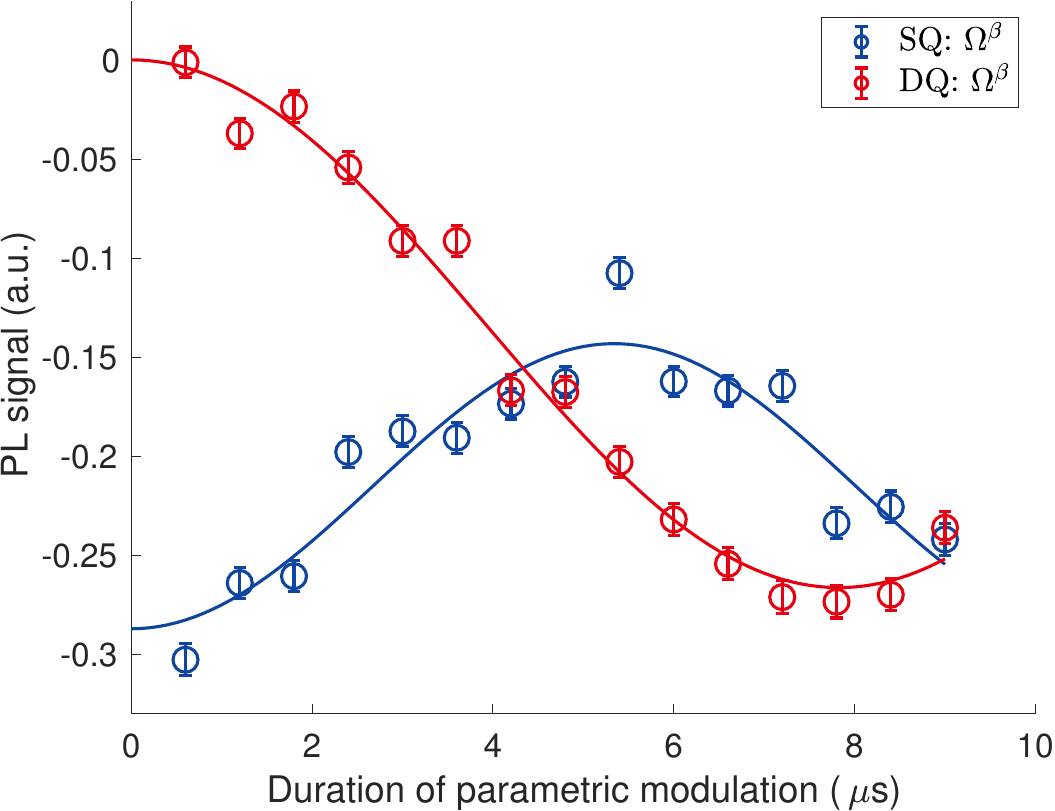}
    \caption{\textbf{Coherent Rabi oscillations for $\Omega^\alpha$ (left) and $\Omega^\beta$ (right)}. The experiment is performed at $(\alpha=5\pi/16, \beta=\phi=0)$. In each plot, blue is for SQ transition and red for DQ transition. The circles are experimental data and solid lines are sinusoidal fits. 
    }
    \label{fig:rabi_data1}
\end{figure*}

\begin{figure*}[ht] 
    \centering
    \includegraphics[width=0.4\textwidth]{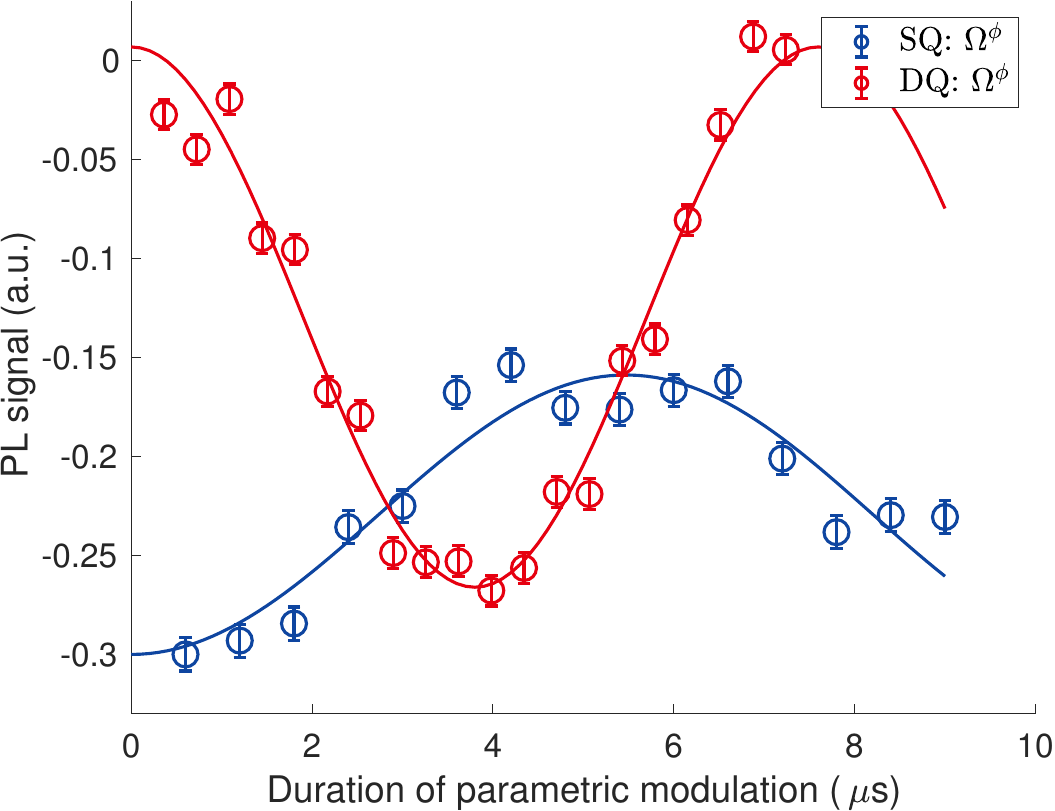}
    \includegraphics[width=0.4\textwidth]{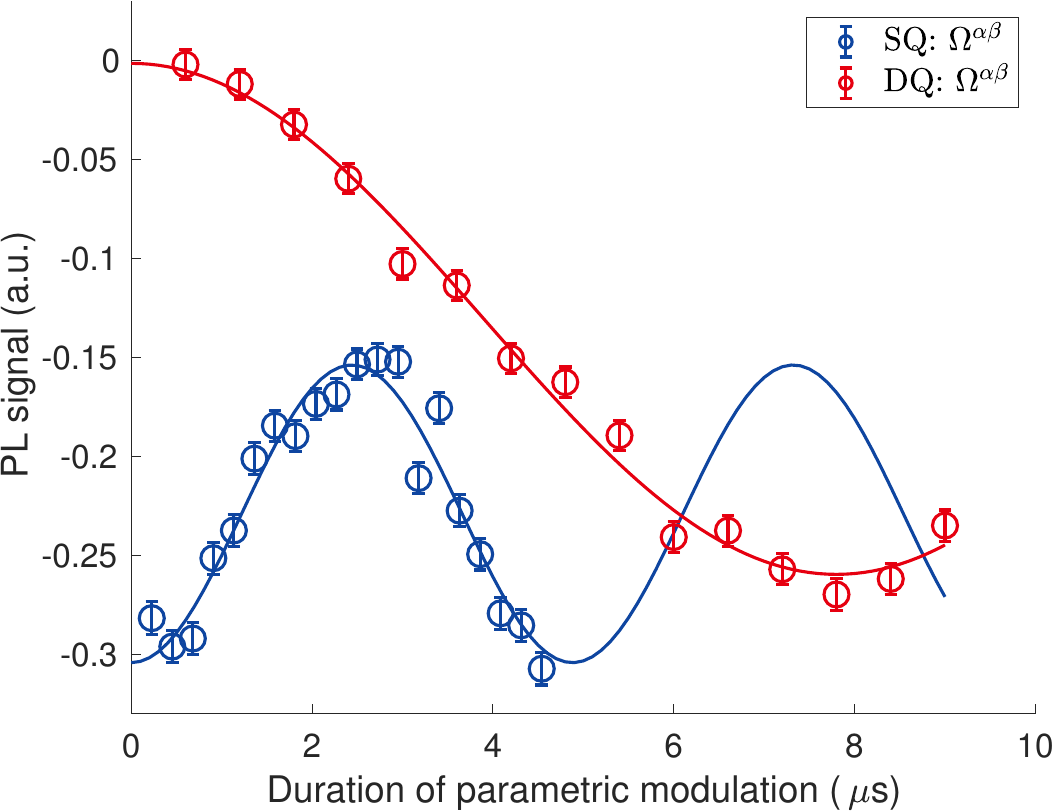}
    \caption{\textbf{Coherent Rabi oscillations for $\Omega^\phi$ (left) and $\Omega^{\alpha\beta}$ (right)}. The experiment is performed at $(\alpha=5\pi/16, \beta=\phi=0)$. In each plot, blue is for SQ transition and red for DQ transition. The circles are experimental data and solid lines are sinusoidal fits. 
    }
    \label{fig:rabi_data2}
\end{figure*}

\begin{figure*}[ht] 
    \centering
    \includegraphics[width=0.4\textwidth]{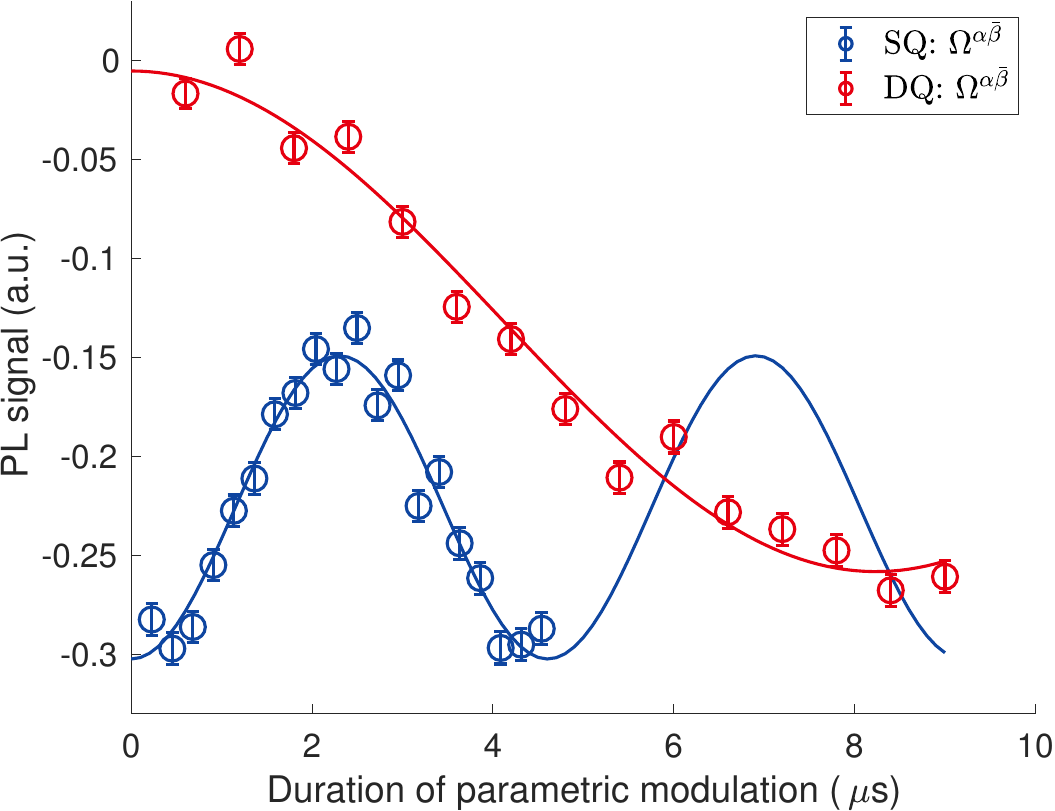}
    \includegraphics[width=0.4\textwidth]{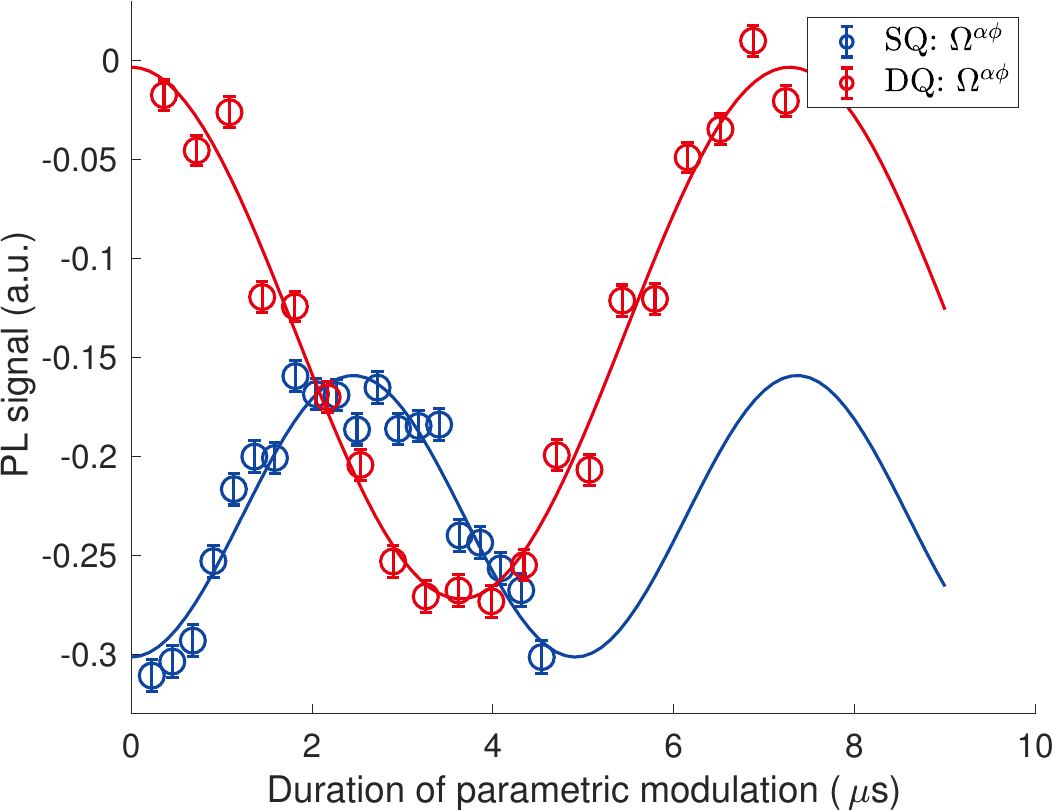}
    \caption{\textbf{Coherent Rabi oscillations for $\Omega^{\alpha\bar{\beta}}$ (left) and $\Omega^{\alpha\phi}$ (right)}. The experiment is performed at $(\alpha=5\pi/16, \beta=\phi=0)$. In each plot, blue is for SQ transition and red for DQ transition. The circles are experimental data and solid lines are sinusoidal fits. 
    }
    \label{fig:rabi_data3}
\end{figure*}

\begin{figure*}[ht] 
    \centering
    \includegraphics[width=0.4\textwidth]{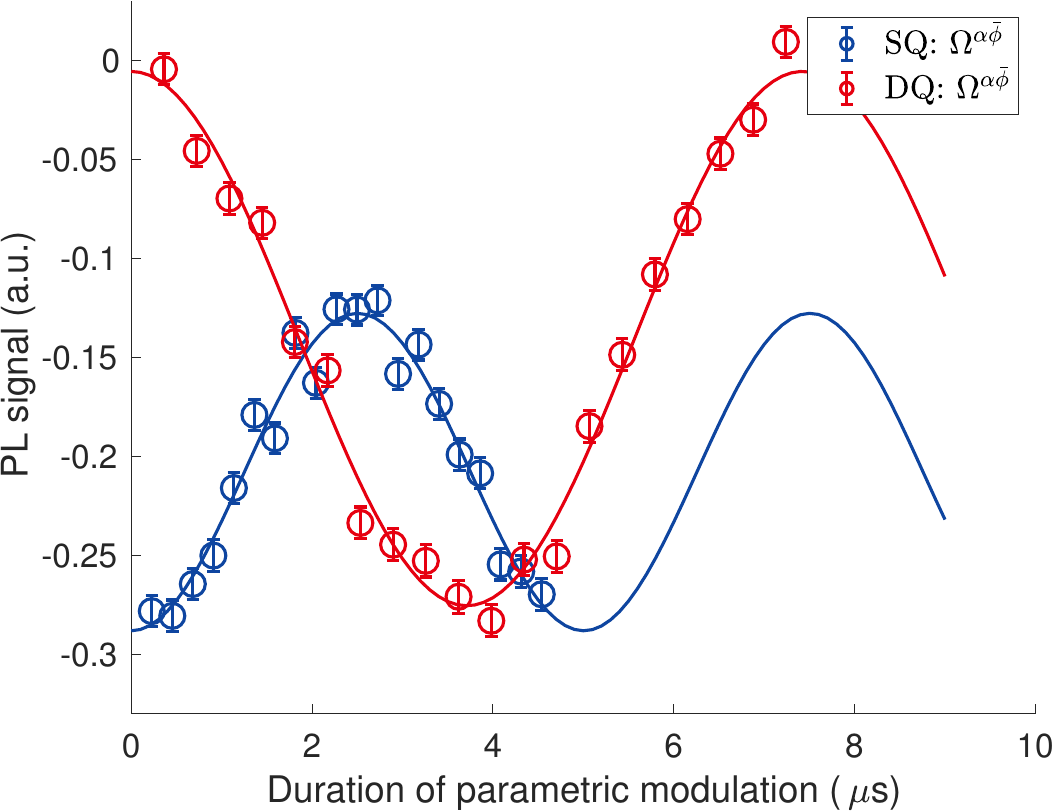}
    \includegraphics[width=0.4\textwidth]{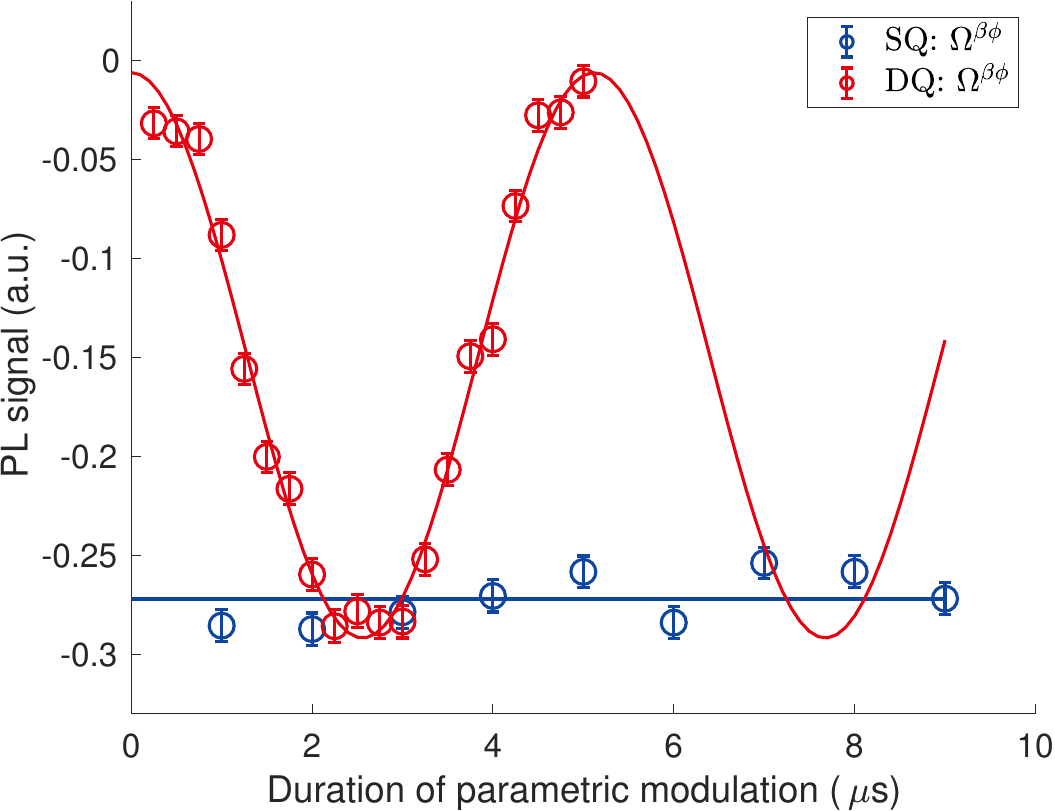}
    \caption{\textbf{Coherent Rabi oscillations for $\Omega^{\alpha\bar{\phi}}$ (left) and $\Omega^{\beta\phi}$ (right)}. The experiment is performed at $(\alpha=5\pi/16, \beta=\phi=0)$. In each plot, blue is for SQ transition and red for DQ transition. The circles are experimental data and solid lines are sinusoidal fits. 
    }
    \label{fig:rabi_data4}
\end{figure*}

\begin{figure*}[ht] 
    \centering
    \includegraphics[width=0.4\textwidth]{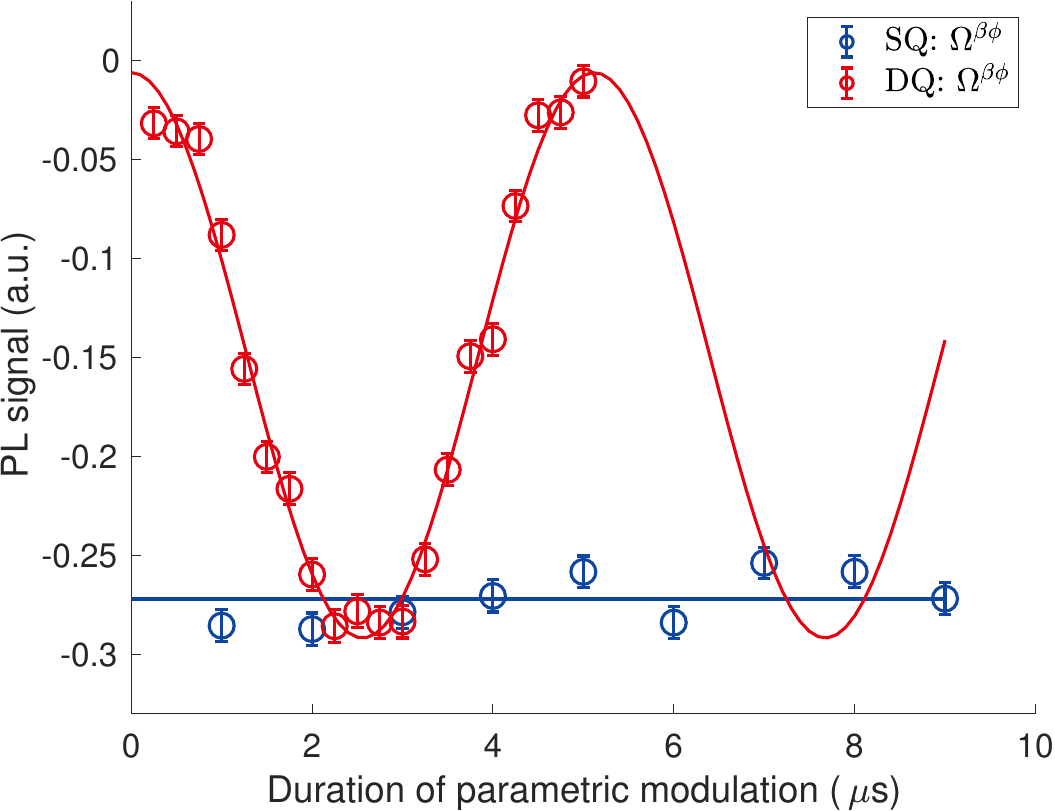}
    \caption{\textbf{Coherent Rabi oscillations for $\Omega^{\beta\bar{\phi}}$}. The experiment is performed at $(\alpha=5\pi/16, \beta=\phi=0)$. In each plot, blue is for SQ transition and red for DQ transition. The circles are experimental data and solid lines are sinusoidal fits. 
    }
    \label{fig:rabi_data5}
\end{figure*}

With the capability of state preparation and readout described above, we now show the coherent Rabi oscillations observed in experiments under appropriate linear parametric modulations. In Fig.~\ref{fig:rabi_data1}-~\ref{fig:rabi_data5}, we plot all 18 Rabi oscillations (9 for SQ transitions and 9 for DQ transitions) measured for $\alpha=5\pi/16,\beta=\phi=0$, which are in turn used to obtain the matrix elements $\Gamma$ (Fig.~1 in main text) to extract all 6 independent metric tensor components $g_{\mu\nu}$ (Fig.~2 in main text), and ultimately yields the $\mathcal{DD}$ invariant, as described in previous sections and the main text.

\subsection{Experimental verification of rotation symmetry about $\beta, \phi$}
\begin{figure*}[ht] 
    \centering
    \includegraphics[width=0.4\textwidth]{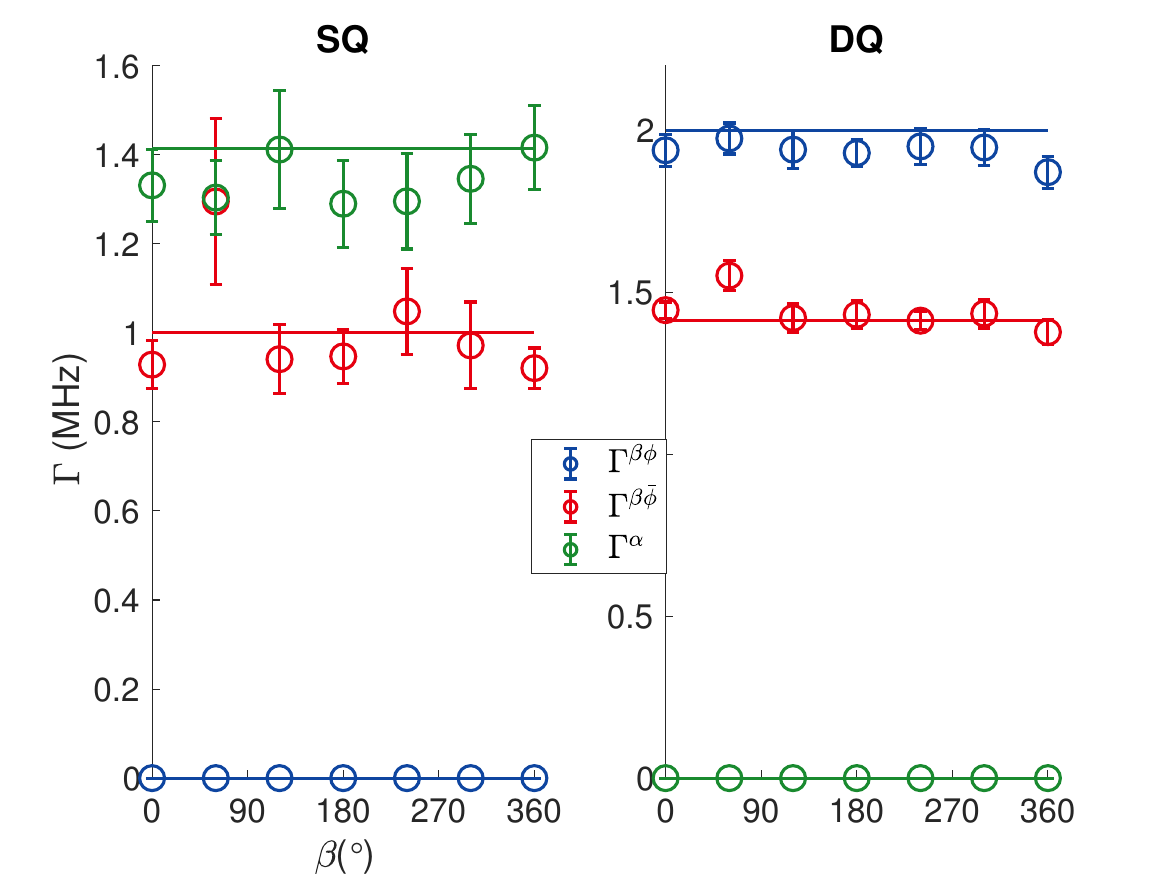}
    \includegraphics[width=0.4\textwidth]{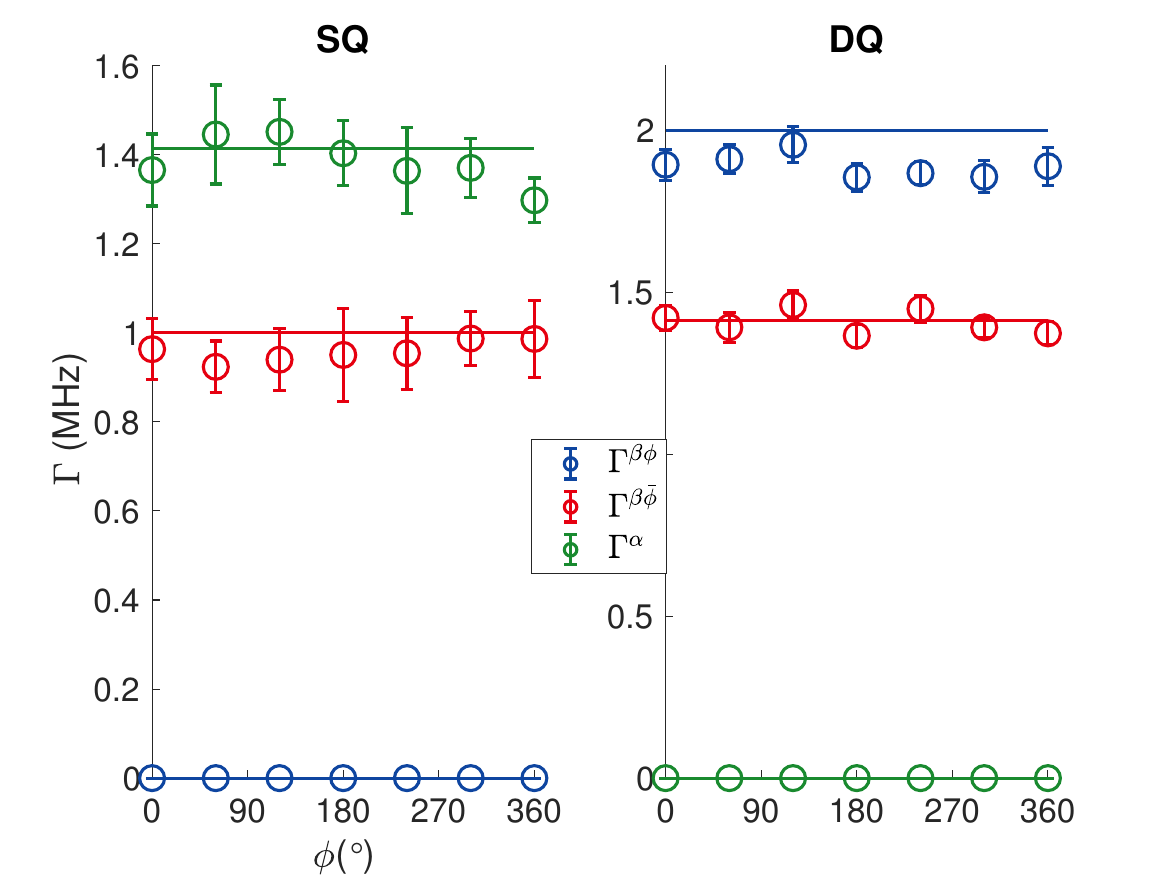}
    \includegraphics[width=0.4\textwidth]{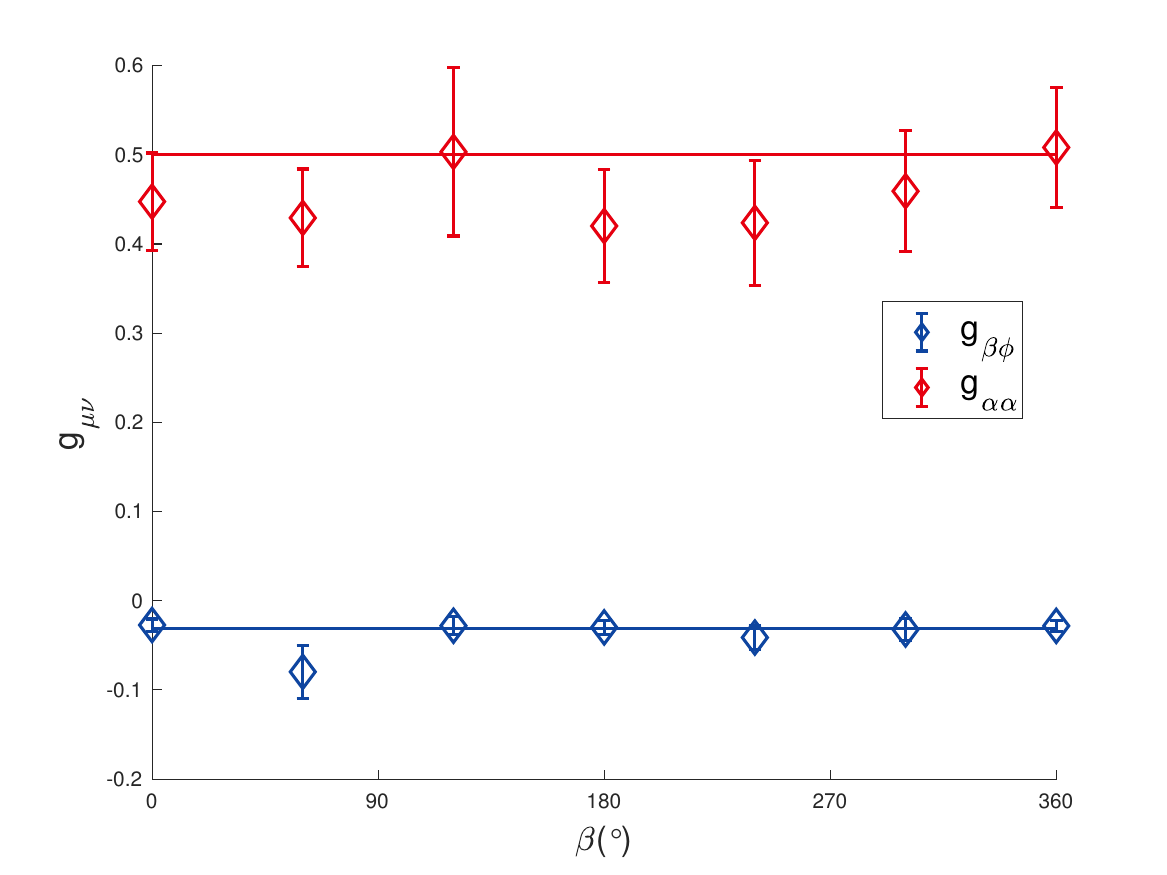}
    \includegraphics[width=0.4\textwidth]{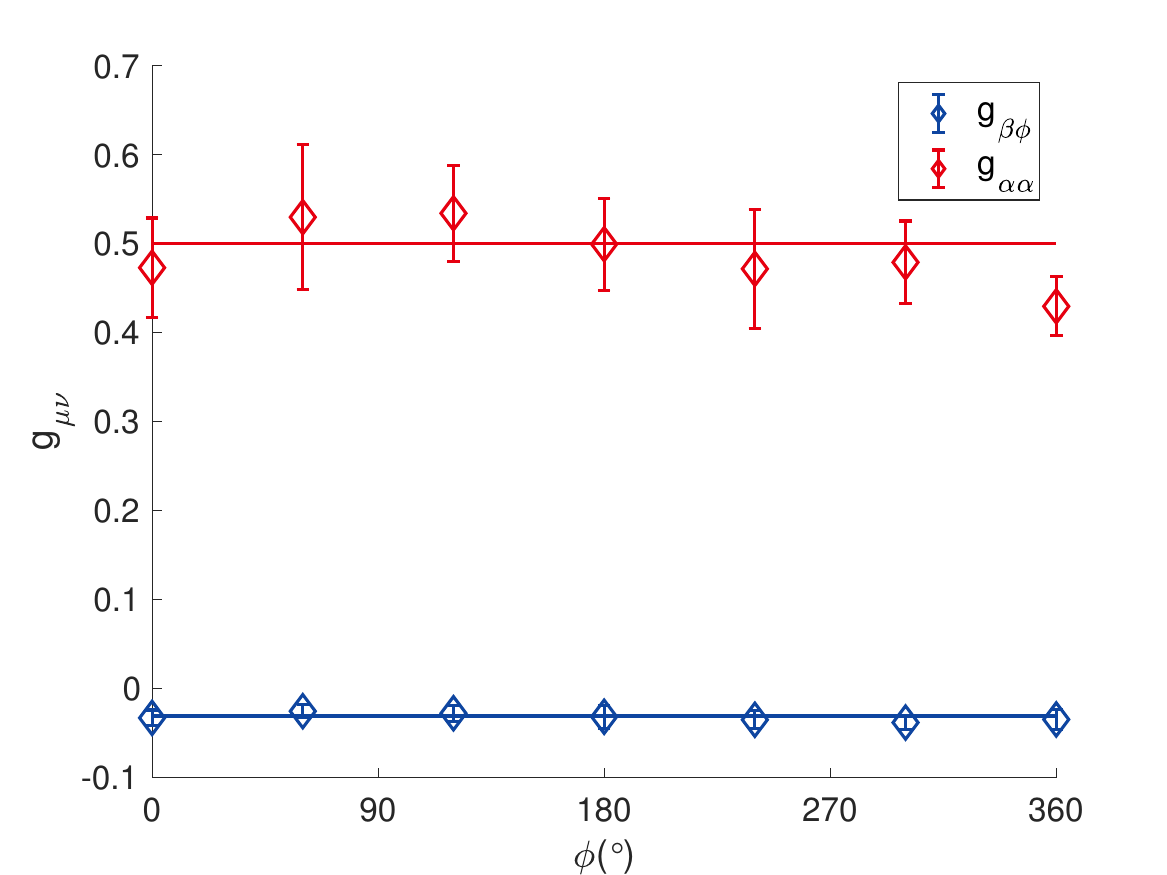}\caption{\textbf{Rotation symmetry of $\beta, \phi$} We perform experiments at $\alpha=\pi/8$, sweeping either $\beta$ or $\phi$ and measuring $g_{\alpha\alpha}, g_{\beta\phi}$ to verify that the metric tensor is independent of $\beta, \phi$ due to the rotation symmetry of the Hamiltonian. On the left (right) we show results when fixing $\phi=0$ ($\beta=0$) and sweeping $\beta$ ($\phi$). The top panel shows relevant matrix element measurements $\Gamma^{\alpha}$ (green), $\Gamma^{\beta\phi}$ (blue), $\Gamma^{\beta\bar{\phi}}$ (red) in circles, and theoretical values in solid lines. The bottom panel shows extracted metric tensor components. Within experimental error they stay constant over $\beta, \phi$.
    }
    \label{fig:rotation_symmetry}
\end{figure*}

The Weyl-type Hamiltonian is rotationally symmetric about $\beta, \phi$ under our parametrization of $(H_0, \alpha, \beta, \phi)$. As a result, the metric tensor and generalized 3-form Berry curvature is independent of $\beta, \phi$. To show that this indeed occurs in our experiments, we fix $\alpha=\pi/8$ and sweep $\beta, \phi\in[0,2\pi]$, and measure the corresponding matrix element $\Gamma$ for the metric tensor components $g_{\alpha\alpha}, g_{\beta\phi}$. The results are shown in Fig.~\ref{fig:rotation_symmetry}, where we indeed see these measurements remain constant within experimental error for different $\beta, \phi$.

\subsection{Matrix element measurements for elliptical parametric modulation}
\begin{figure*}[ht] 
    \centering
    \includegraphics[width=0.32\textwidth]{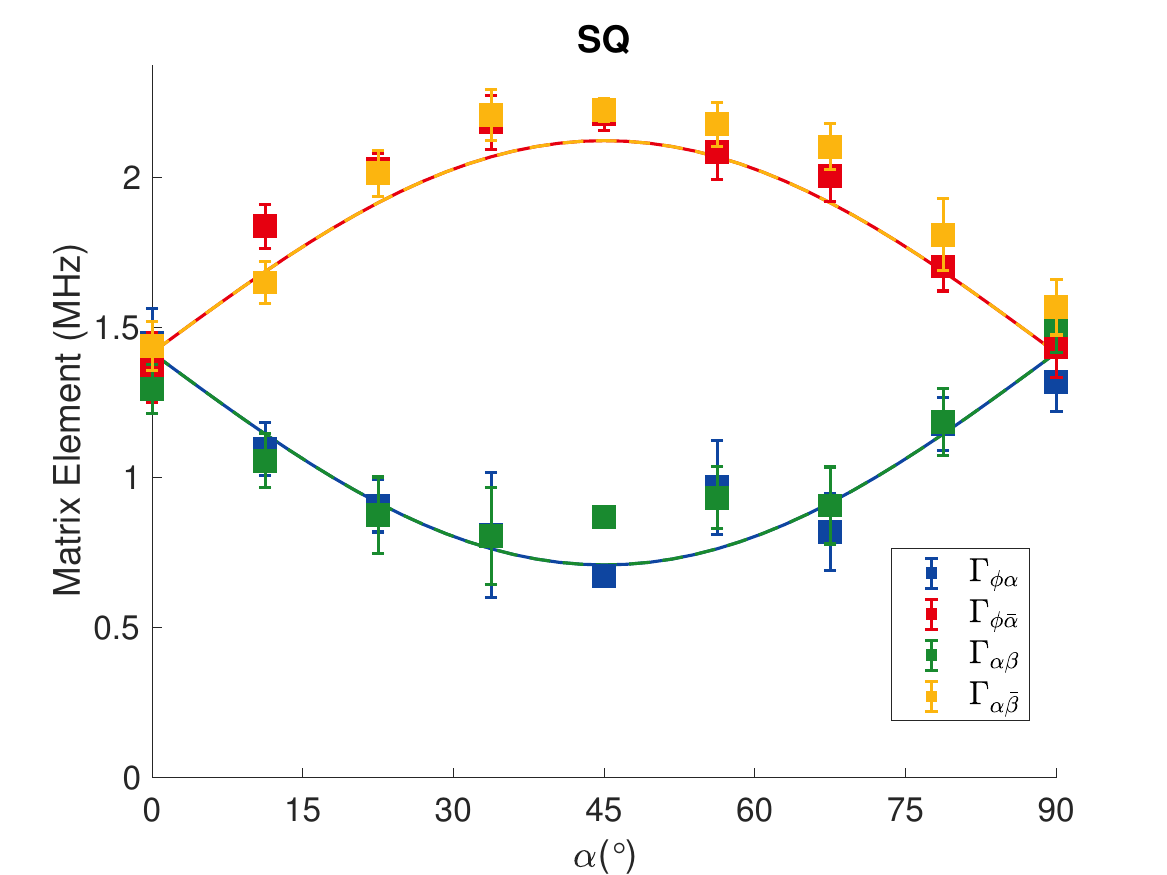}
    \includegraphics[width=0.32\textwidth]{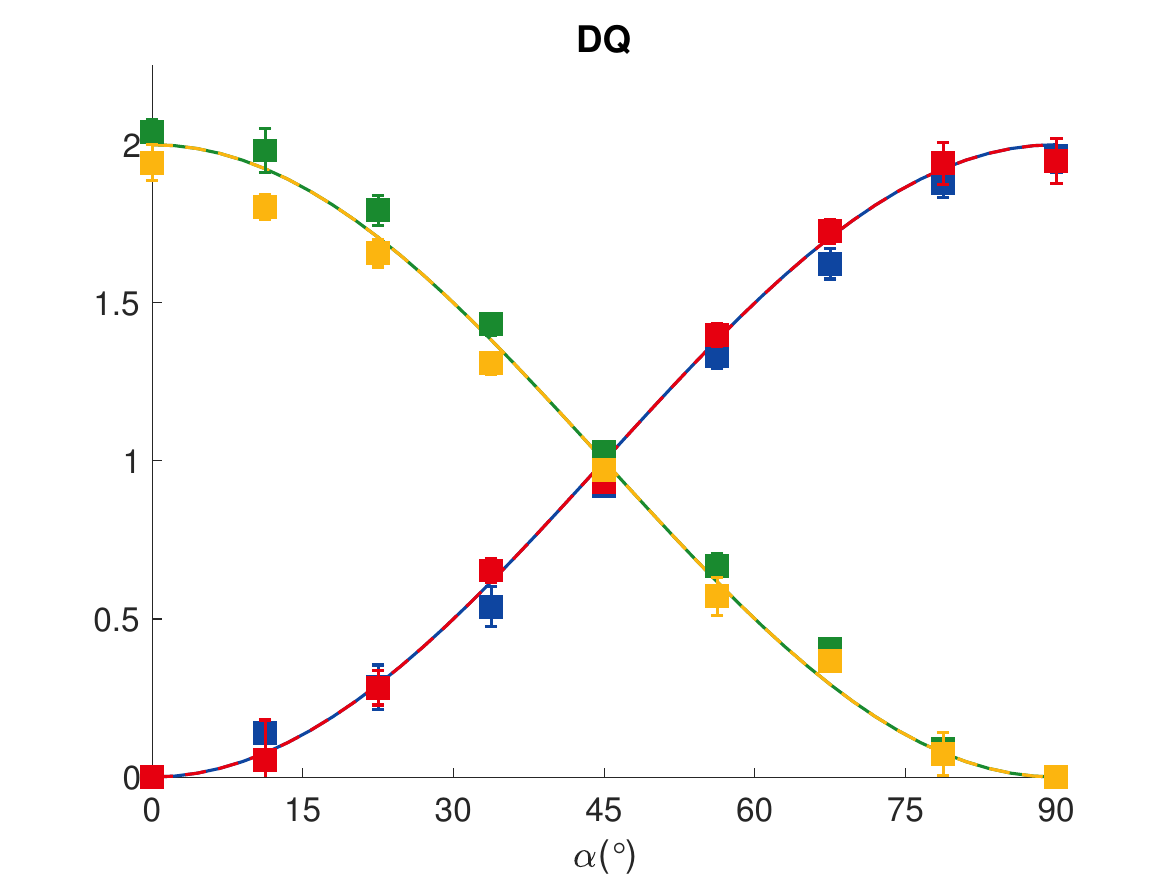}
    \caption{\textbf{Matrix elements under elliptical parametric modulations when $B_z=0$.} 
    } 
    \label{fig:matrix_element_elliptical_TM}
\end{figure*}

In this section we show the measured matrix elements for SQ and DQ transitions using elliptical parametric modulation under $B_z=0$ in Fig.~\ref{fig:matrix_element_elliptical_TM}. We extract the Berry curvatures from them, which are shown in Fig.~2 (c) in the main text, and eventually reveal the tensor monopole.

\subsection{Radial field distribution}
In addition to the quantized topological charge, the monopole can be fully characterized by its field distribution~\cite{Nepomechie85,Palumbo2018}
\begin{equation}
\mathcal{H}_{\mu \nu \lambda} (\ve{q}) = \epsilon_{\mu \nu \lambda \gamma} q_{\gamma}  \, / (q_x^2+q_y^2+q_z^2+q_w^2)^2,
\end{equation}
which reflects the fact that the curvature field radially emanates from the topological defect in 4D parameter space. As a consequence, the monopole field has a characteristic inverse-cube dependence on the radial coordinate, $\mathcal{H}\sim(1/H_0)^3$. We now verify this additional signature of the tensor monopole in our experiment. For convenience, we choose to view the curvature field in  cartesian coordinates, $(q_x, q_y, q_z, q_w)$. The radial components of the curvature field satisfy 
\begin{equation}
    \mathcal{H}_{xyzw}^\perp dq_x dq_y dq_z dq_w = \mathcal{H}_{\alpha\beta\phi}\, \det[J] ~dq_x dq_y dq_z dq_w = \frac{1}{H_0^3} dq_x dq_y dq_z dq_w ,
\end{equation}
where $J$ is the Jacobian and we have
\begin{equation}
    \det J = \frac{1}{H_0^3} \frac{2}{\sin2\alpha}.
\end{equation}

\begin{figure*}[ht]
\centering
\includegraphics[width=0.4\textwidth]{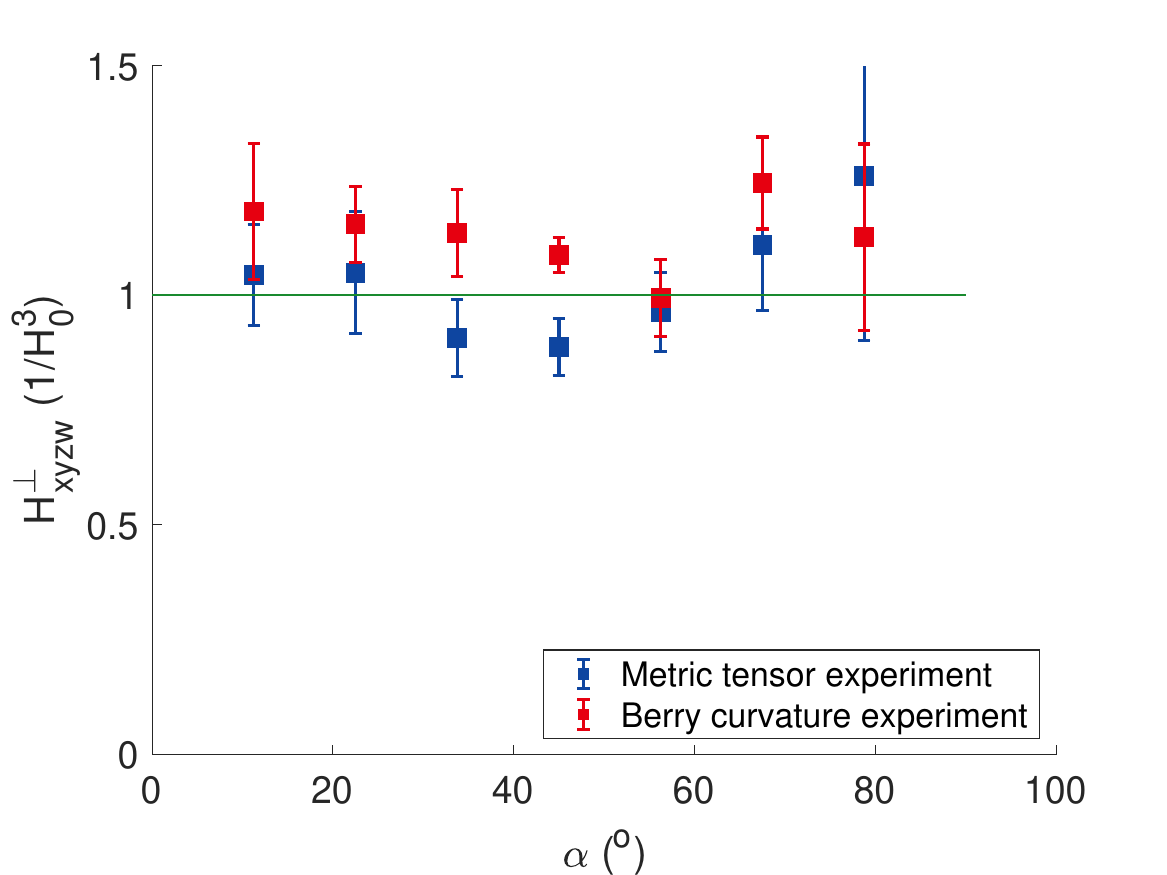}
\caption{\textbf{Radial field distribution.} The Radial field component $\mathcal{H}_{xyzw}^\perp$ extracted from the quantum metric (blue) and Berry curvature (red) experiments. Both display good agreement with the expected inverse-cube dependence on the radial distance.}\label{fig:inverse_cube}
\end{figure*}

Our experimentally measured curvature field displays an inverse-cube dependence on $H_0$, matching well with the theory (Fig.~\ref{fig:inverse_cube}). In analogy with the field emanating from electric/magnetic monopoles in 3D, we only expect this behavior from a monopole source in 4D, namely the tensor monopole.

\subsection{Spectral transition induced by an external field}\label{sec:exp_phase_transition}
As we discussed in the main text, detunings in the dual-frequency microwave pulse induce diagonal terms in our engineered Hamiltonian, acting as an external z field (with the same form as a spin-1 $B_zS_z$ field operator). When $B_z=0$, our observable, $\mathcal G=8\int\epsilon_{\mu\nu\lambda}\sqrt{\det g_{\bar{\mu}\bar{\nu}}}\,d\alpha$, is equivalent to the $\mathcal{DD}$ invariant. 
The corresponding measured data for $\mathcal{G}$ and the metric tensor are shown in Fig.~1 and Fig.~2 of the main text. 
When $B_z\neq0$, the diagonal terms break the chiral symmetry of our system, 
thus breaking the relationship between 
$\det g_{\bar{\mu}\bar{\nu}}$ and 
$\mathcal H_{\mu\nu\lambda}$.   
Surprisingly, similar to the observable $\mathcal{B}$ described in Sec.~\ref{sec:analytical_DD}, we find that the chosen observable $\mathcal G$ can still be used to investigate the behavior of the nodal surfaces  when varying $B_z$.

\begin{figure*}[ht] 
    \centering
    \includegraphics[width=0.32\textwidth]{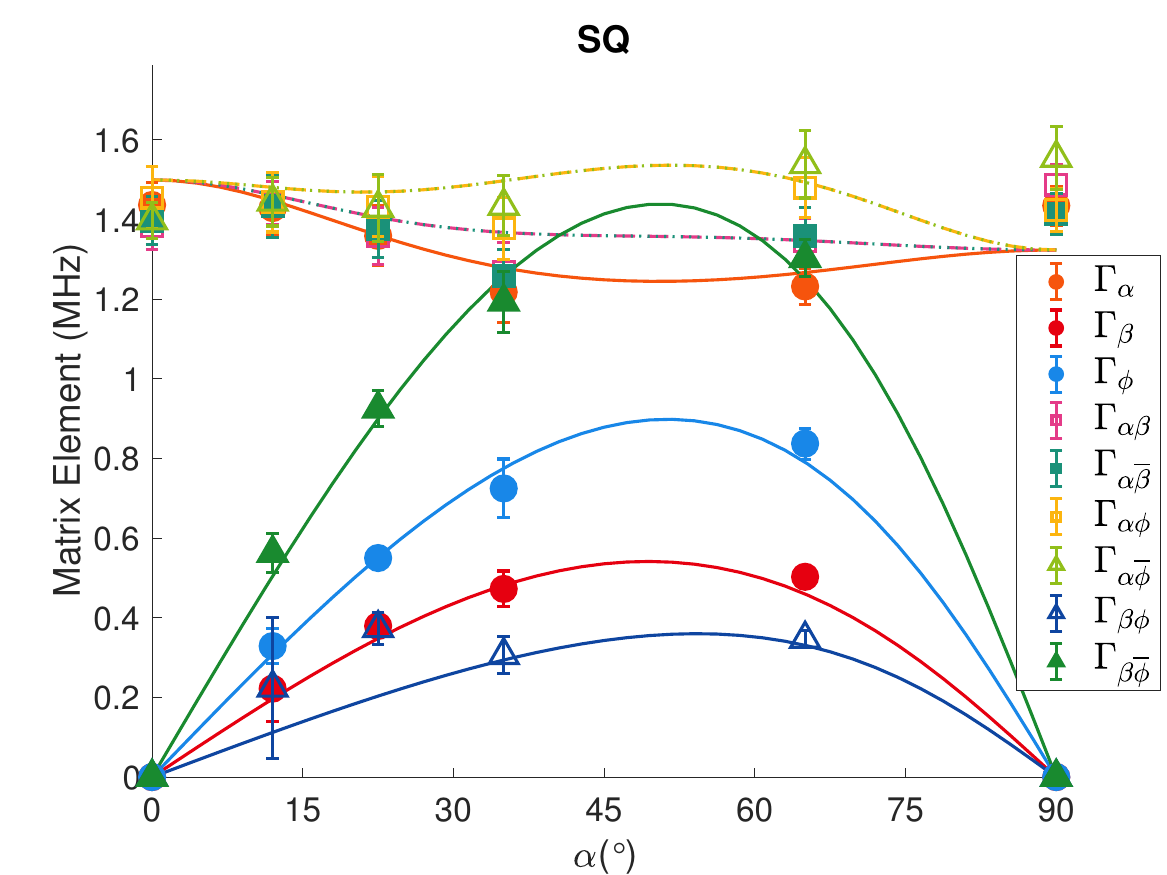}
    \includegraphics[width=0.32\textwidth]{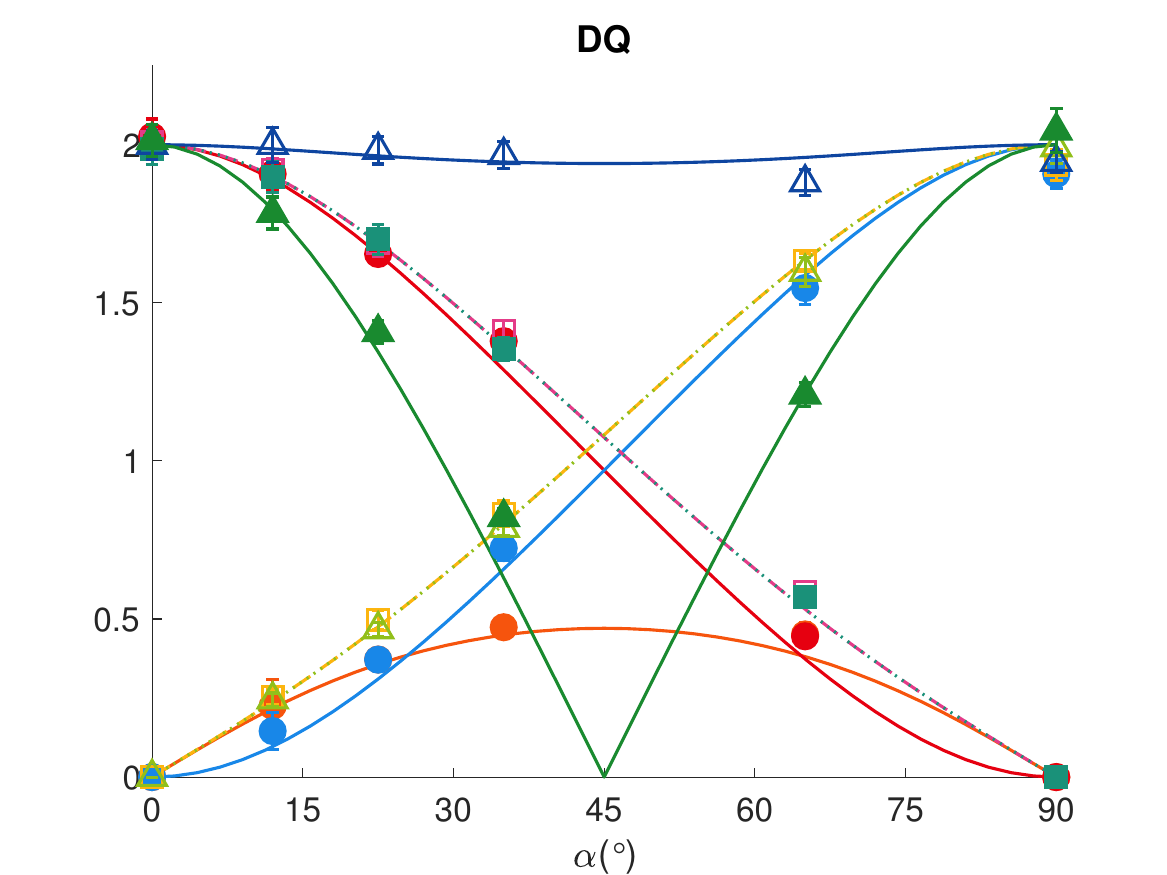}
    \includegraphics[width=0.32\textwidth]{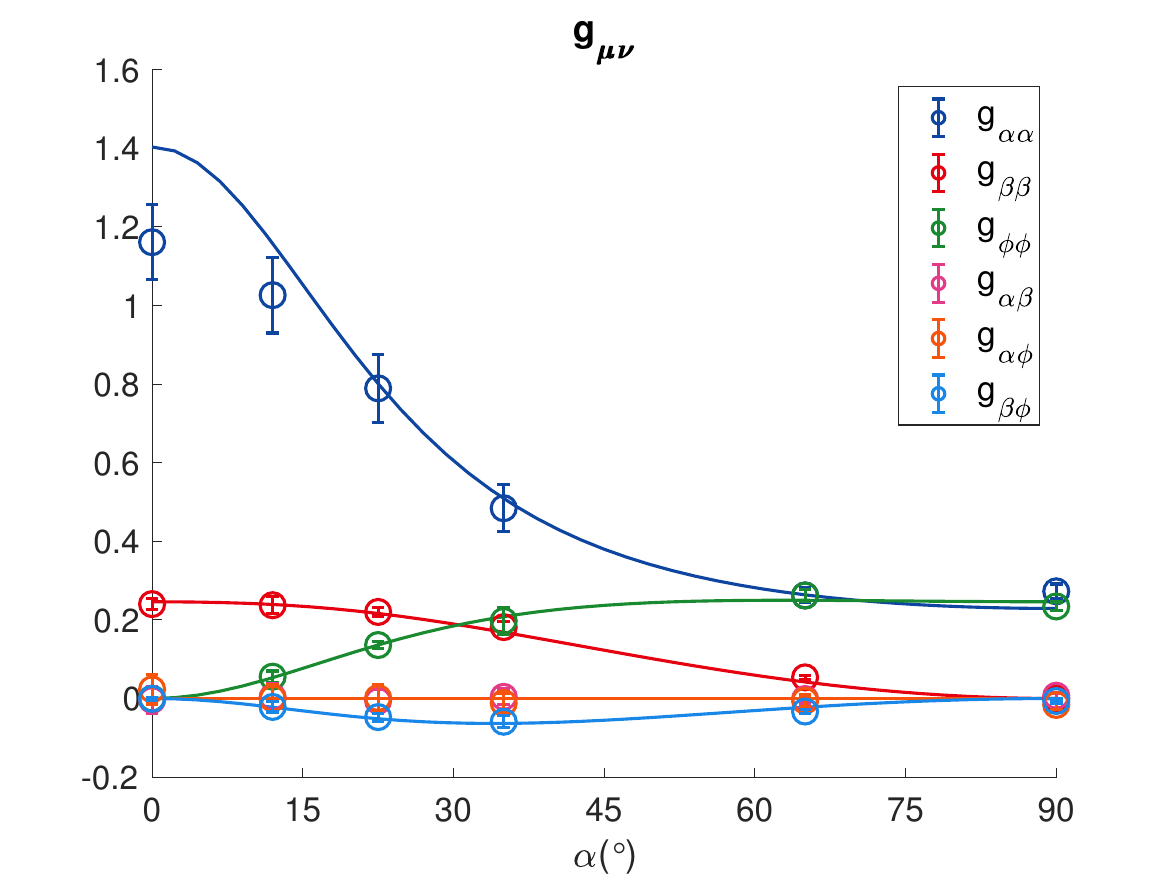}
    \caption{\textbf{Metric tensor measurements for $B_z/H_0=0.25\sqrt{2}$.} Matrix elements $\vert \Gamma_{-, 0}^{\mu(\nu)}\vert$ measured for SQ transitions at $\omega=\omega_r/2$ (left) and Matrix elements $\vert \Gamma_{-, +}^{\mu(\nu)}\vert$ measured for DQ transitions at $\omega=\omega_r$ (middle). On the right we show all 6 independent components of the metric tensor as functions of $\alpha$. Circles are experimental data and solid lines are numerical simulations.}
    \label{fig:h025}
\end{figure*}

\begin{figure*}[ht] 
    \centering
    \includegraphics[width=0.32\textwidth]{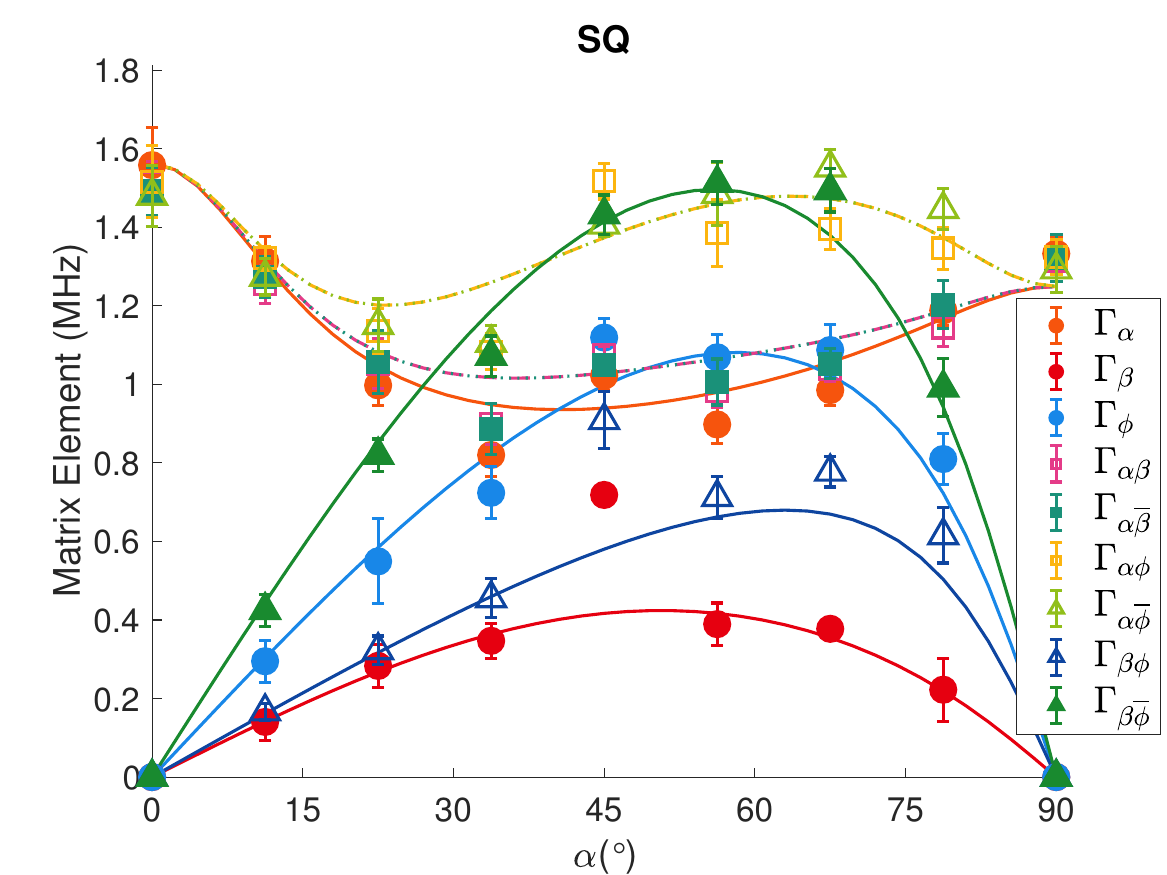}
    \includegraphics[width=0.32\textwidth]{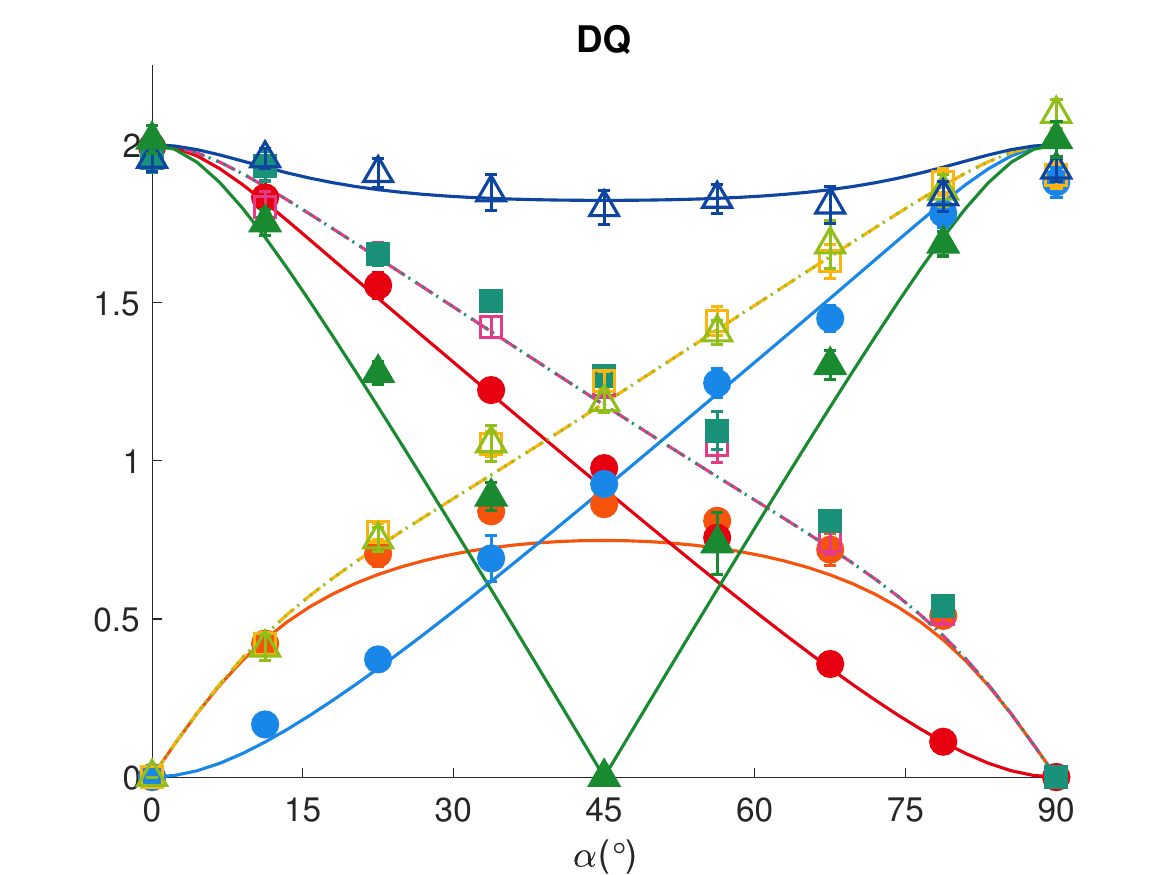}
    \includegraphics[width=0.32\textwidth]{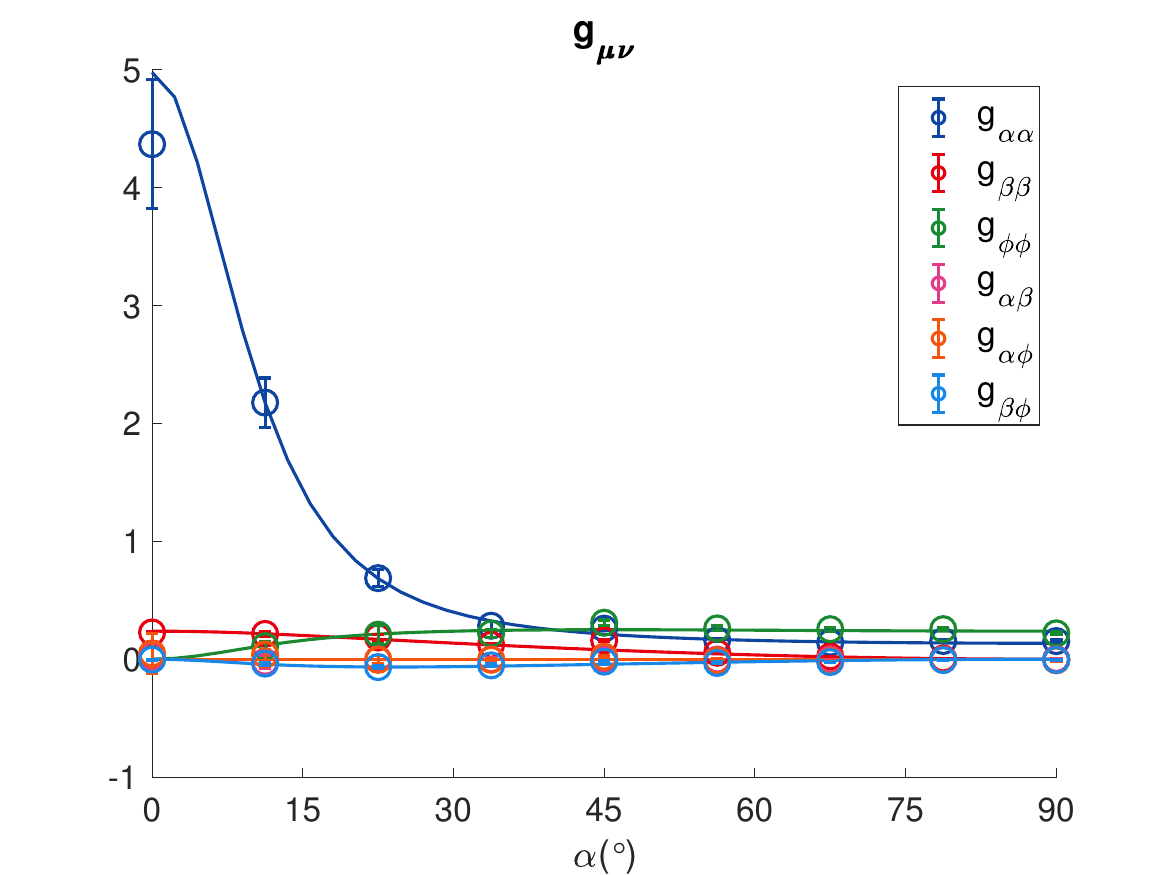}
    \caption{\textbf{Metric tensor measurements for $B_z/H_0=0.45\sqrt{2}$}. Matrix elements $\vert \Gamma_{-, 0}^{\mu(\nu)}\vert$ measured for SQ transitions at $\omega=\omega_r/2$ (left) and Matrix elements $\vert \Gamma_{-, +}^{\mu(\nu)}\vert$ measured for DQ transitions at $\omega=\omega_r$ (middle). On the right we show all 6 independent components of the metric tensor as functions of $\alpha$. Circles are experimental data and solid lines are numerical simulations.}
    \label{fig:h045}
\end{figure*}

\begin{figure*}[ht] 
    \centering
    \includegraphics[width=0.32\textwidth]{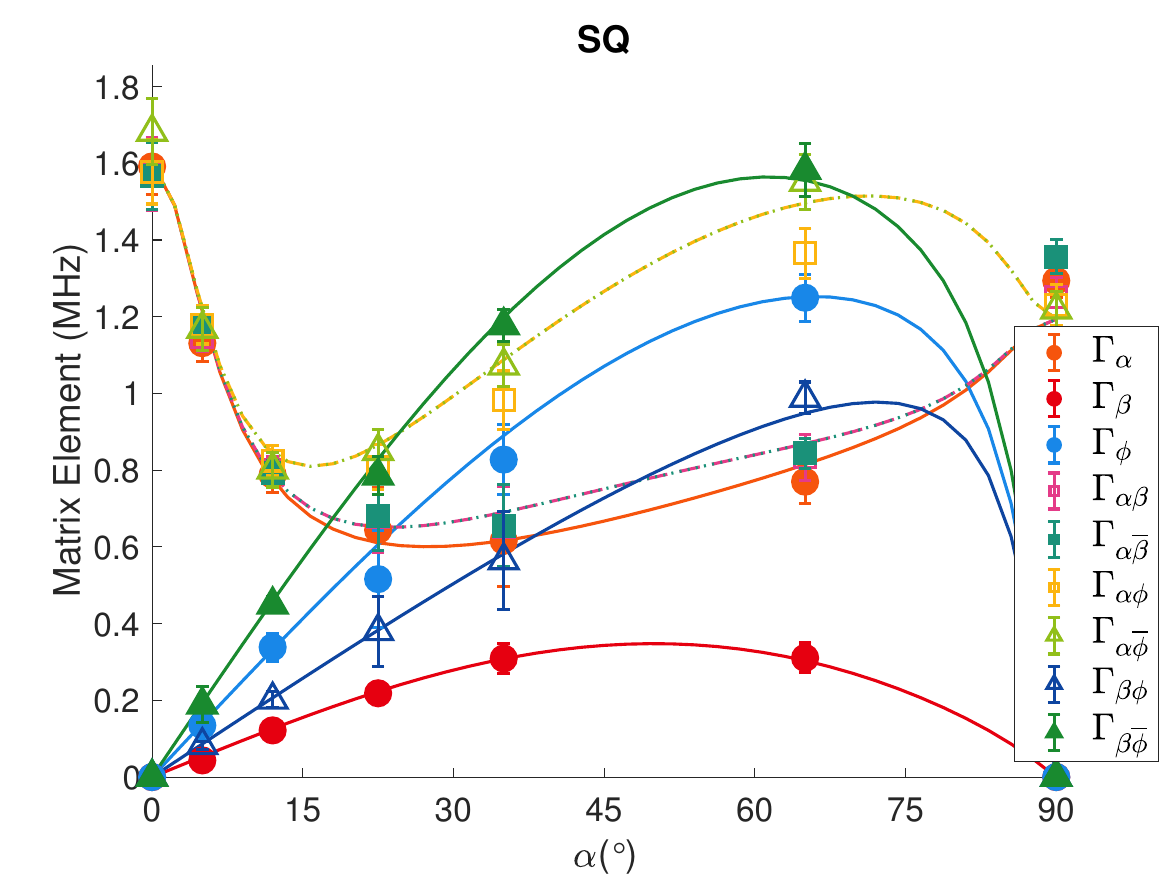}
    \includegraphics[width=0.32\textwidth]{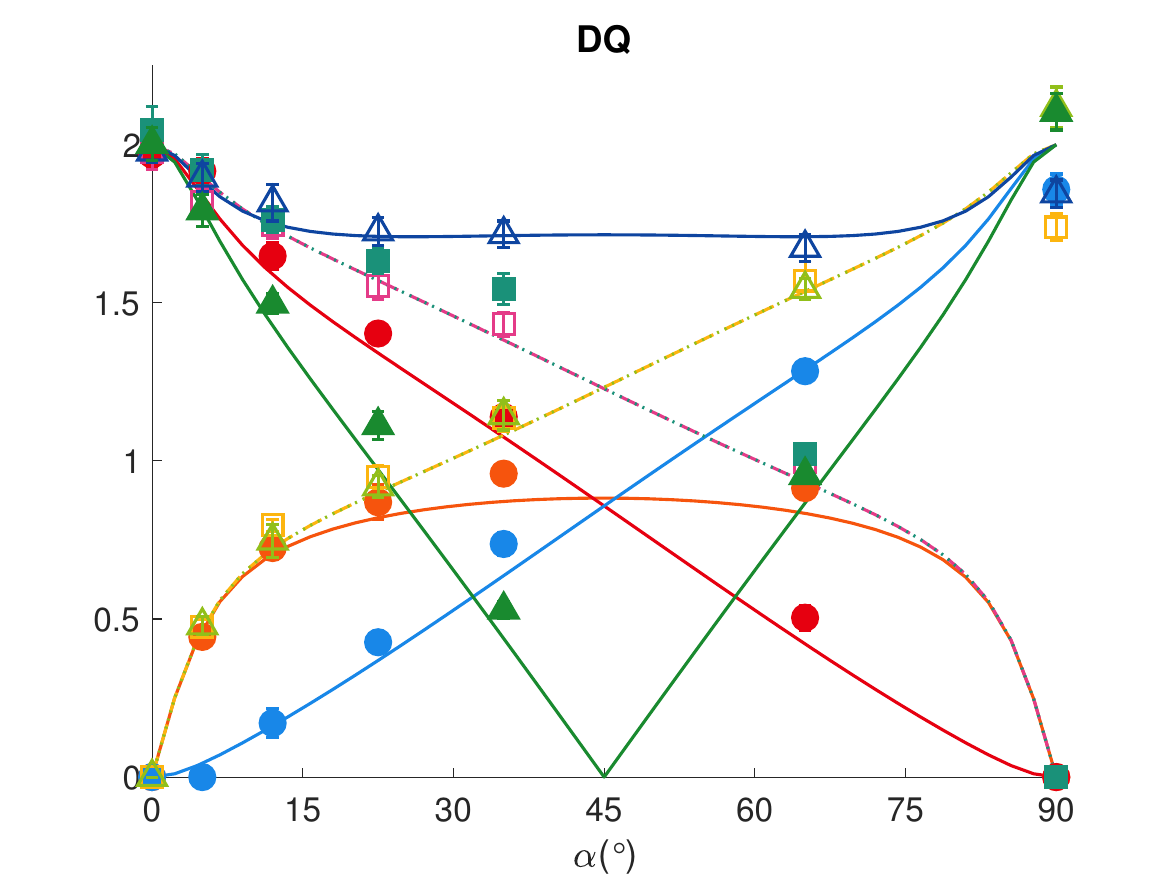}
    \includegraphics[width=0.32\textwidth]{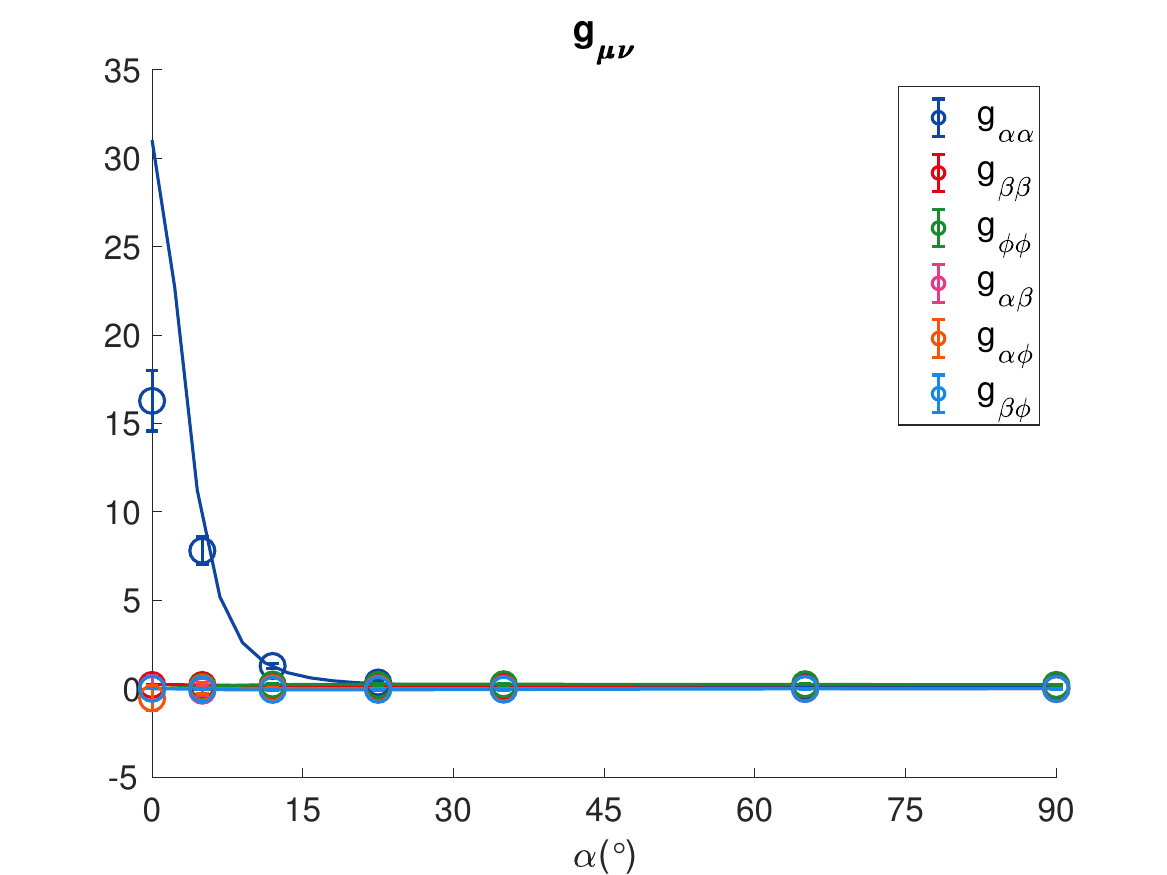}
    \caption{\textbf{Metric tensor measurements for $B_z/H_0=0.6\sqrt{2}$}. Matrix elements $\vert \Gamma_{-, 0}^{\mu(\nu)}\vert$ measured for SQ transitions at $\omega=\omega_r/2$ (left) and Matrix elements $\vert \Gamma_{-, +}^{\mu(\nu)}\vert$ measured for DQ transitions at $\omega=\omega_r$ (middle). On the right we show all 6 independent components of the metric tensor as functions of $\alpha$. Circles are experimental data and solid lines are numerical simulations.}
    \label{fig:h06}
\end{figure*}

\begin{figure*}[ht] 
    \centering
     \includegraphics[width=0.32\textwidth]{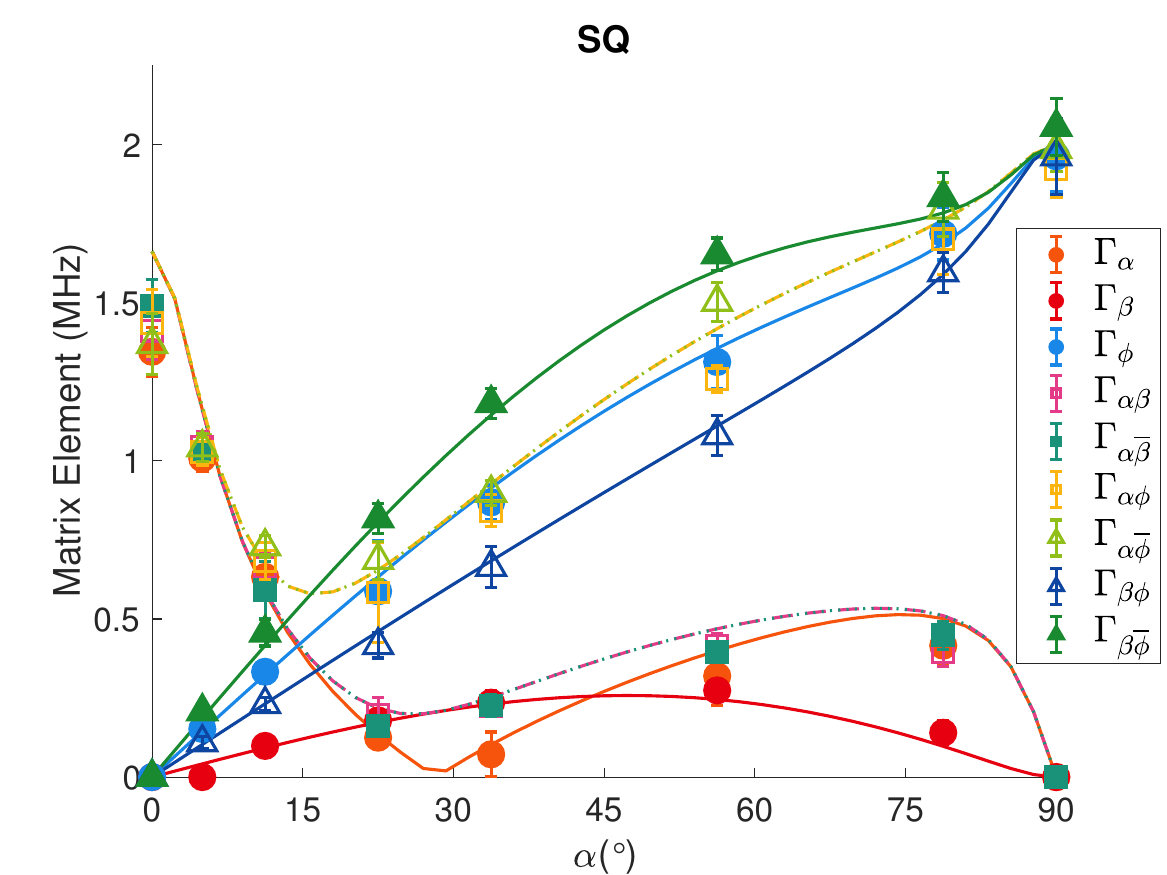}
    \includegraphics[width=0.32\textwidth]{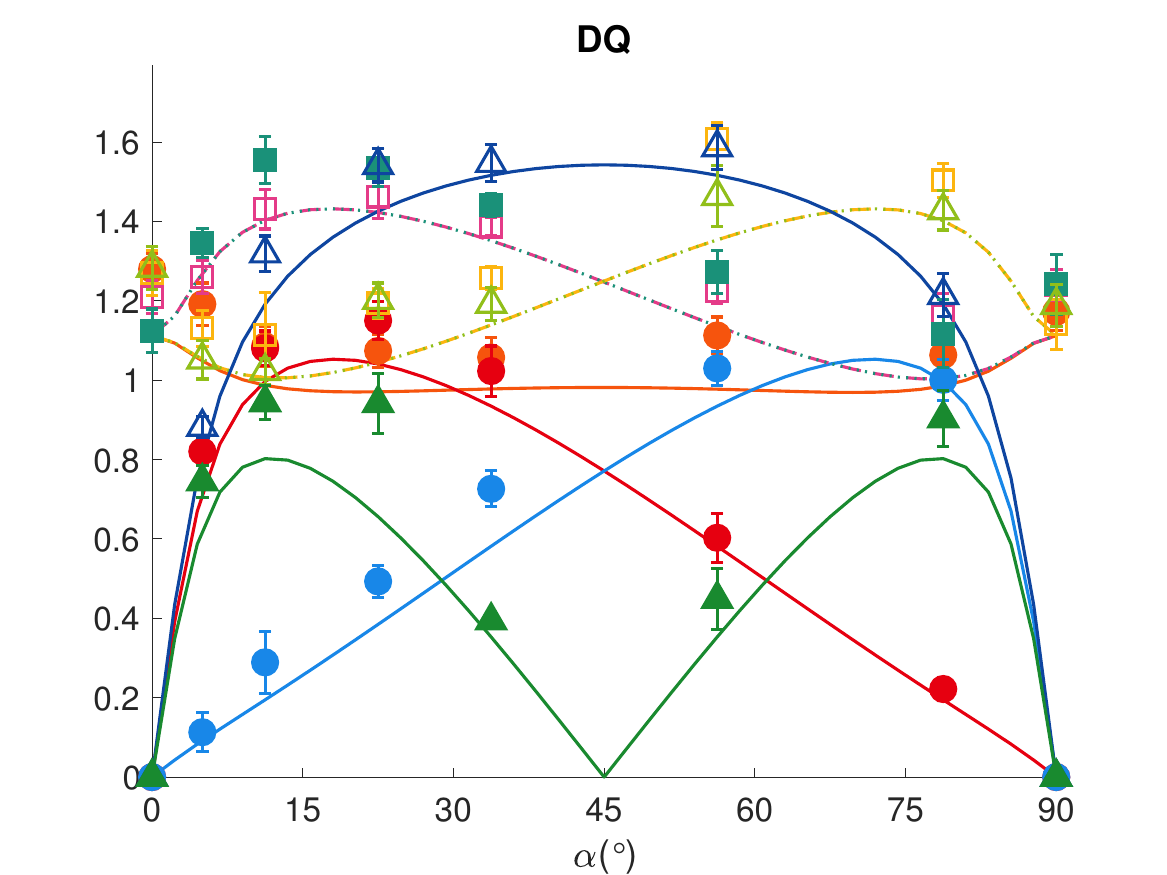}
    \includegraphics[width=0.32\textwidth]{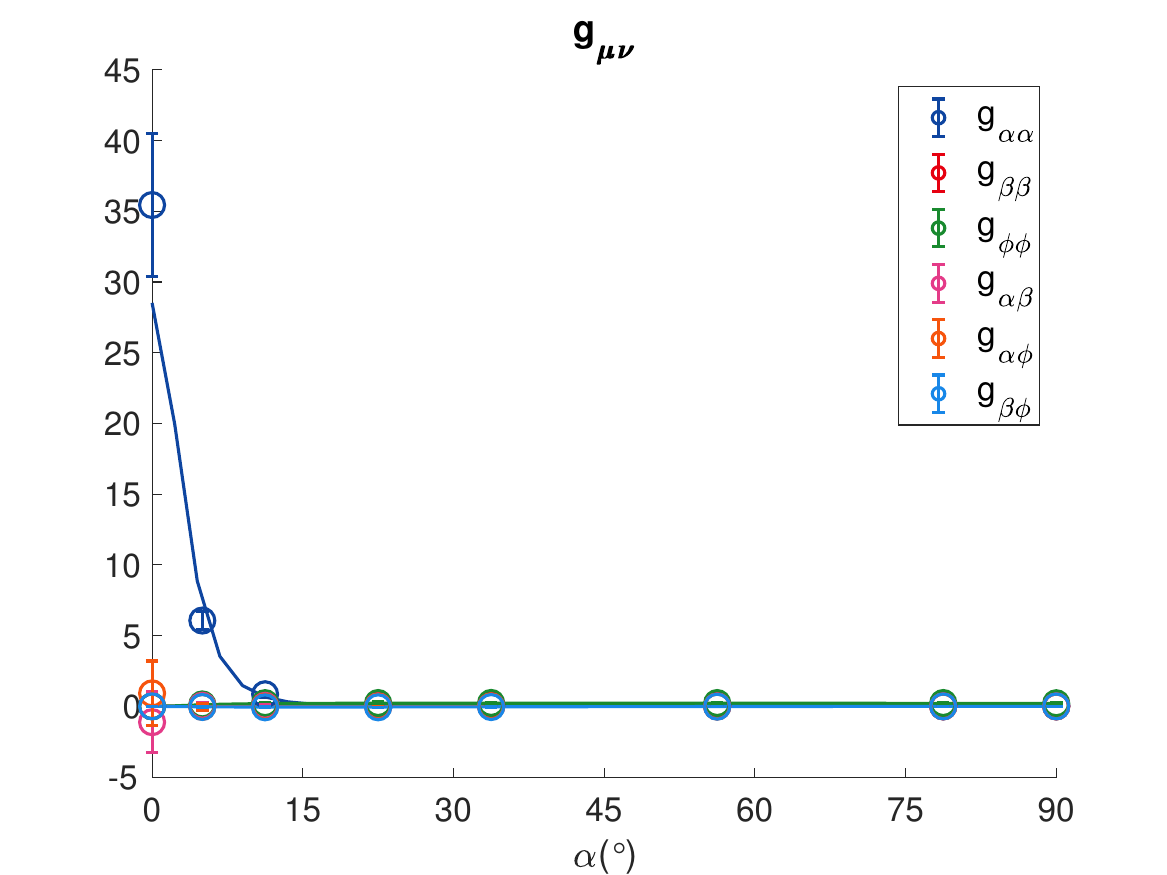}
    \caption{\textbf{Metric tensor measurements for $B_z/H_0=0.825\sqrt{2}$}. Matrix elements $\vert \Gamma_{-, 0}^{\mu(\nu)}\vert$ measured for SQ transitions at $\omega=\omega_r/2$ (left) and Matrix elements $\vert \Gamma_{-, +}^{\mu(\nu)}\vert$ measured for DQ transitions at $\omega=\omega_r$ (middle). On the right we show all 6 independent components of the metric tensor as functions of $\alpha$. Circles are experimental data and solid lines are numerical simulations.}
    \label{fig:h0825}
\end{figure*}

\begin{figure*}[ht] 
    \centering
    \includegraphics[width=0.32\textwidth]{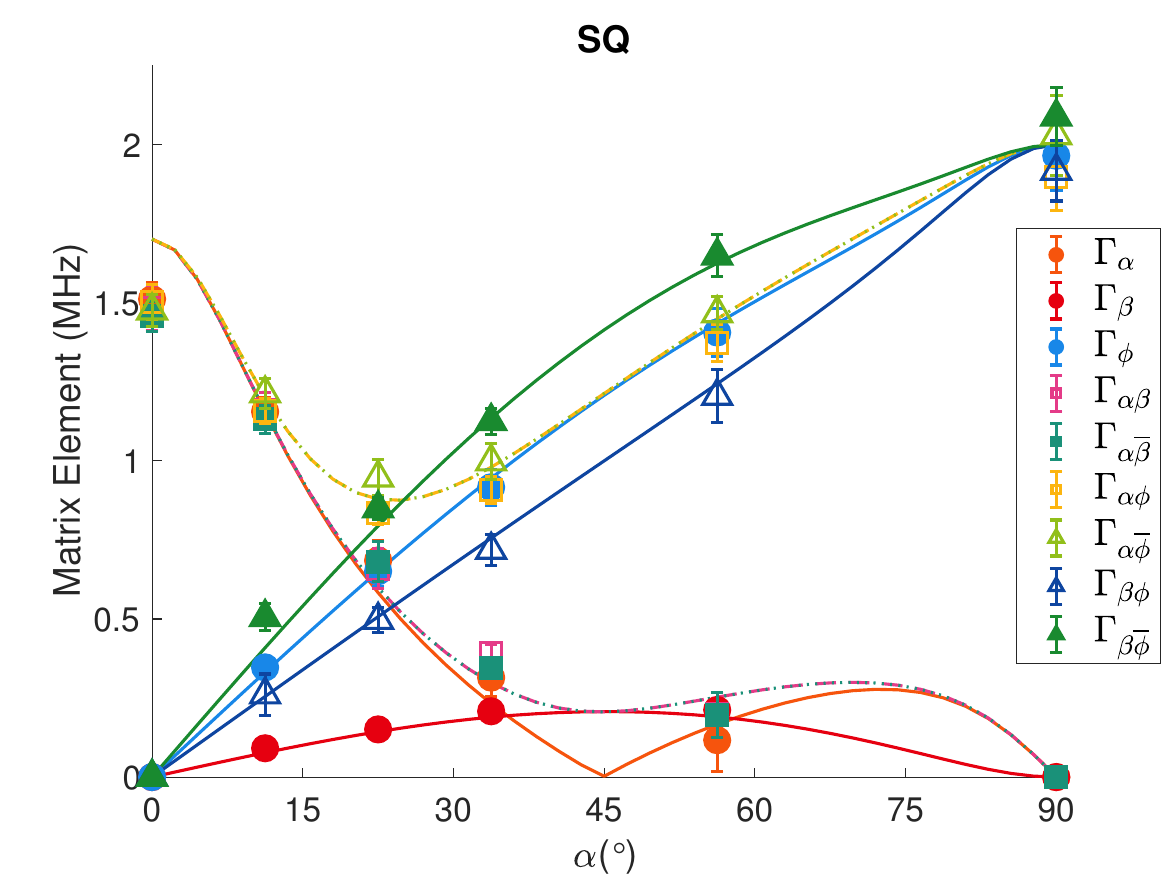}
    \includegraphics[width=0.32\textwidth]{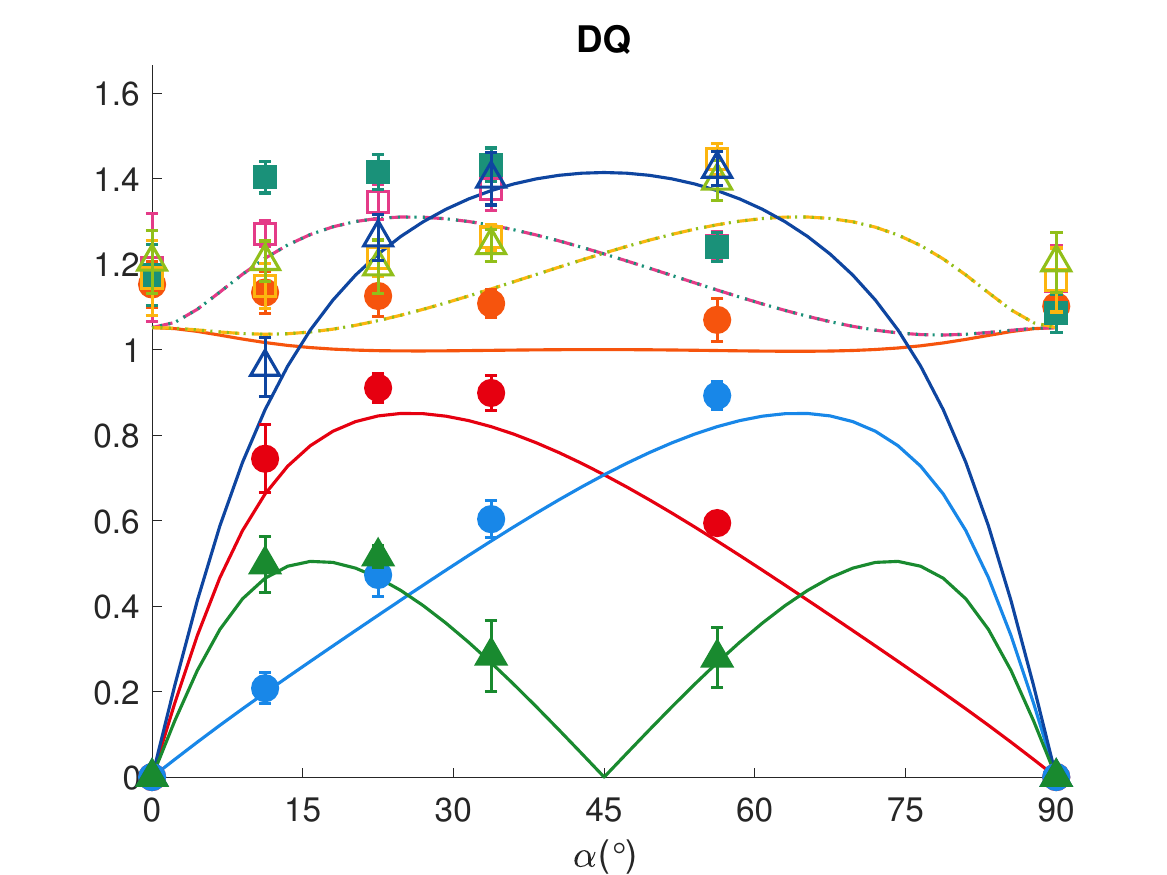}
    \includegraphics[width=0.32\textwidth]{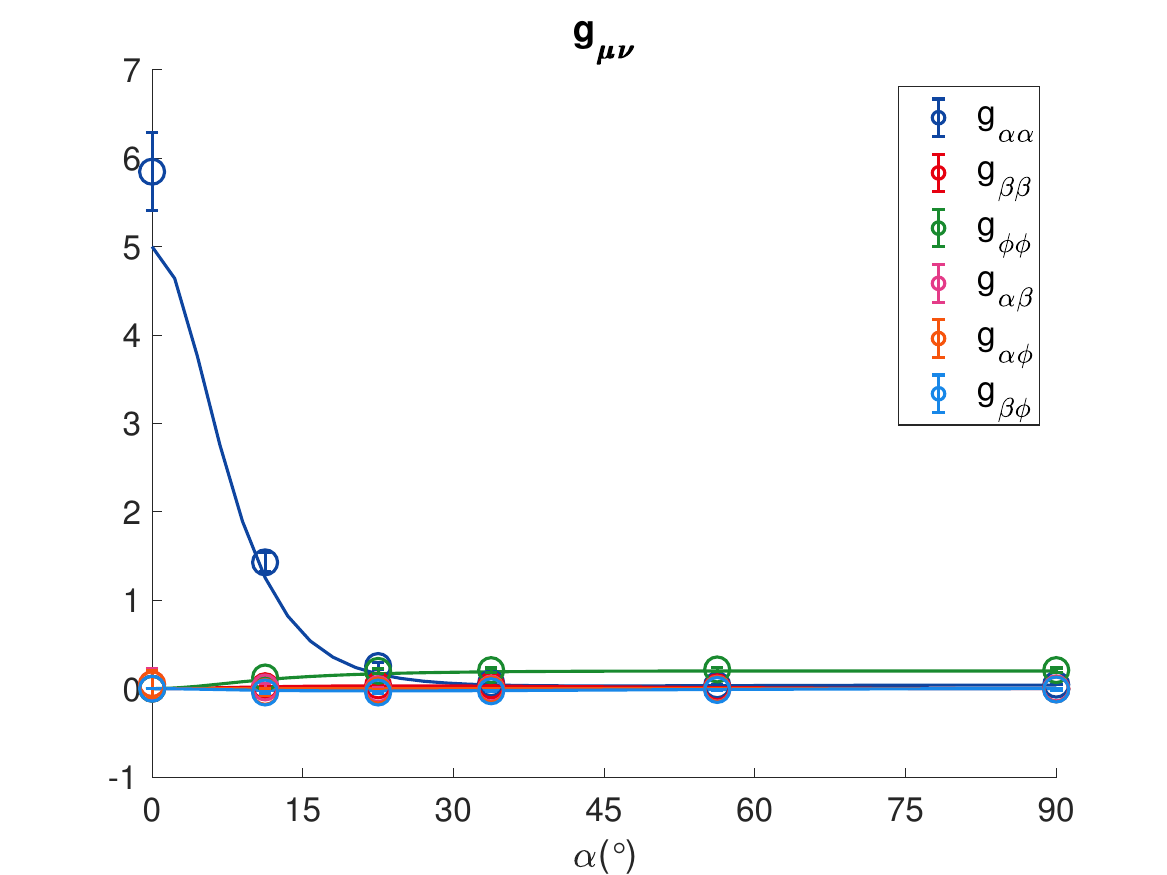}
    \caption{\textbf{Metric tensor measurements for $B_z/H_0=\sqrt{2}$}. Matrix elements $\vert \Gamma_{-, 0}^{\mu(\nu)}\vert$ measured for SQ transitions at $\omega=\omega_r/2$ (left) and Matrix elements $\vert \Gamma_{-, +}^{\mu(\nu)}\vert$ measured for DQ transitions at $\omega=\omega_r$ (middle). On the right we show all 6 independent components of the metric tensor as functions of $\alpha$. Circles are experimental data and solid lines are numerical simulations.}
    \label{fig:h1}
\end{figure*}

\begin{figure*}[ht] 
    \centering
    \includegraphics[width=0.32\textwidth]{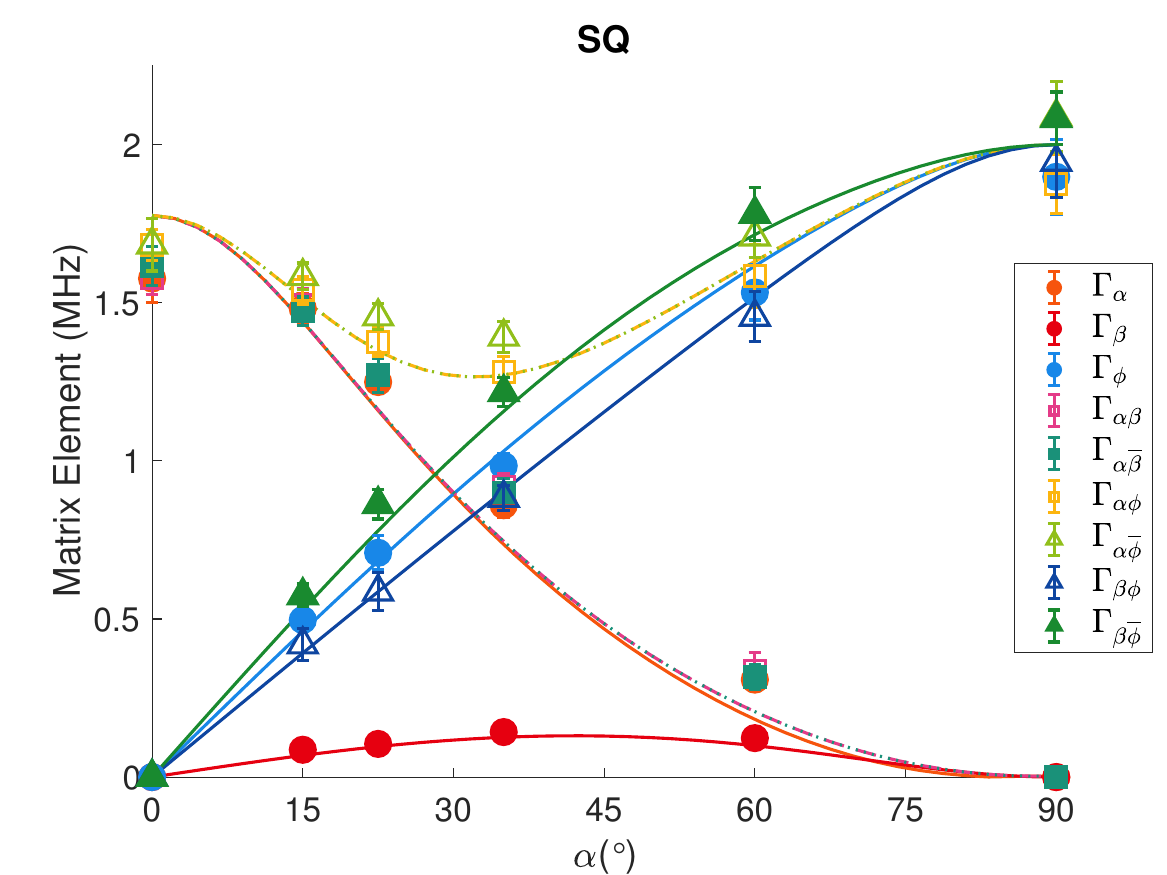}
    \includegraphics[width=0.32\textwidth]{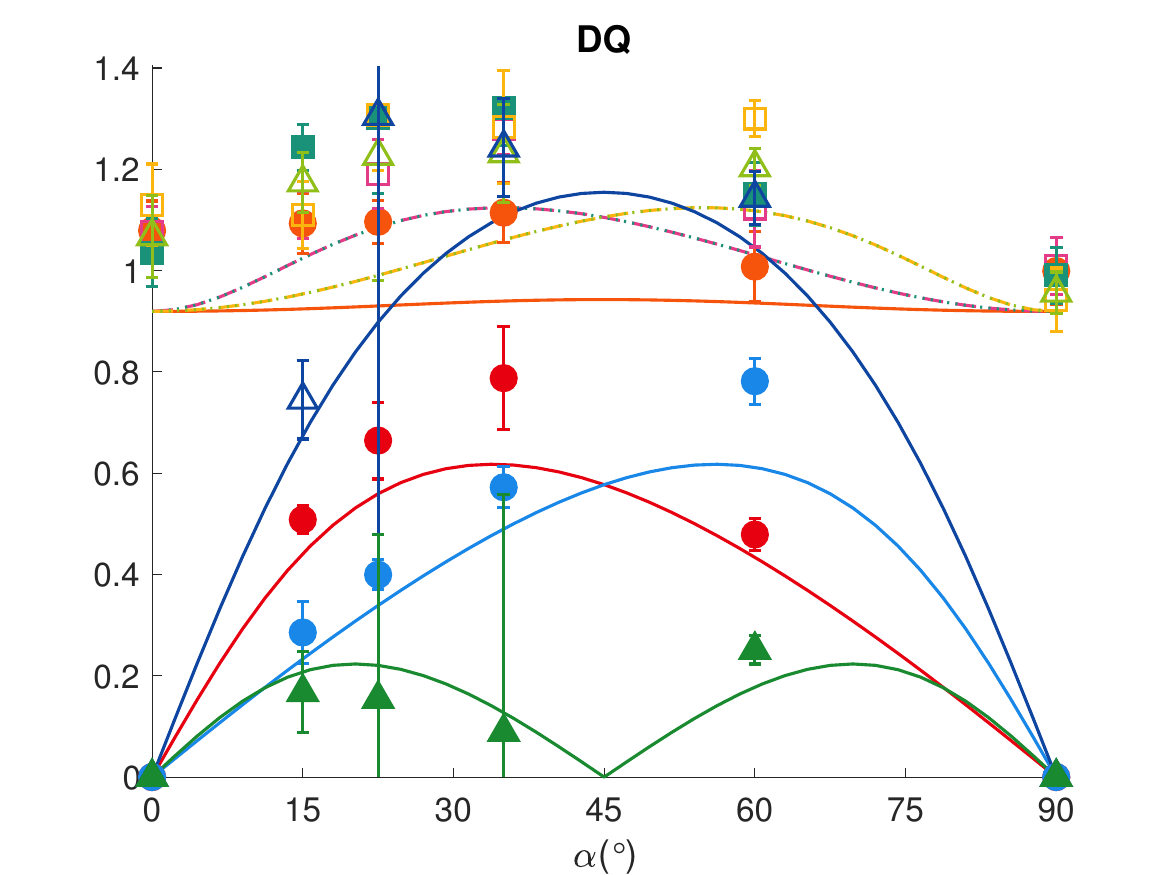}
    \includegraphics[width=0.32\textwidth]{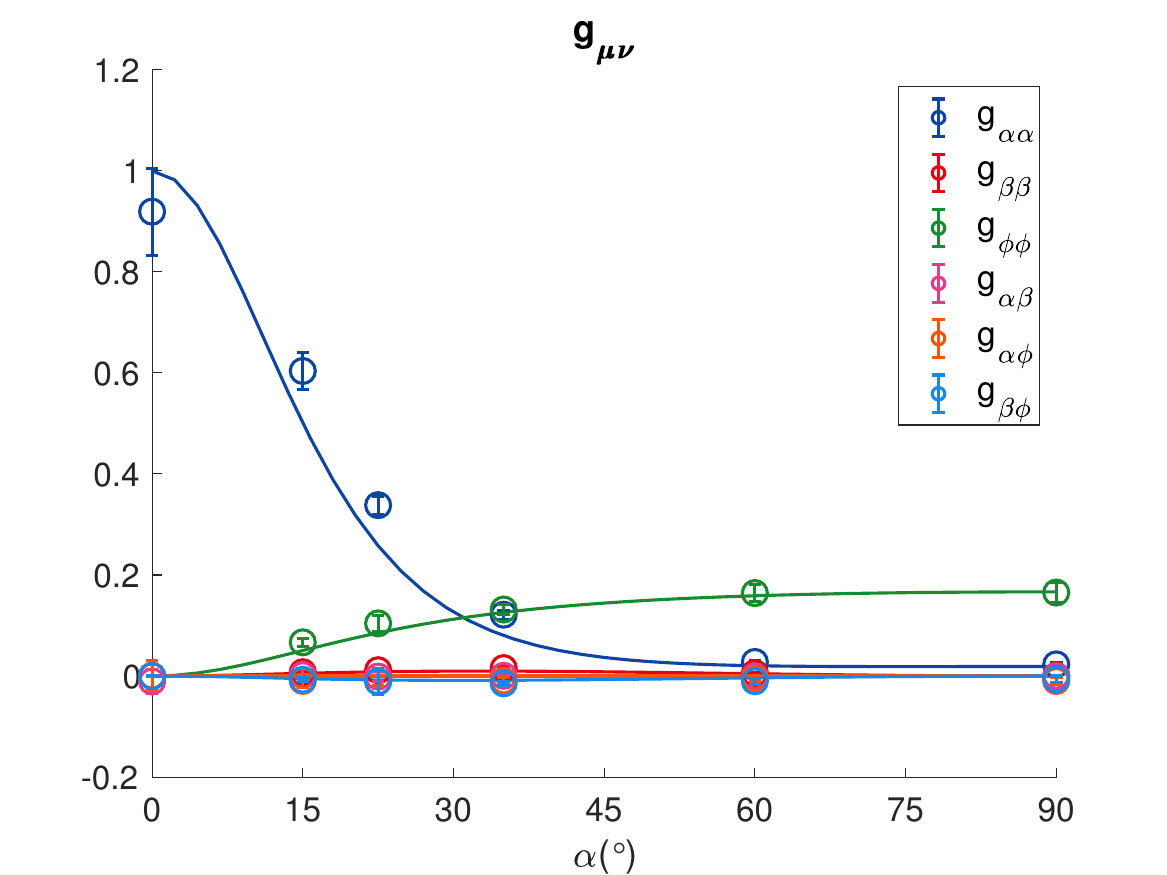}
    
    \caption{\textbf{Metric tensor measurements for $B_z/H_0=2$}. Matrix elements $\vert \Gamma_{-, 0}^{\mu(\nu)}\vert$ measured for SQ transitions at $\omega=\omega_r/2$ (left) and Matrix elements $\vert \Gamma_{-, +}^{\mu(\nu)}\vert$ measured for DQ transitions at $\omega=\omega_r$ (middle). On the right we show all 6 independent components of the metric tensor as functions of $\alpha$. Circles are experimental data and solid lines are numerical simulations.}
    \label{fig:hsqrt2}
\end{figure*}

We have shown that it is experimentally feasible to measure the metric tensor through parametric modulation, which yields $\mathcal{G}$. Ref.~\cite{Palumbo2018} has shown that when $B_z=0$, $\mathcal{G}$ is equivalent to the $\mathcal{DD}$ invariant. Here we numerically simulate the case when $B_z>0$, where the two are no longer equivalent. The simulation result is shown in green triangles in Fig.~3 in the main text. We see that $\mathcal{G}$ is a good approximation to the $\vert\mathcal{B}\vert$. When $B_z<H_0$, they quantitatively match well. At $B_z=H_0$, $\mathcal{G}$ correctly characterizes the transition when the spectral rings cross the boundary of the enclosed manifold. We therefore experimentally measure $\mathcal{G}$ to reveal the phase transition, as shown in Fig.~3 in the main text. 
We show additional experimental data relevant to the phase transition in Fig.~\ref{fig:h025}-\ref{fig:hsqrt2}.

\begin{figure*}[ht] 
    \centering
    \includegraphics[width=0.32\textwidth]{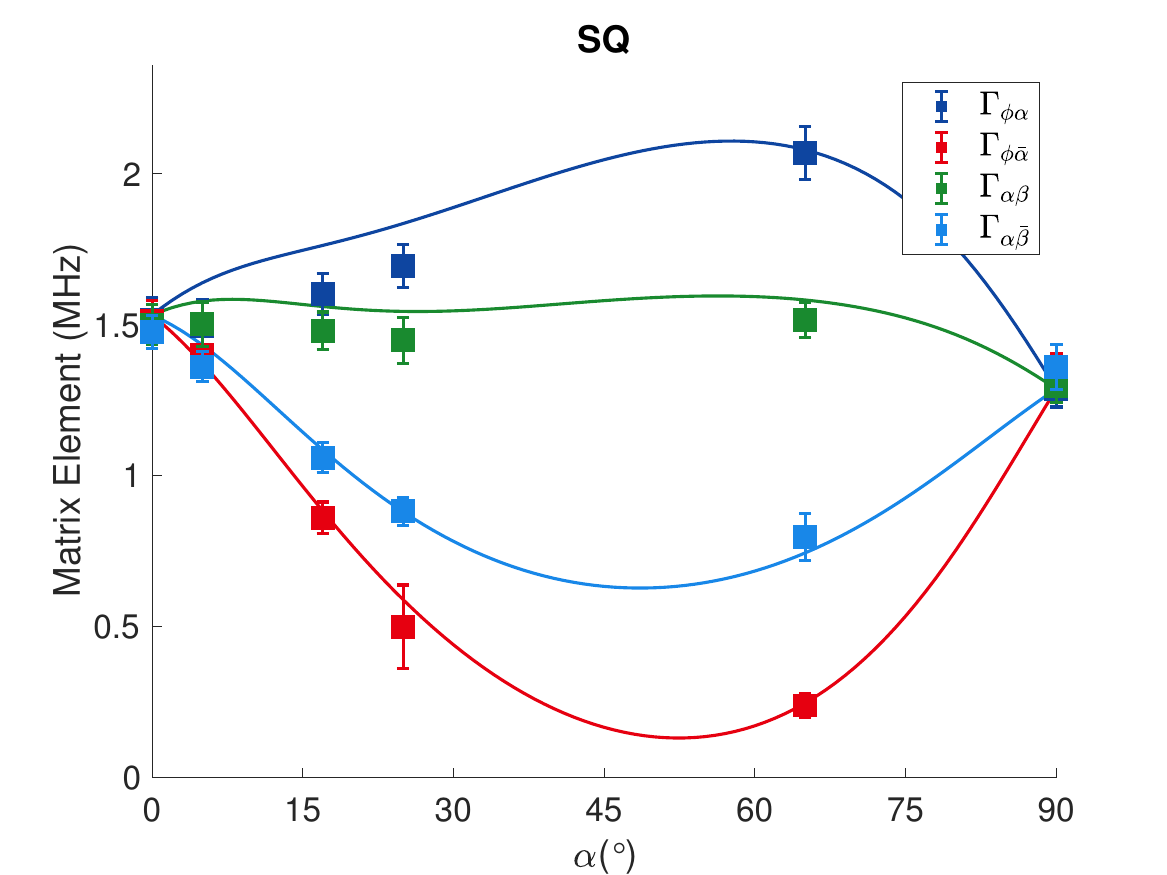}
    \includegraphics[width=0.32\textwidth]{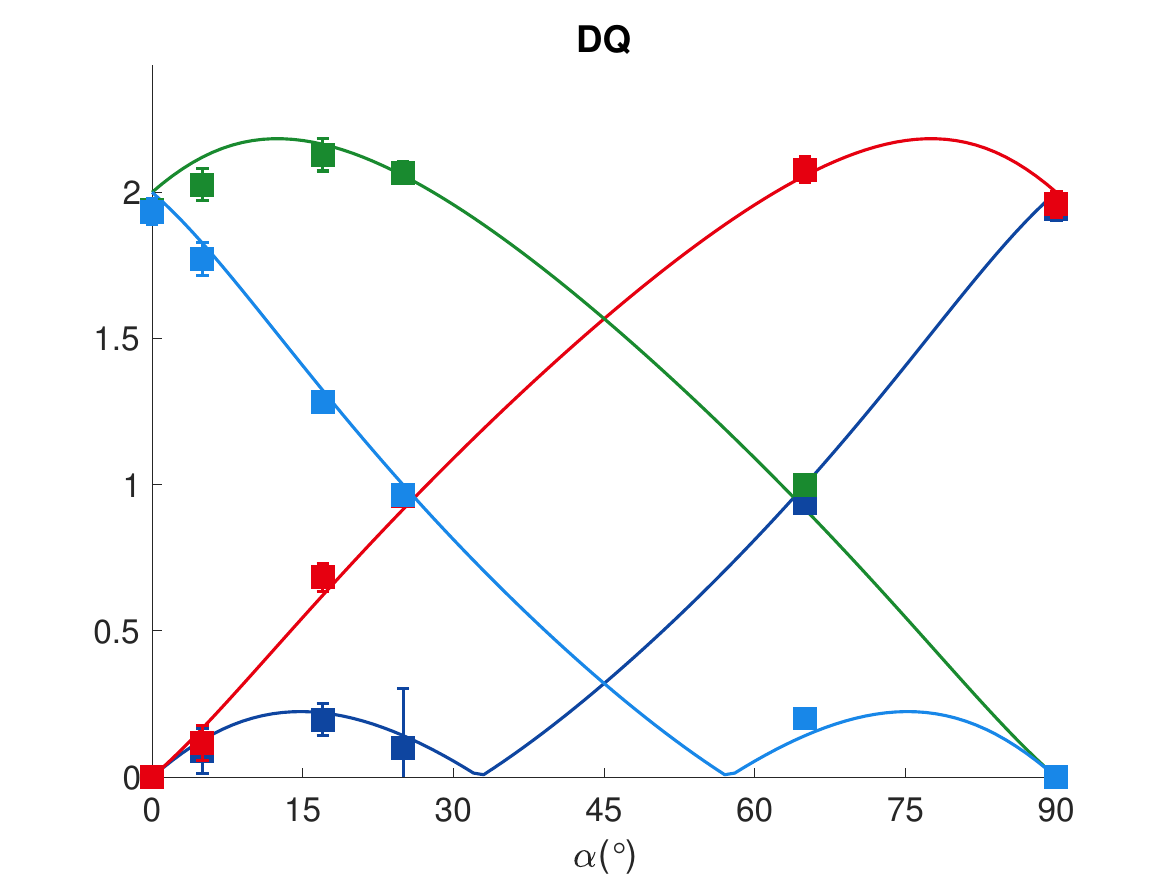}
    \includegraphics[width=0.32\textwidth]{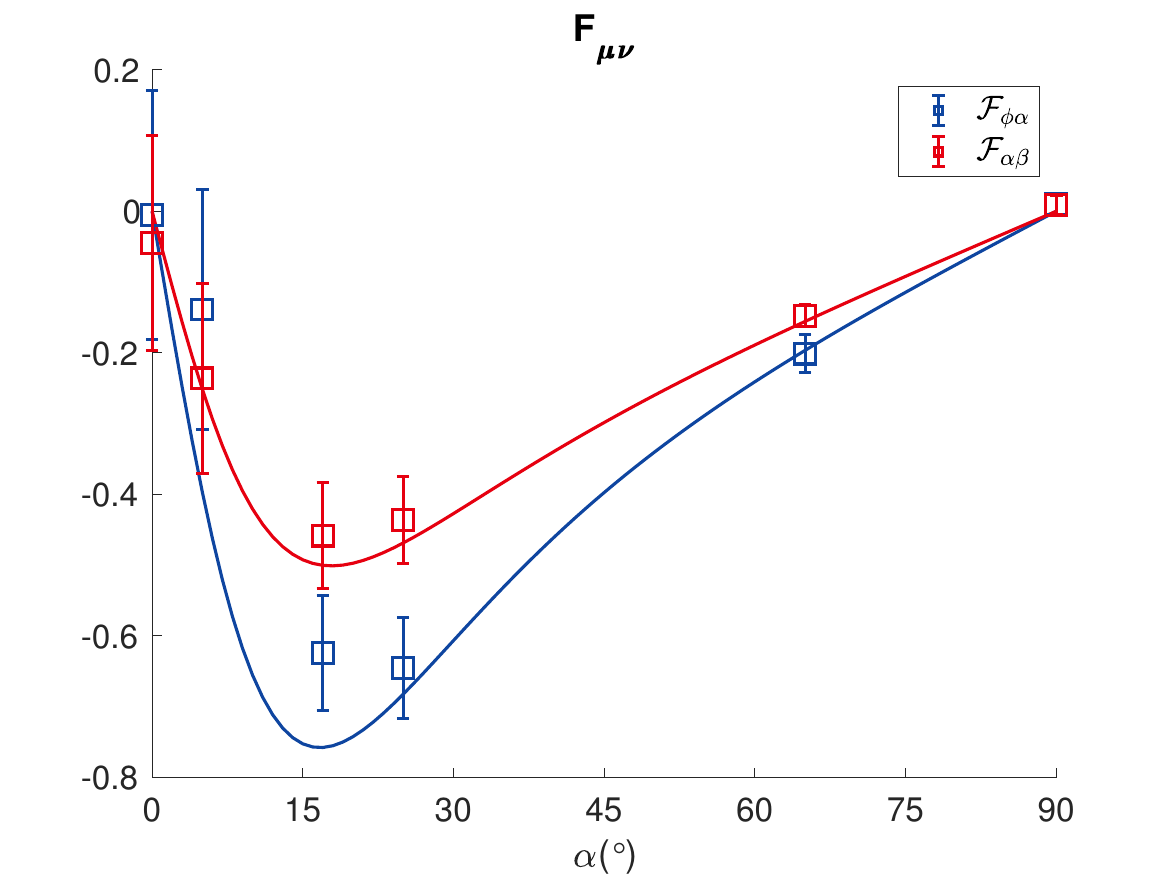}
    
    \caption{\textbf{Berry curvature measurements for $B_z/H_0=0.35\sqrt{2}$}. Matrix elements $\vert \Gamma_{-, 0}^{\mu(\nu)}\vert$ measured for SQ transitions at $\omega=\omega_r/2$ (left) and Matrix elements $\vert \Gamma_{-, +}^{\mu(\nu)}\vert$ measured for DQ transitions at $\omega=\omega_r$ (middle) using elliptical modulation. On the right we show the two relevant Berry curvatures as functions of $\alpha$. Squares are experimental data and solid lines are numerical simulations.}
    \label{fig:F_035}
\end{figure*}

\begin{figure*}[ht] 
    \centering
    \includegraphics[width=0.32\textwidth]{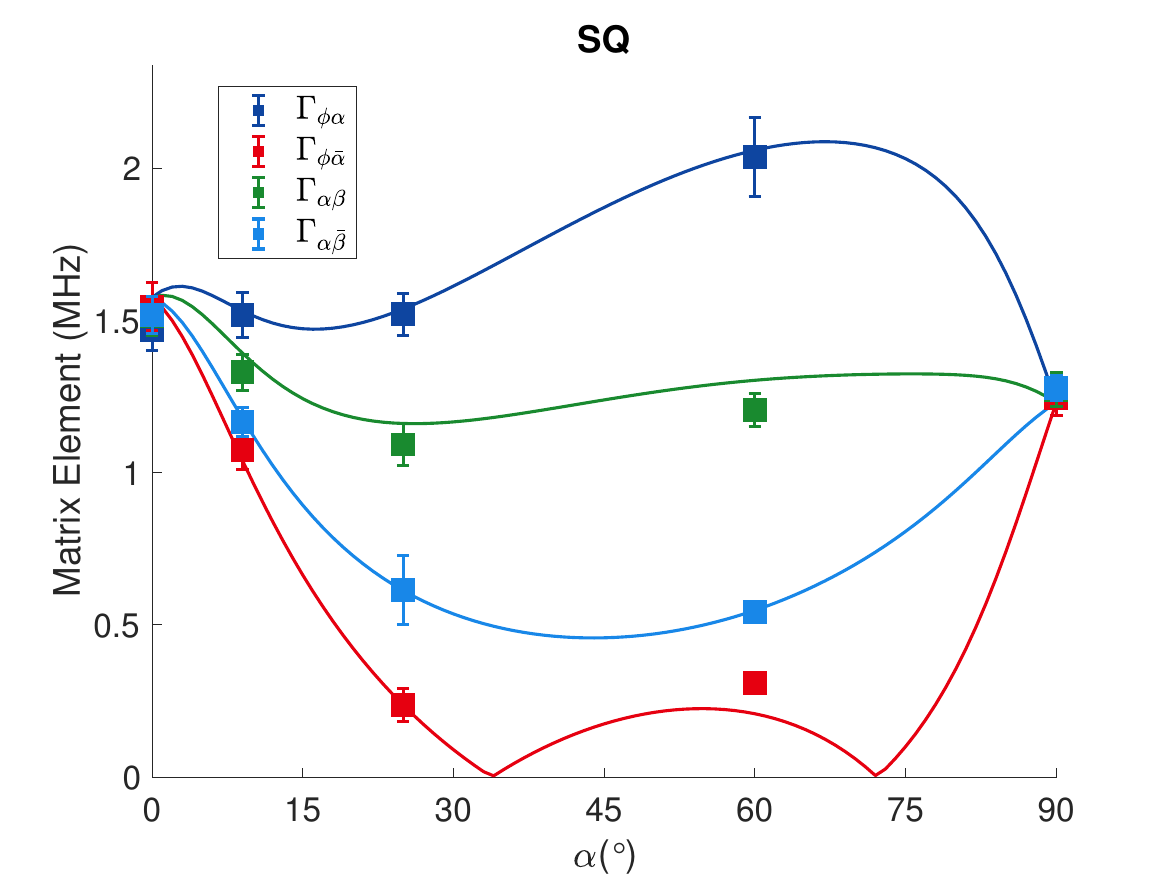}
    \includegraphics[width=0.32\textwidth]{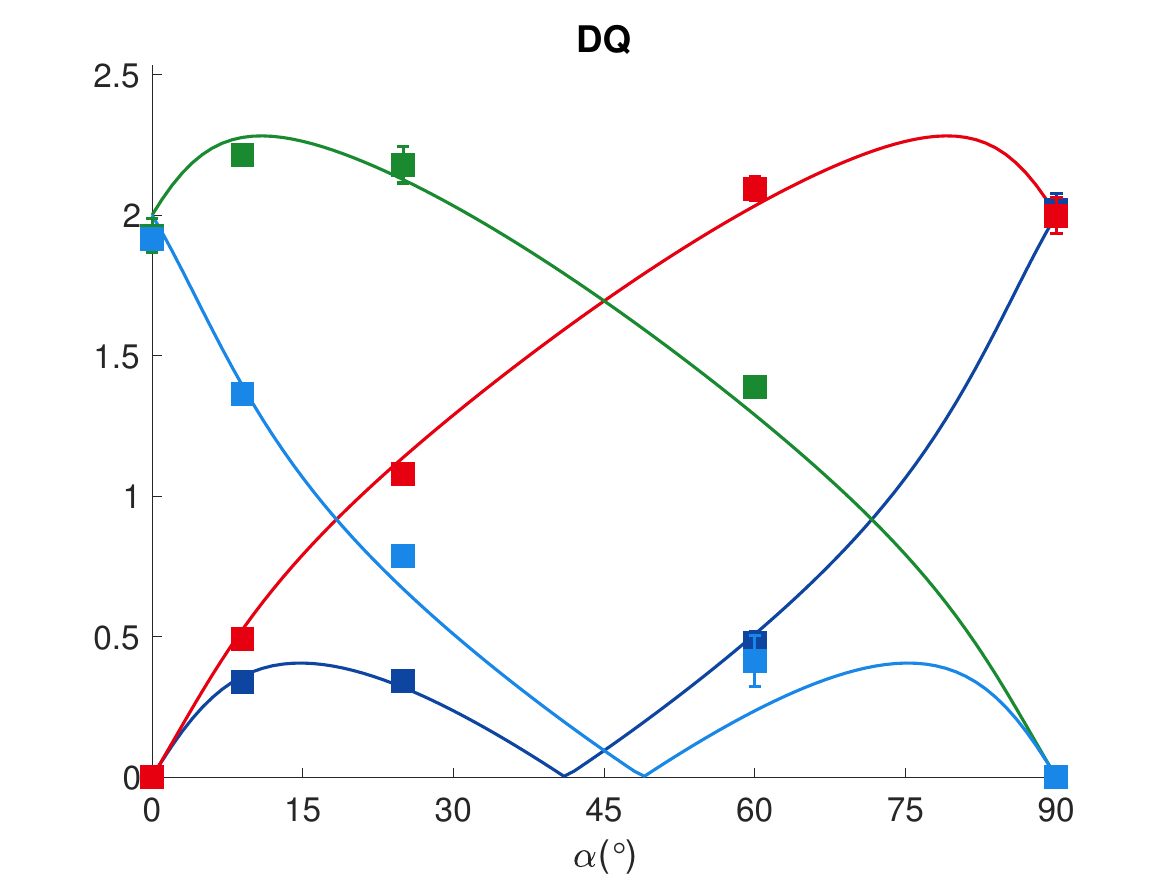}
    \includegraphics[width=0.32\textwidth]{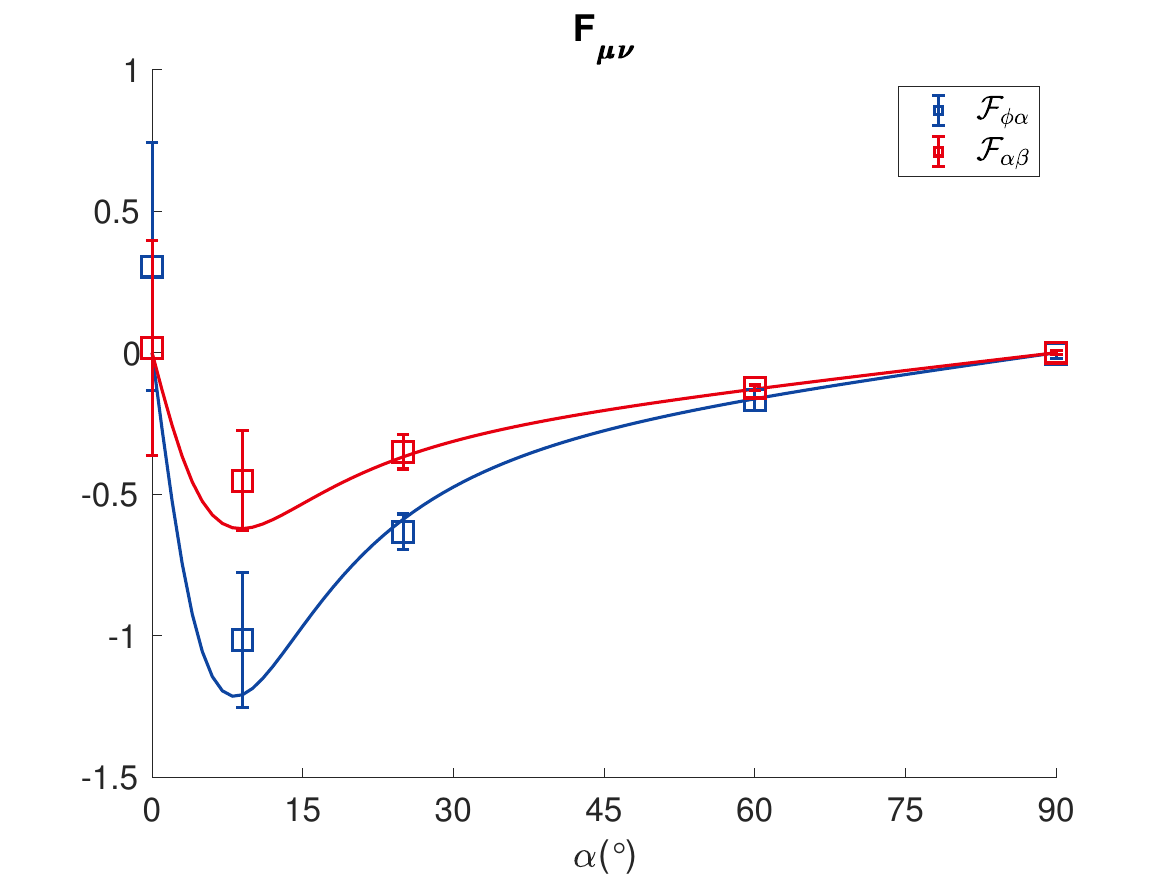}
    
    \caption{\textbf{Berry curvature measurements for $B_z/H_0=0.5\sqrt{2}$}. Matrix elements $\vert \Gamma_{-, 0}^{\mu(\nu)}\vert$ measured for SQ transitions at $\omega=\omega_r/2$ (left) and Matrix elements $\vert \Gamma_{-, +}^{\mu(\nu)}\vert$ measured for DQ transitions at $\omega=\omega_r$ (middle) using elliptical modulation. On the right we show the two relevant Berry curvatures as functions of $\alpha$. Squares are experimental data and solid lines are numerical simulations.}
    \label{fig:F_05}
\end{figure*}

\begin{figure*}[ht] 
    \centering
    \includegraphics[width=0.32\textwidth]{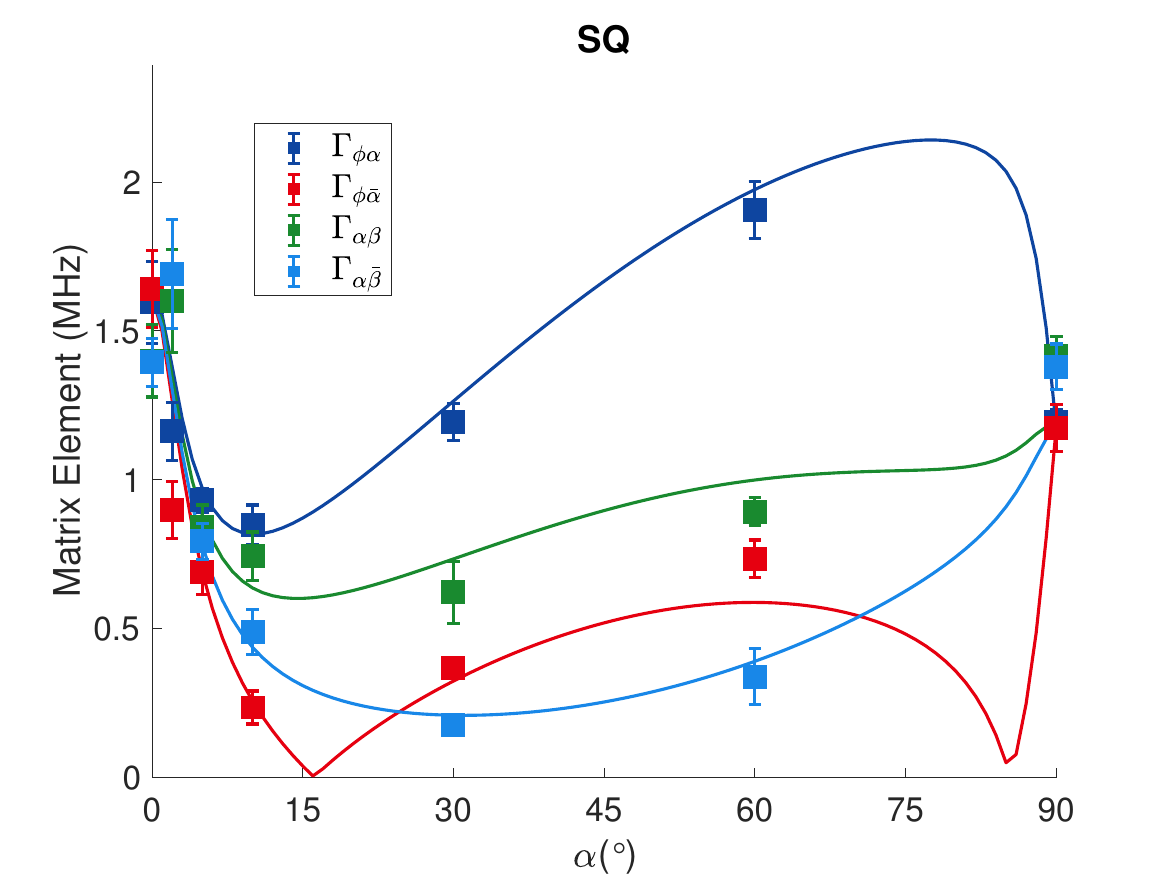}
    \includegraphics[width=0.32\textwidth]{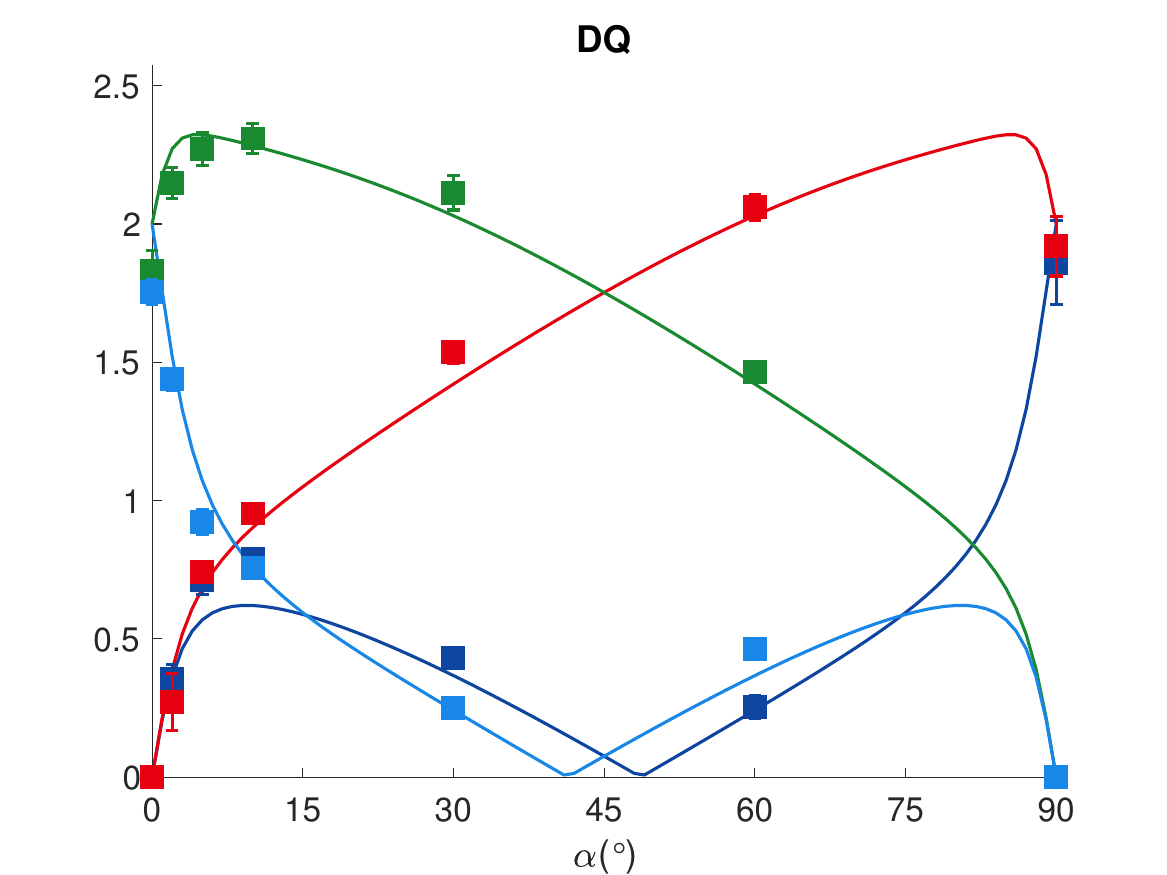}
    \includegraphics[width=0.32\textwidth]{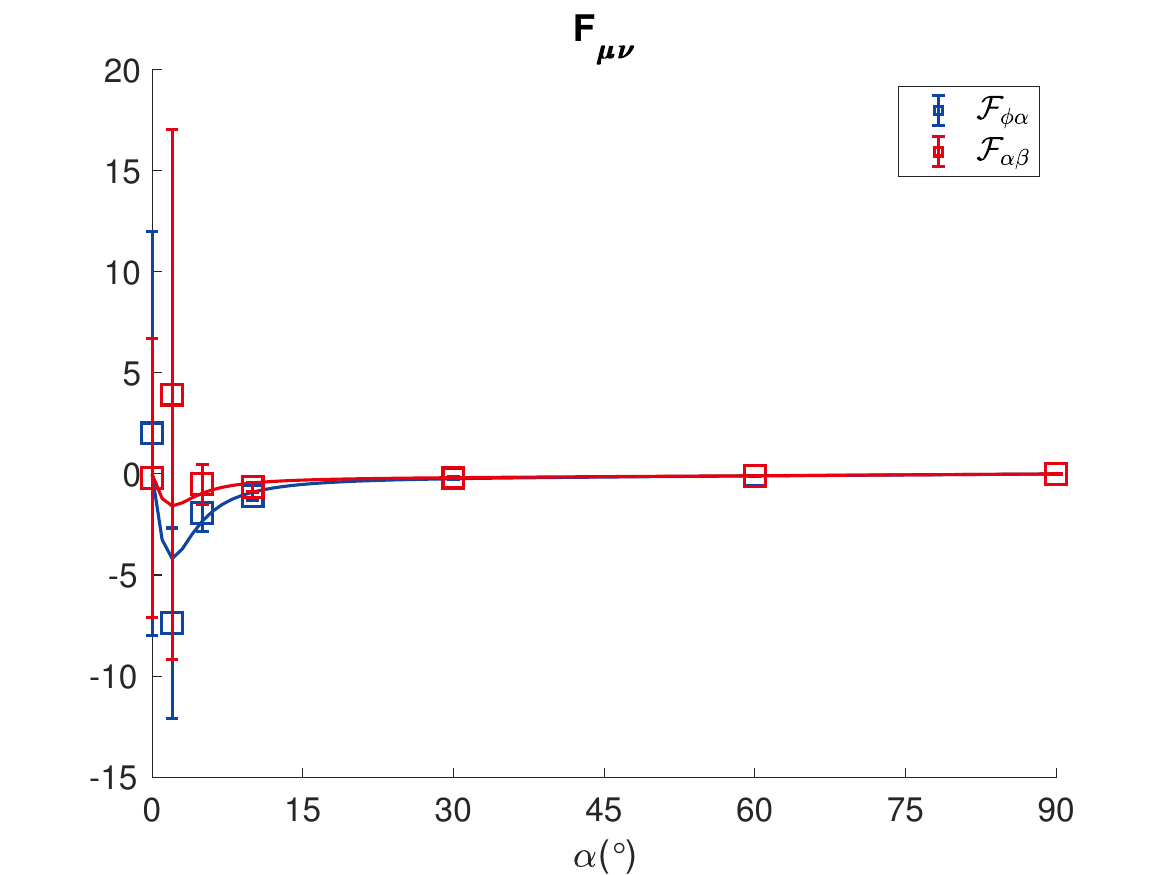}
    \caption{\textbf{Berry curvature measurements for $B_z/H_0=0.65\sqrt{2}$}. Matrix elements $\vert \Gamma_{-, 0}^{\mu(\nu)}\vert$ measured for SQ transitions at $\omega=\omega_r/2$ (left) and Matrix elements $\vert \Gamma_{-, +}^{\mu(\nu)}\vert$ measured for DQ transitions at $\omega=\omega_r$ (middle) using elliptical modulation. On the right we show the two relevant Berry curvatures as functions of $\alpha$. Squares are experimental data and solid lines are numerical simulations.}
    \label{fig:F_065}
\end{figure*}

\begin{figure*}[ht] 
    \centering
    \includegraphics[width=0.32\textwidth]{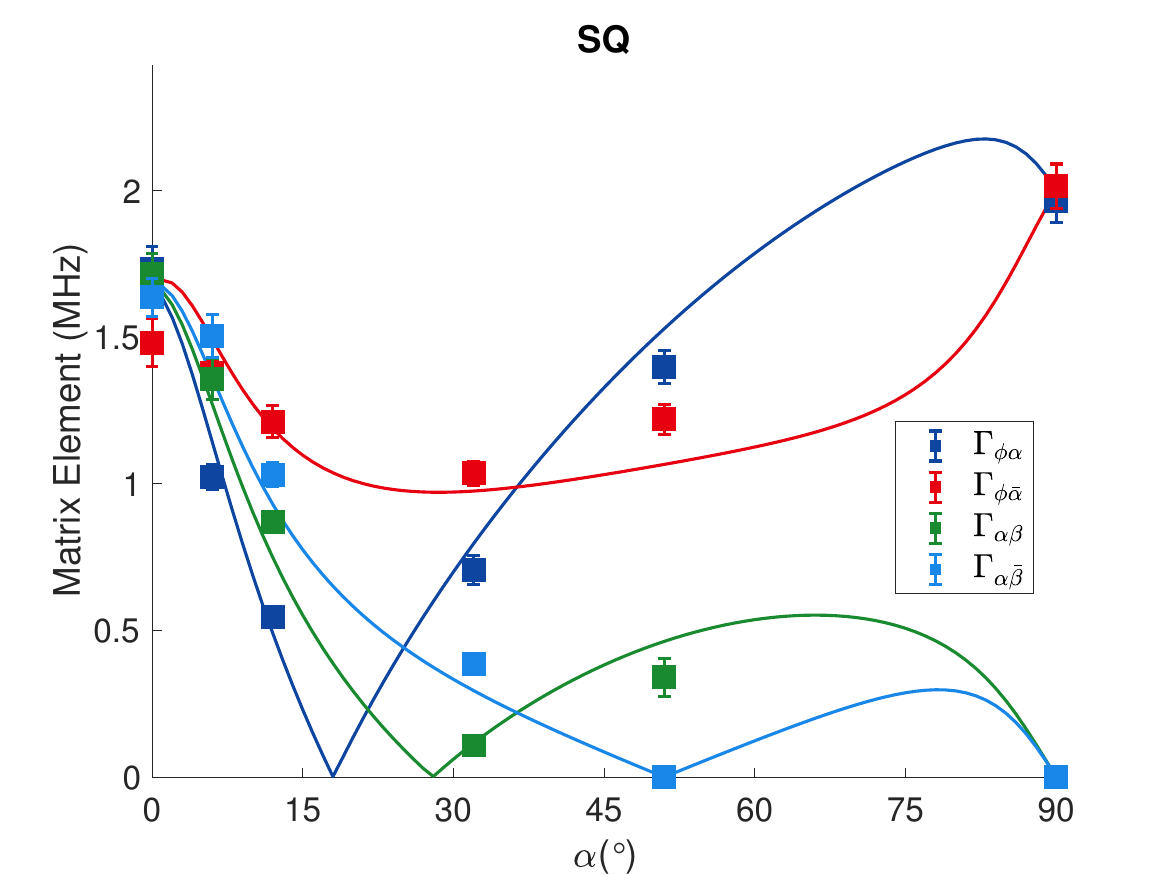}
    \includegraphics[width=0.32\textwidth]{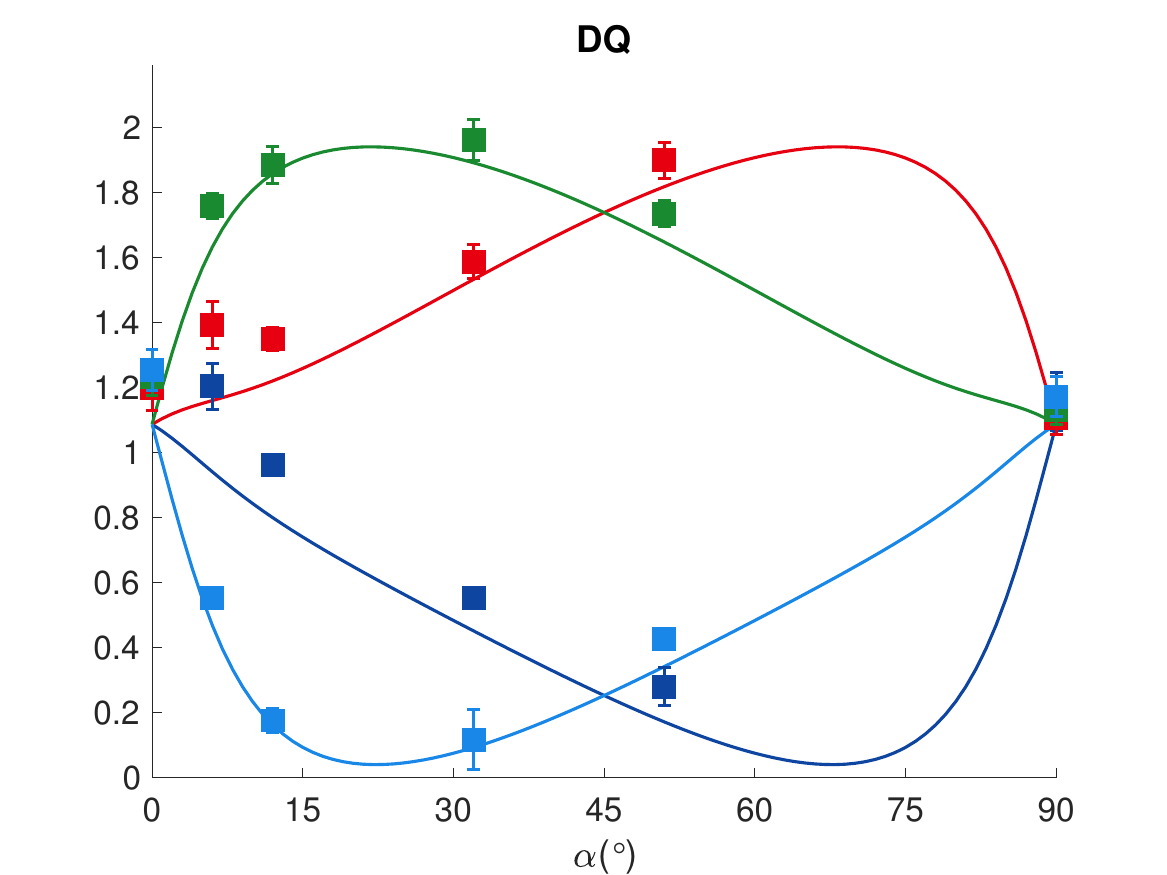}
    \includegraphics[width=0.32\textwidth]{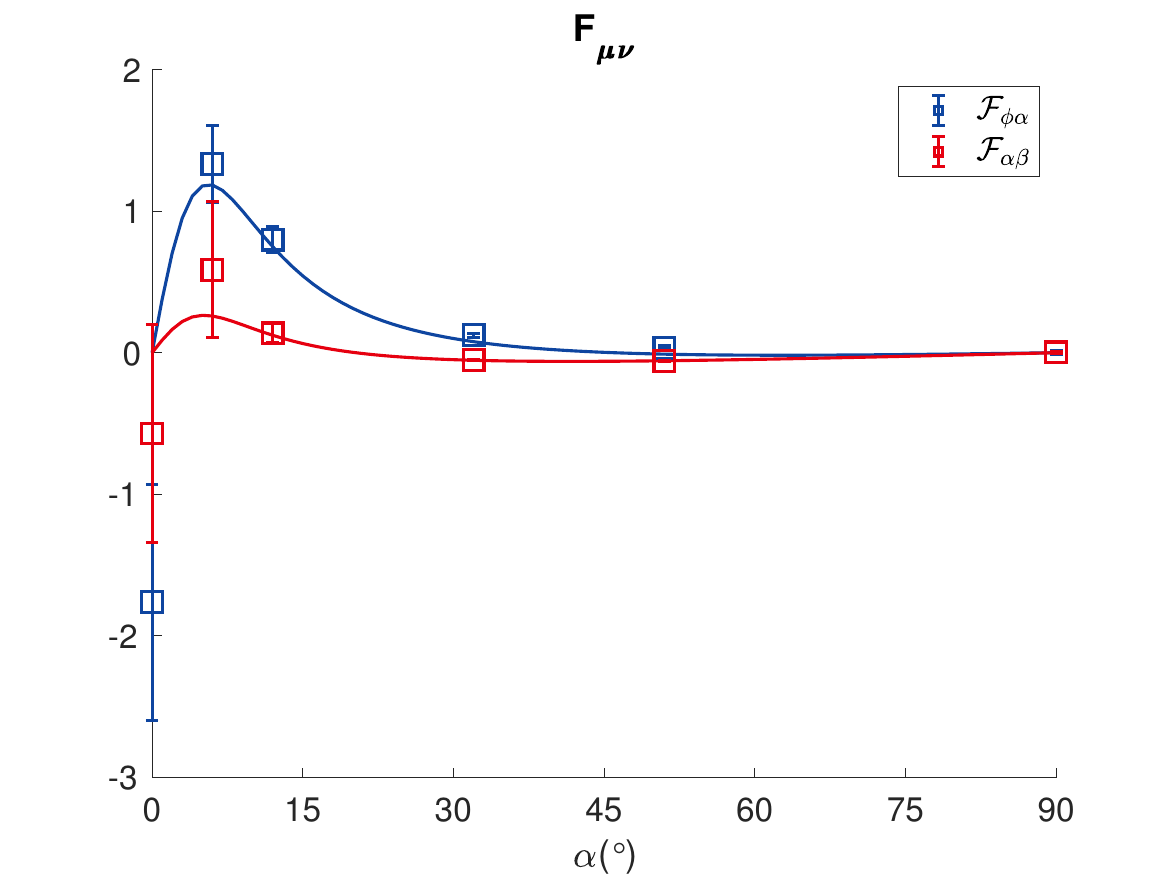}
    
    \caption{\textbf{Berry curvature measurements for $B_z/H_0=0.9\sqrt{2}$}. Matrix elements $\vert \Gamma_{-, 0}^{\mu(\nu)}\vert$ measured for SQ transitions at $\omega=\omega_r/2$ (left) and Matrix elements $\vert \Gamma_{-, +}^{\mu(\nu)}\vert$ measured for DQ transitions at $\omega=\omega_r$ (middle) using elliptical modulation. On the right we show the two relevant Berry curvatures as functions of $\alpha$. Squares are experimental data and solid lines are numerical simulations.}
    \label{fig:F_09}
\end{figure*}

\begin{figure*}[ht] 
    \centering
    \includegraphics[width=0.32\textwidth]{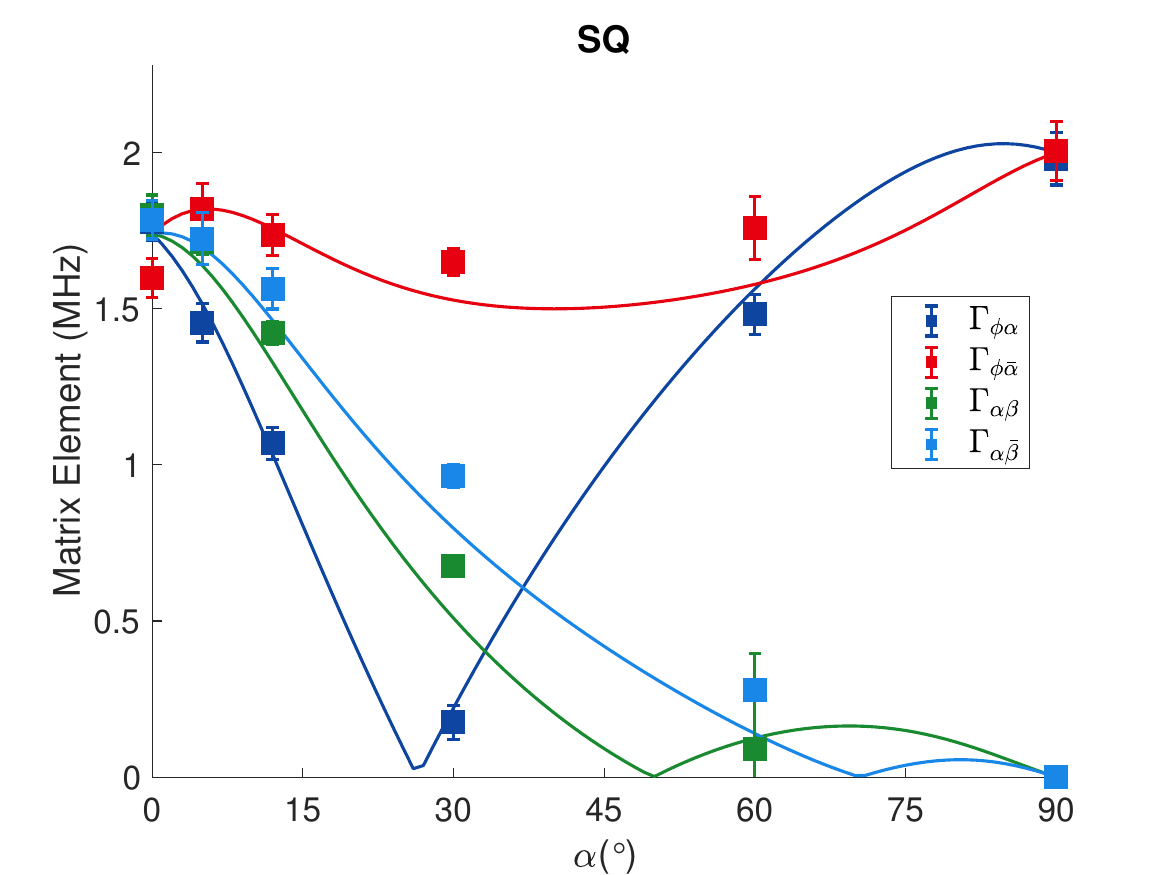}
    \includegraphics[width=0.32\textwidth]{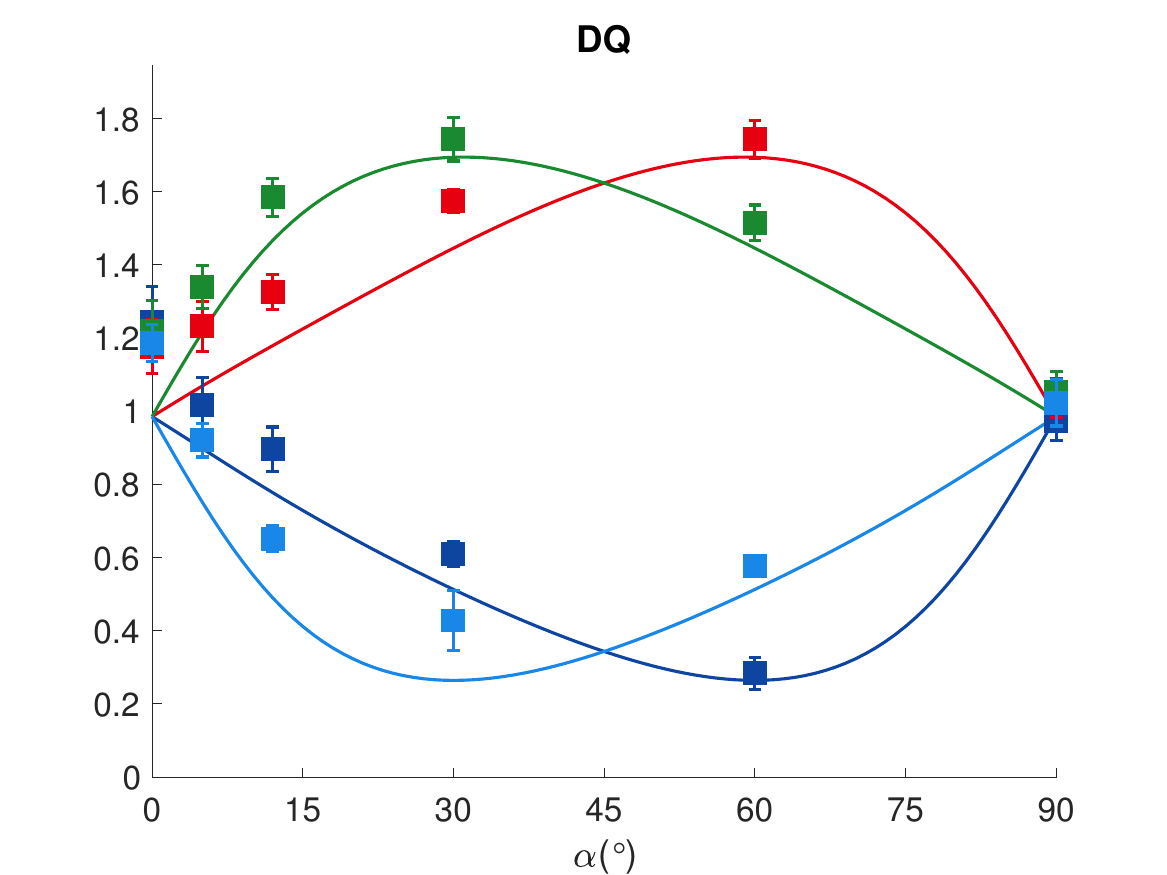}
    \includegraphics[width=0.32\textwidth]{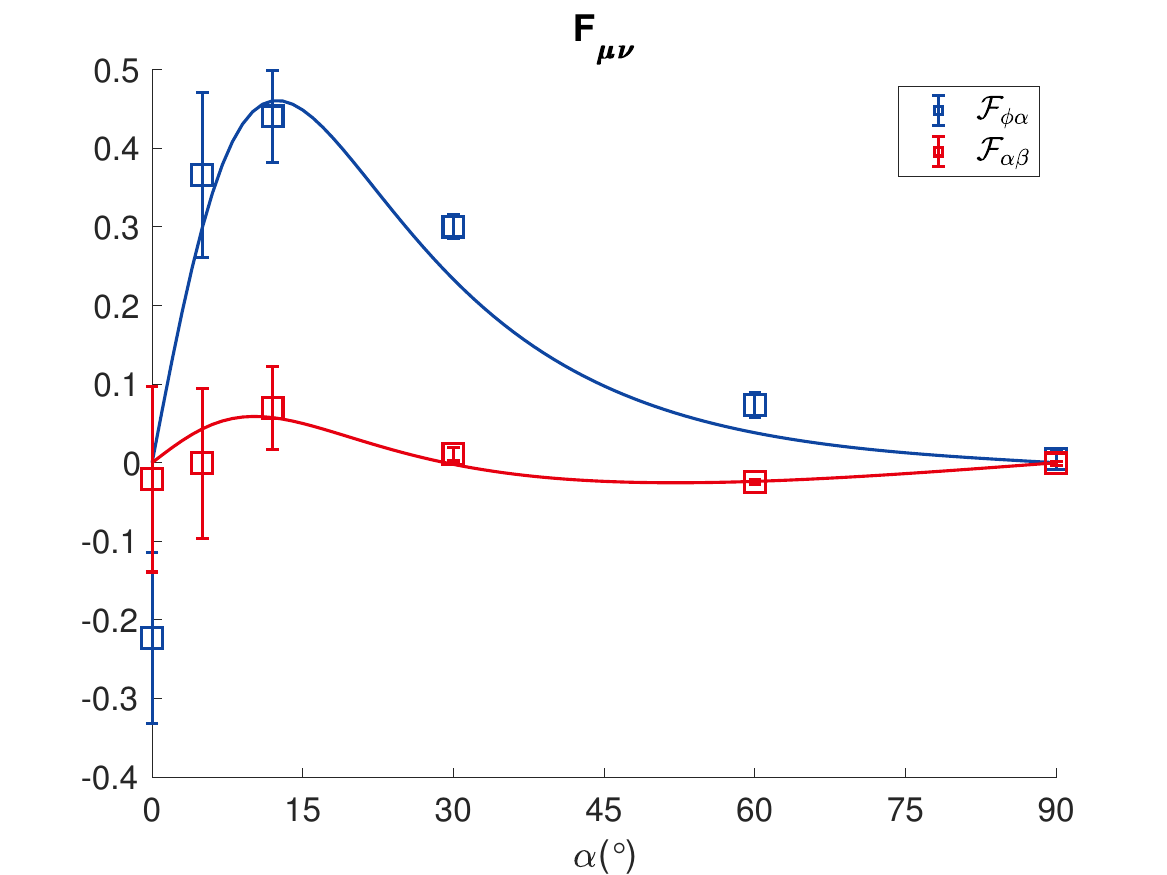}
    
    \caption{\textbf{Berry curvature measurements for $B_z/H_0=1.2\sqrt{2}$}. Matrix elements $\vert \Gamma_{-, 0}^{\mu(\nu)}\vert$ measured for SQ transitions at $\omega=\omega_r/2$ (left) and Matrix elements $\vert \Gamma_{-, +}^{\mu(\nu)}\vert$ measured for DQ transitions at $\omega=\omega_r$ (middle) using elliptical modulation. On the right we show the two relevant Berry curvatures as functions of $\alpha$. Squares are experimental data and solid lines are numerical simulations.}
    \label{fig:F_12}
\end{figure*}

\begin{figure*}[ht] 
    \centering
    \includegraphics[width=0.32\textwidth]{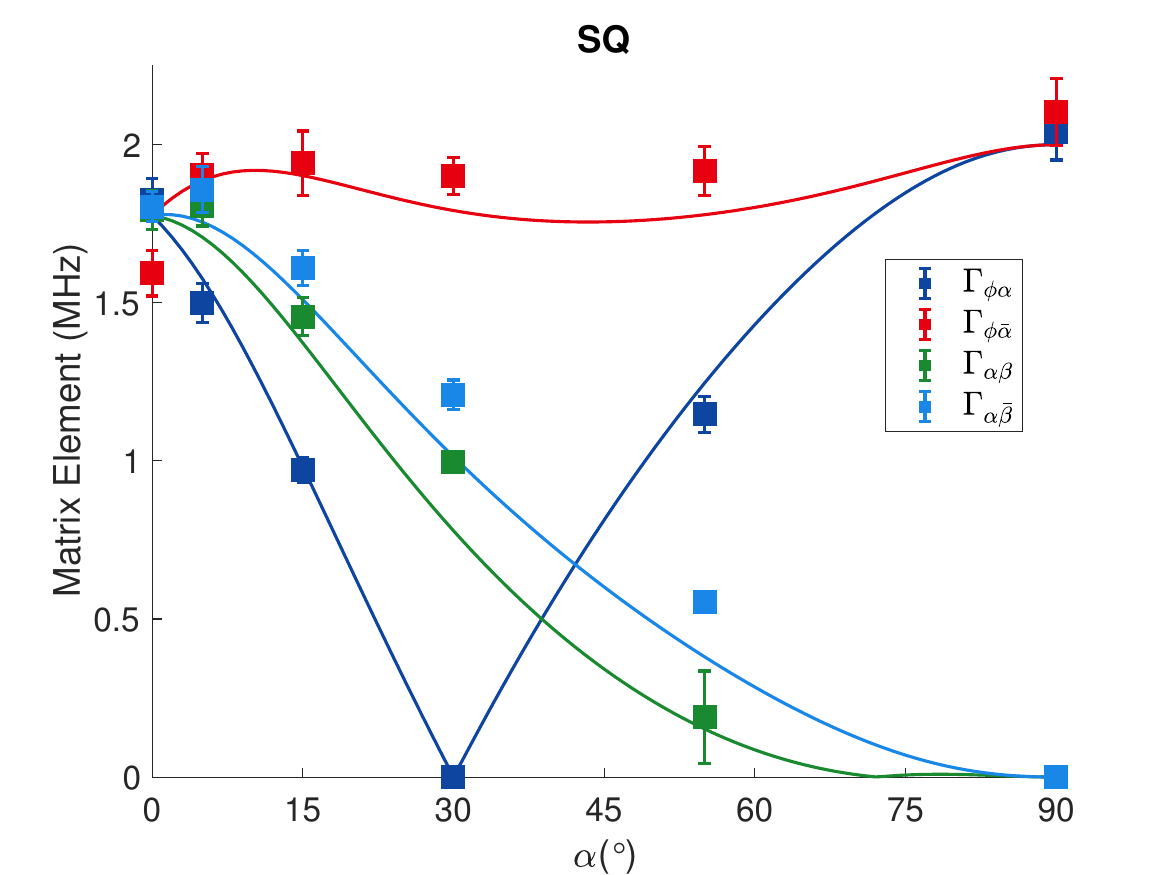}
    \includegraphics[width=0.32\textwidth]{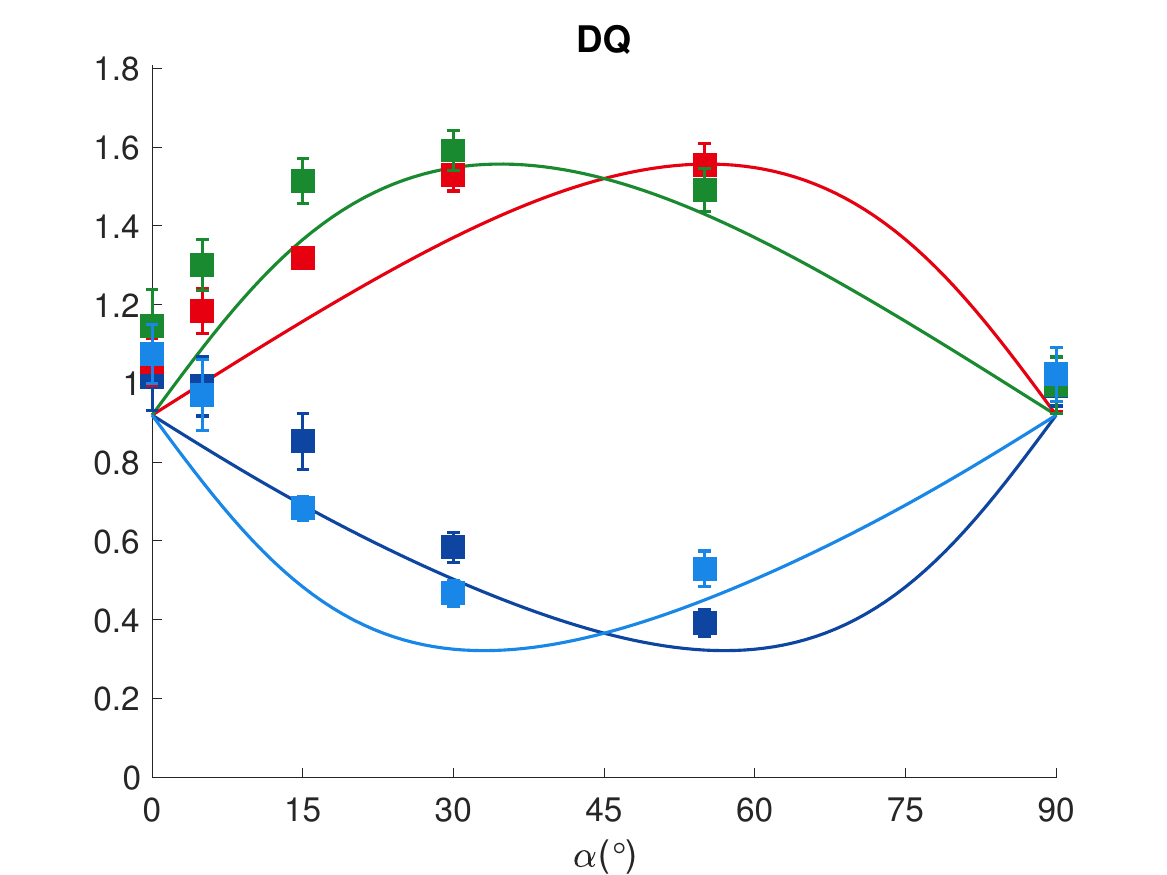}
    \includegraphics[width=0.32\textwidth]{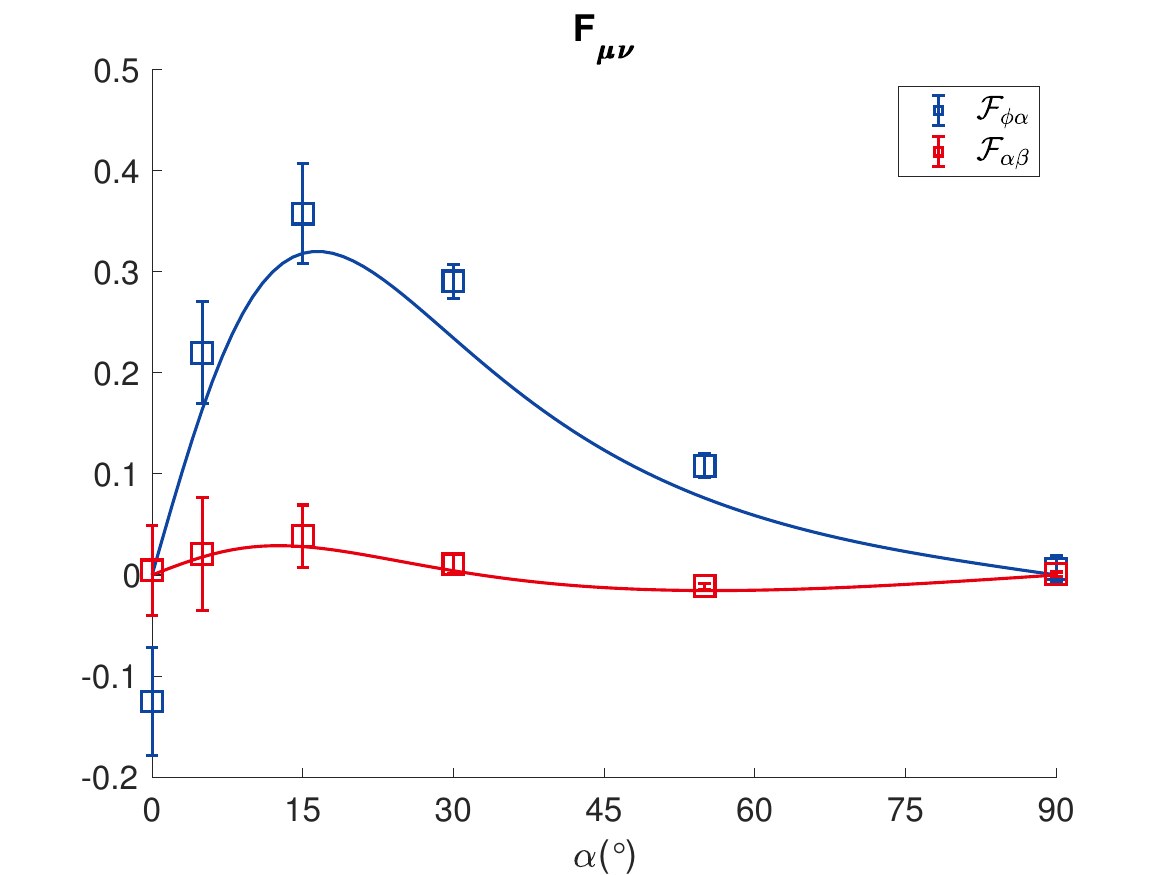}
    
    \caption{\textbf{Berry curvature measurements for $B_z/H_0=2$}. Matrix elements $\vert \Gamma_{-, 0}^{\mu(\nu)}\vert$ measured for SQ transitions at $\omega=\omega_r/2$ (left) and Matrix elements $\vert \Gamma_{-, +}^{\mu(\nu)}\vert$ measured for DQ transitions at $\omega=\omega_r$ (middle) using elliptical modulation. On the right we show the two relevant Berry curvatures as functions of $\alpha$. Squares are experimental data and solid lines are numerical simulations.}
    \label{fig:F_sqrt2}
\end{figure*}

\begin{figure*}[ht] 
    \centering
    \includegraphics[width=0.32\textwidth]{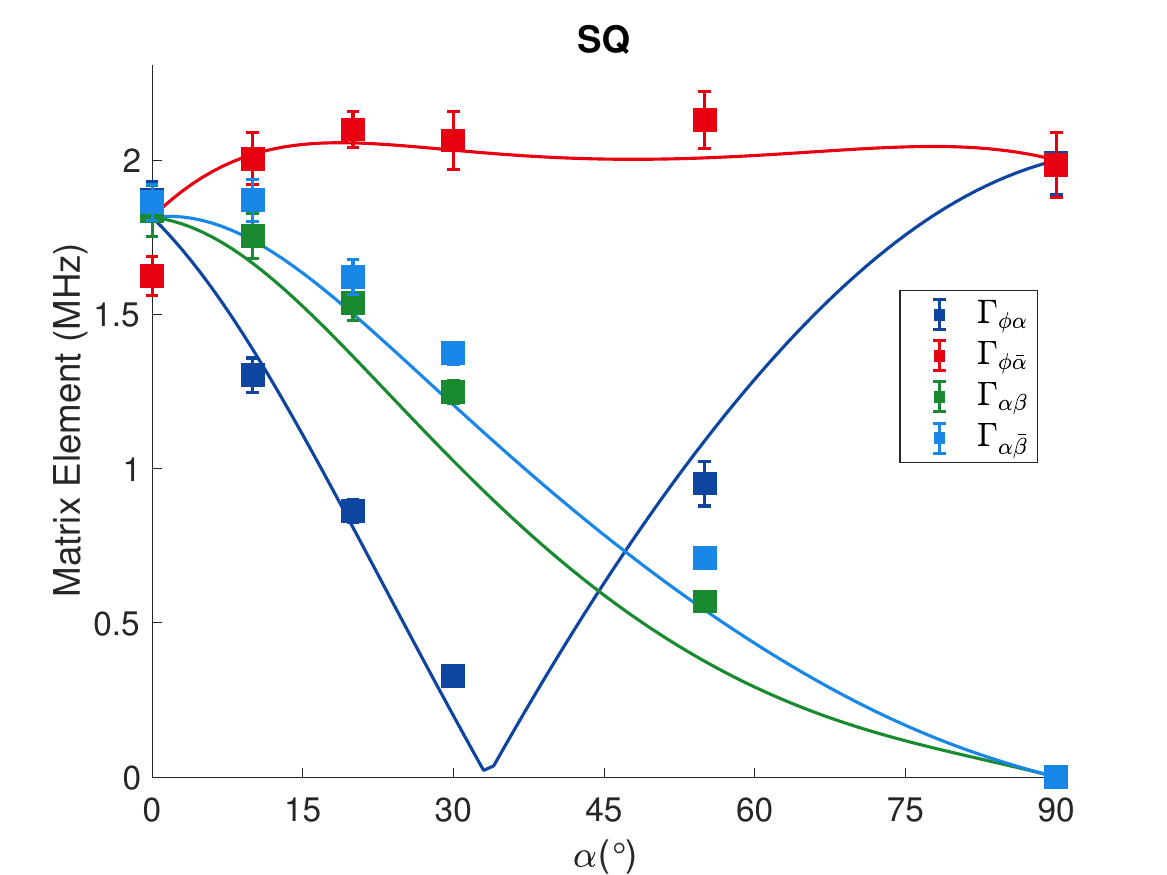}
    \includegraphics[width=0.32\textwidth]{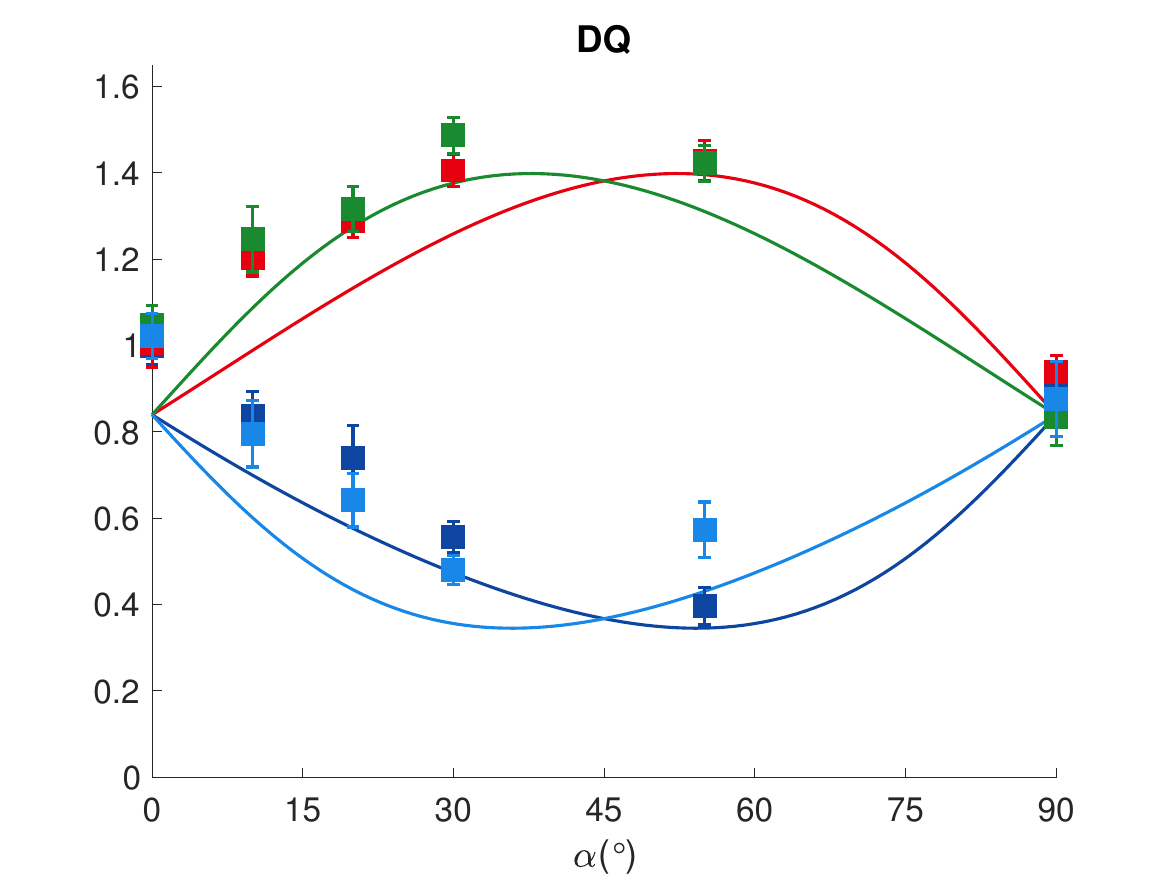}
    \includegraphics[width=0.32\textwidth]{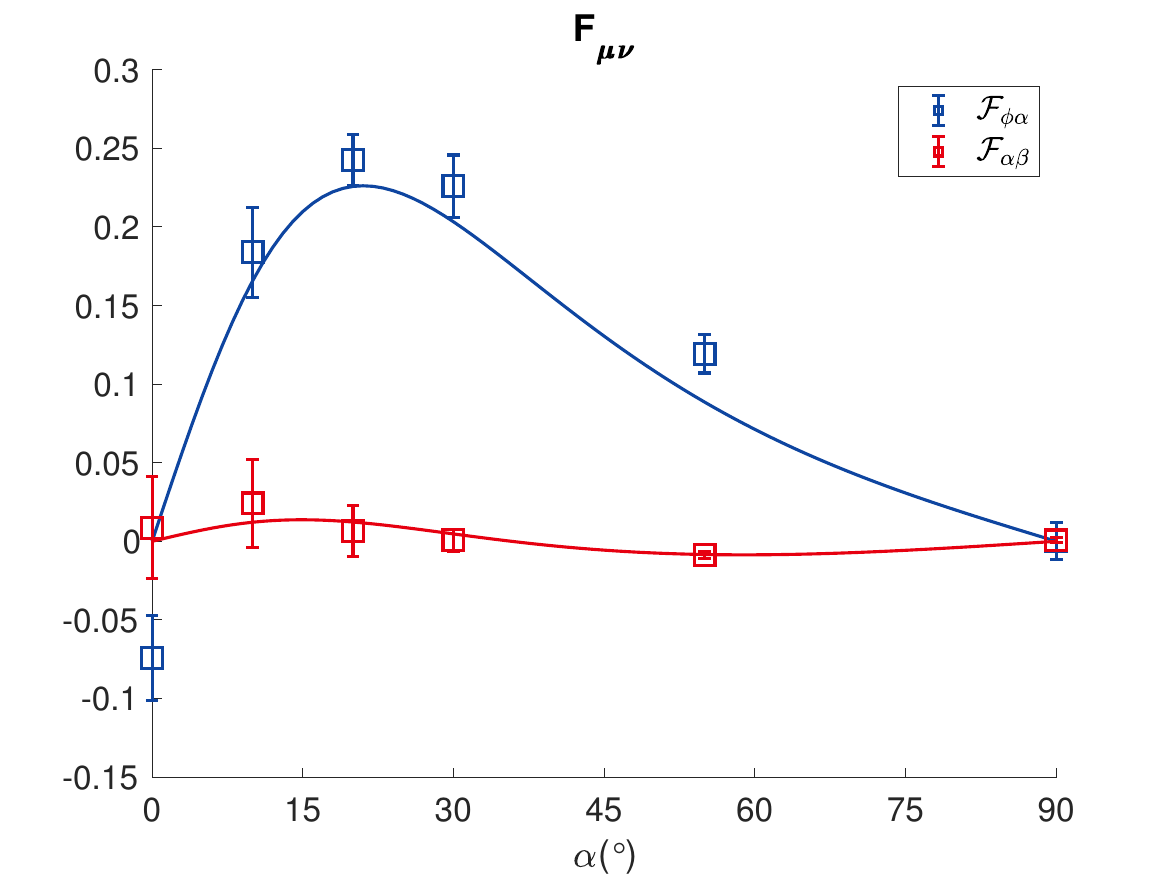}
    
    \caption{\textbf{Berry curvature measurements for $B_z/H_0=1.7\sqrt{2}$}. Matrix elements $\vert \Gamma_{-, 0}^{\mu(\nu)}\vert$ measured for SQ transitions at $\omega=\omega_r/2$ (left) and Matrix elements $\vert \Gamma_{-, +}^{\mu(\nu)}\vert$ measured for DQ transitions at $\omega=\omega_r$ (middle) using elliptical modulation. On the right we show the two relevant Berry curvatures as functions of $\alpha$. Squares are experimental data and solid lines are numerical simulations.}
    \label{fig:F_17}
\end{figure*}

While measuring the Berry connection generally involves quantum state tomography and is time-consuming, we note that thanks to the chosen parametrization for our particular Hamiltonian, measurement of $\mathcal B$ does not require state tomography, as we have shown in Section ~\ref{sec:analytical_DD}. To this end, we calculate and measure the observable $\mathcal{B}$ from the Berry connection using Eq.~\ref{eq:relation_H_B},~\ref{eq:connection_state_components},~\ref{eq:H_in_F}, as shown in yellow (analytical result) and red squares (experiment) in Fig.~3 of the main text.  The Berry curvature $\mathcal{F}$ is measured by the parametric modulation, similar to the measurement of the metric tensor, as discussed in previous sections. The experimentally measured $\mathcal B$ is shown in red squares in Fig.~3 of the main text, and additional experimental data relevant to the measurements are presented in Fig.~\ref{fig:F_035}-\ref{fig:F_17}.
$\mathcal{B}$ stays constant $\mathcal{B}=1$ when $B_z<H_0$. At $B_z=H_0$, the two degenerate nodal rings are at the boundary of our enclosed manifold, as indicated by a sharp change in $\mathcal{B}$ to $\mathcal{B}(B_z/H_0=1)=-1/3$, as expected.

\bibliography{SI}
\bibliographystyle{apsrev4-1}